\documentclass[preprint,floatfix,aps,prd,showpacs,nofootinbib,amsmath,amssymb,amsfonts,superscriptaddress]{revtex4-1}
\usepackage{bm}
\usepackage{graphicx,color}
\usepackage{physics}
\usepackage{dsfont}
\usepackage[normalem]{ulem}
\usepackage{hyperref,url}
\usepackage{mathrsfs}
\usepackage{calrsfs}
\usepackage{bbold}
\usepackage{subfig}

\usepackage{varioref}
\usepackage[active]{srcltx}

\expandafter\ifx\csname package@font\endcsname\relax\else
\expandafter\expandafter
\expandafter\usepackage
\expandafter\expandafter
\expandafter{\csname package@font\endcsname}%
\fi
\hypersetup{
	colorlinks=true,       % false: boxed links; true: colored links
	linkcolor=blue,          % color of internal links
	citecolor=blue,        % color of links to bibliography
}

\usepackage[section]{placeins}
\usepackage{enumitem}
\usepackage{fancybox}
\labelformat{figure}{Fig.~#1} 
\def\be {\begin{equation}}
\def\ee {\end{equation}}
\def\bea {\begin{eqnarray}}
\def\eea {\end{eqnarray}}
\def\f {\phi}
\def\la {\label}
\def\fr {\frac}
\def\le {\left}
\def\ri {\right}
\def\O  {\Omega}

\def\r  {\rho}
\def\pa {\partial}
\def\nn {\nonumber}
\def\th {\theta}

\newcommand{\Lim}[1]{\raisebox{0.5ex}{\scalebox{0.8}{$\displaystyle \lim_{#1}\;$}}}

\begin{document}
\title{Dynamical scaling symmetry and asymptotic quantum correlations for time-dependent scalar fields}

\author{S. Mahesh Chandran} 
\email{maheshchandran@iitb.ac.in}
\affiliation{Department of Physics, Indian Institute of Technology Bombay, Mumbai 400076, India}
\author{S. Shankaranarayanan}
\email{shanki@phy.iitb.ac.in}
\affiliation{Department of Physics, Indian Institute of Technology Bombay, Mumbai 400076, India}
%%%
\begin{abstract}
In time-independent quantum systems, entanglement entropy possesses an inherent scaling symmetry that the energy of the system does not have. The symmetry also assures that entropy divergence can be associated with the zero modes. We generalize this symmetry to time-dependent systems all the way from a coupled harmonic oscillator with a time-dependent frequency, to quantum scalar fields with time-dependent mass. We show that such systems have dynamical scaling symmetry that leaves the evolution of various measures of quantum correlations invariant --- entanglement entropy, GS fidelity, Loschmidt echo, and circuit complexity. Using this symmetry, we show that several quantum correlations are related at late-times when the system develops instabilities. We then quantify such instabilities in terms of scrambling time and Lyapunov exponents. The delayed onset of exponential decay of the Loschmidt echo is found to be determined by the largest inverted mode in the system. On the other hand, a zero-mode retains information about the system for a considerably longer time, finally resulting in a power-law decay of the Loschmidt echo. We extend the analysis to time-dependent massive scalar fields in $(1 + 1)-$dimensions and discuss the implications of zero-modes and inverted modes occurring in the system at late-times. We explicitly show the entropy scaling oscillates between the \emph{area-law} and \emph{volume-law} for a scalar field with stable modes or zero-modes. We then provide a qualitative discussion of the above effects for scalar fields in cosmological and black-hole space-times. 
\end{abstract}

\maketitle
\section{Introduction}

Entanglement, popularly measured in terms of von Neumann entropy, is a fundamental property that captures non-trivial correlations in interacting bipartite quantum systems~\cite{2008Amico.etalRev.Mod.Phys.,2009-Horodecki.etal-RMP,2009-Latorre.Riera-JPA,2010-Eisert.etal-Rev.Mod.Phys.,2010VieiraJournalofPhysicsConferenceSeries}. While the wave-function can be used to describe the system as a whole, quantum correlations preclude us from constructing a separate wave-function for each subsystem. Entanglement has been widely studied in literature over the last three decades, in simple quantum systems such as the hydrogen atom all the way to black holes. Some relevant applications of entanglement include diagnosing quantum chaos~\cite{2002-Bandyopadhyay.Lakshminarayan-PRL}, identifying signatures of quantum crossovers and phase transitions~\cite{2008Li.HaldanePhys.Rev.Lett.,2009-Latorre.Riera-JPA,2017-Kumar.Shankaranarayanan-SRep}, testing eigenstate thermalization~\cite{2018-Muralidharan.etal-PRE,2018-Byju.etal-Arxiv} understating the thermodynamics of space-time horizons~\cite{Das2010,2020Chandran.ShankaranarayananPhys.Rev.D}, and analyzing quantum fluctuations during cosmological inflation~\cite{2022-Martin.etal-JCAP}.

However, measuring entanglement entropy in quantum fields is problematic as it is divergent. It is cumbersome to extract valuable information about the quantum field without utilizing regularization, such as by employing an ultra-violet (UV) cutoff~\cite{2017-Unruh.Wald-ReportsonProgressinPhysics}. 
While this divergence is commonly attributed to the UV (high energy) limit, it was recently shown that there is a more general criterion for entropy divergence --- the generation of zero-modes~\cite{2014-Mallayya.etal-Phys.Rev.D,2019-Chandran.Shankaranarayanan-Phys.Rev.D,2020Chandran.ShankaranarayananPhys.Rev.D}. 
This is made possible through an inherent scaling symmetry of the entanglement entropy that connects the UV, and the IR (infrared)~\cite{2014-Mallayya.etal-Phys.Rev.D}. 
The current authors have shown that such a scaling symmetry exists in time-independent quantum systems such as the hydrogen atom to quantum fields in asymptotically flat and non-flat space-times~\cite{2014-Mallayya.etal-Phys.Rev.D,2019-Chandran.Shankaranarayanan-Phys.Rev.D,2020Chandran.ShankaranarayananPhys.Rev.D}.
The key advantage of mapping the entropy divergence 
to the occurrence of zero modes is isolating the divergence part from the non-divergence part. More importantly, in the rescaled variables, the entanglement entropy is not sensitive to UV physics.

Given these advantages, it is natural to ask whether the same scaling symmetry applies to time-dependent systems. Specifically, what new information does such a scaling symmetry provide about quantum correlations and divergence in time-dependent systems? Unlike time-independent systems,
the entanglement entropy of time-dependent systems has explicit time-dependence. This leads us to ask: Knowing the asymptotic behavior of entanglement entropy, can we infer the kind of instabilities in the system and help quantify them? This work addresses these questions by considering coupled harmonic oscillator with time-dependent frequency. We then extend the analysis to time-dependent Bosonic quadratic Hamiltonian and 2-D  scalar field.

%\red{While this was previously worked out for the time-independent case, its generalization to a dynamical quantum system will be discussed in this paper. As a consequence of this symmetry, understanding zero modes and their effects on quantum systems can directly help shed some light on the physics of quantum fields, both in the UV as well as IR.}

Entanglement dynamics of continuous-variable quantum systems are relatively less explored. They have become increasingly relevant over the last decade~\cite{2017Ho.AbaninPhys.Rev.B,2017Ghosh.etalEPLEurophysicsLetters,2018MakarovPhys.Rev.E,2021HabArrih.etalInternationalJournalofGeometricMethodsinModernPhysics,2022AndrzejewskiQuantumInformationProcessing}. The simplest models involve studying subsystem dynamics in response to a quench in the quantum system, wherein the free parameters of the global Hamiltonian are made time-dependent. Its wave-function is allowed to evolve unitarily with time. For a general bosonic quadratic Hamiltonian, the time evolution of entanglement entropy has been shown to take the following form~\cite{2018Hackl.etalPhys.Rev.A}:
\begin{equation}\label{rigol}
	S_{A}(t)=\Lambda_{A} t+C_{A} \ln (t)+X_{A}(t),
\end{equation}
where $\Lambda_A$ is a real number, $C_A$ is an integer, and $X_A$ is a bounded function. Whereas linear growth is a characteristic feature of unstable systems, logarithmic growth arises from metastable systems. Such instabilities can occur in a variety of scenarios, ranging from inverted harmonic oscillators to momentum modes of quantum fluctuations that exit the Hubble radius during cosmic inflation~\cite{2020Bhattacharyya.etalPhys.Rev.D}. It is also to be noted that while saturation of entropy occurs following a linear growth for finite systems, it can grow unbounded in systems with continuous degrees of freedom. Here, we show that the logarithmic growth of entanglement entropy is related to the appearance of zero-modes, and $C_A$ is related to the number of zero-modes..

As demonstrated in Eq. \eqref{rigol}, entropy can serve as a diagnostic tool for stability. However, it is still a single number and can not capture the processes in detail. This necessitates using alternative measures to quantify the instabilities in the system systematically. For example, while OTOCs (Out-of-Time-Order-Correlator) have been widely used to identify quantum chaos in various many-body systems~\cite{2018-Byju.etal-Arxiv}, measures like Loschmidt Echo~\cite{2001Jalabert.PastawskiPhys.Rev.Lett.,2006Gorin.etalPhysicsReports,Pal2022} and circuit complexity \cite{2017Jefferson.MyersJournalofHighEnergyPhysics,2020Ali.etalPhys.Rev.D,2019Ali.etalJournalofHighEnergyPhysics,Jaiswal2021} are more recent, valuable additions to the toolbox when it comes to measuring the scrambling time (how quickly information is disseminated throughout the system) and Lyapunov exponents (a measure of exponential sensitivity to initial conditions)~\cite{2020Bhattacharyya.etalPhys.Rev.D,Qu2021}. However, a key difference is that, unlike entanglement entropy, a subsystem quantity, the measures above are properties of the global wave-function. This brings us to the following question: Since all of these different quantities contain information about the quantum correlations in the systems, are these measures related to each other in the asymptotic limit? Interestingly, we demonstrate that asymptotically in time and in the presence of instabilities, all these measures are interrelated through simple expressions.

In this work, we study the subsystem dynamics of entanglement entropy in quadratic Hamiltonians, where the system initially in its ground state undergoes a quantum quench. We first generalize the scaling symmetry of entanglement entropy to dynamical systems. We then outline an analytical proof in the real-space for Eq.~\eqref{rigol} while also fixing the unknown coefficients for harmonic chains. Along with entanglement entropy, we also test the presence of both zero modes and unstable modes with the help of quantum fidelity, Loschmidt Echo, and circuit complexity of the evolving wave-function and obtain relations in the asymptotic limit. In the case of unstable modes, we can characterize both the scrambling time and Lyapunov exponents from the above measures. Finally, we also run numerical simulations of entanglement dynamics in a lattice-regularized quantum scalar field in $(1+1)-$dimensions upon performing (i) a global mass quench and (ii) a boundary condition quench. In both cases, the quench results in ``entanglement ripples" traveling throughout the system, whereas in mass quenches we further observe subsystem scaling of entanglement featuring \emph{area-law to volume-law} oscillations. 

Quantum fields in time-dependent backgrounds are ideal settings to study these effects. Especially in gravity, there are two settings where quantum physics at short distances (high energies) influences the physics at long distances (low energies). These are: (i) Cosmological inflation, where the putative exponential expansion of the very early ($\approx 10^{-34}$ seconds after the Big-Bang) and very small ($\approx 10^{-26}$ m) Universe causes quantum effects in that epoch to show up in current
observations, such as the Cosmic Microwave Background Radiation (CMBR)~\cite{1984-Kodama.Sasaki-PTPS,1992-Mukhanov.etal-PRep}, and
(ii) Black Holes, where outgoing low-energy quantum modes from the
horizon evolve from high energy modes due to high gravitational
red-shifts~\cite{1973-Bekenstein-Phys.Rev.D,1975-Hawking-CMP}. Such exotic physical scenarios can in principle be studied by looking at the global quench dynamics of the massive scalar field in corresponding background space-times.

The boundary quench is useful in simulating the Dynamical Casimir effect (DCE)~\cite{1989YablonovitchPhys.Rev.Lett.,1997Golestanian.KardarPhys.Rev.Lett.}, which serves as a heuristic model for Hawking radiation, the Unruh effect, and various other phenomena~\cite{1996DaviesNature}. The DCE is brought about by moving mirrors in the vacuum that leads to a dissipative force on the plate, resulting in field excitations~\cite{Dalvit2011,1991Braginsky.KhaliliPhysicsLettersA,1992Jackel.ReynaudQuantumOpticsJournaloftheEuropeanOpticalSocietyPartB}. 
Alternatively, DCE can be simulated by fixing the boundary and switching to a time-dependent Robin boundary condition instead~\cite{2011Silva.FarinaPhys.Rev.D,2012FARINA.etalInternationalJournalofModernPhysicsConferenceSeries}. Upon implementing this, the ensuing entanglement dynamics help capture the signatures of DCE in massive scalar fields.

The paper is organized as follows: In Section \ref{model} we introduce the model and the quantifying tools employed. We generalize the scaling symmetry of entanglement entropy to dynamical systems and show how the late-time behavior of these systems can be used to understand quantum correlations. We numerically obtain the entanglement entropy and fidelity at all times and show that the analytical results for the asymptotic limit match the numerical results. In Section \ref{sec:scalarfielddyn}, we apply the dynamical scaling symmetry in lattice-regularized time-dependent scalar field theories and discuss the difference in the late-time correlations due to boundary quench and mass quench. In Section \ref{sec:conc}, we conclude by discussing the physical interpretations of this crossover, as well as directions for future research. Throughout this work, we use natural units $\hbar=c=1$.

\section{Dynamical scaling symmetry and entanglement entropy}\label{model}

In theory, the time-dependent Schroedinger equation~\cite{1967LewisPhys.Rev.Lett.,2009-Campbell-JPA}
\begin{equation}
i \hbar \frac{\partial \Psi(x,t)}{\partial t} = H(t) \Psi(x,t)    
\end{equation}
can be solved using the time evolution operator given formally by 
\begin{equation}
\label{eq:UnitaryOper}
U\left(t, t_{0}\right)=\hat{T}\left(\exp \left(-\frac{i}{\hbar} \int_{t_{0}}^{t} H\left(t^{\prime}\right) d t^{\prime}\right)\right)
\end{equation}
where $\hat{T}$ is the time ordering operator which orders operators with larger times to the left. This unitary operator takes a state at time $t_{0}$ to a state at time $t$ so that:
$$
\Psi(x, t)=U\left(t, t_{0}\right) \Psi\left(x, t_{0}\right)
$$
However, explicit construction of \eqref{eq:UnitaryOper} is rarely possible. Nevertheless, it is possible to obtain the time evolution operator for a specific form of a quadratic Hamiltonian~\cite{1999-Song-JPA,1999-Song-PRA,2009-Campbell-JPA}. Since our goal is to quantify quantum correlations in field-theoretic systems, in this section, we will begin by focusing our attention on the two-coupled harmonic oscillators (CHO) with time-dependent frequency. 
The Hamiltonian of this system is:
\begin{equation}
	\label{eq:CHO-Hamil}
	\tilde{H}(\tilde{t}) =\frac{\tilde{p}_1^2}{2}+\frac{\tilde{p}_2^2}{2}+\frac{1}{2}\tilde{\omega}^2(\tilde{t})\left(\tilde{x}_1^2+\tilde{x}_2^2\right)+\frac{\tilde{\alpha}^2}{2}\left(\tilde{x}_1-\tilde{x}_2\right)^2 \, ,
\end{equation}
where $\tilde{\omega}(\tilde{t})$ is the time-dependent frequency and $\tilde{\alpha}$ is the coupling constant. Here, we have used tildes to represent dimensionfull variables and parameters, to distinguish them from their dimensionless counterparts that will appear in subsequent sections. 
Under the transformations $\tilde{x}_{\pm}=(\tilde{x}_1\pm \tilde{x}_2)/\sqrt{2}$, the above Hamiltonian reduces to:
\begin{equation}
	\label{eq:CHO-Hamil02}
	\tilde{H}(\tilde{t}) =\frac{\tilde{p}_+^2}{2}+\frac{\tilde{p}_-^2}{2}+\frac{1}{2}\tilde{\omega}_+^2(\tilde{t})\tilde{x}_+^2+\frac{1}{2}\tilde{\omega}_-^2(\tilde{t})\tilde{x}_-^2,
\end{equation}
where the time-dependent normal modes are:
\begin{equation}
	\tilde{\omega}_-(\tilde{t})=\sqrt{\tilde{\omega}^2(\tilde{t})+2\tilde{\alpha}^2}; \qquad  \tilde{\omega}_+(\tilde{t})=\tilde{\omega}(\tilde{t}).
\end{equation}
We solve the time-dependent Schrodinger equation for each uncoupled oscillator as elaborated in Appendix \ref{App:GaussianState}. For this, we consider the form-invariant Gaussian state, which evolves from an initial ground state (GS) at time $t=0$ and develops excitations in the instantaneous eigenbasis defined at each time-slice. Such a solution takes the form \cite{2008LoheJournalofPhysicsAMathematicalandTheoretical}:
\begin{equation}\label{GS}
\tilde{\Psi}_{\rm GS}(\tilde{x}_+,\tilde{x}_-,t)=\prod_{j=\{+,-\}}\left(\frac{\tilde{\omega}_j(0)}{\pi \tilde{b}_j^2(\tilde{t})}\right)^{1/4}\exp{-\left(\frac{\tilde{\omega}_j(0)}{\tilde{b}_j^2(\tilde{t})}-i\frac{\dot{\tilde{b}}_j(\tilde{t})}{\tilde{b}_j(\tilde{t})}\right)\frac{\tilde{x}_j^2}{2}-\frac{i}{2}\tilde{\omega}_j(0)\tilde{\tau}_j(\tilde{t})},
\end{equation}
where $\tilde{\tau}_j =\int \tilde{b}_j^{-2}(\tilde{t}) dt$. The scaling parameters $\tilde{b}_j$ are solutions of the non-linear Ermakov-Pinney equation \cite{Pinney_1950,1967LewisPhys.Rev.Lett.,1968LewisJournalofMathematicalPhysics,2008LoheJournalofPhysicsAMathematicalandTheoretical}:
\begin{equation}\label{ermakov}
	\ddot{\tilde{b}}_j(\tilde{t})+\tilde{\omega}_j^2(\tilde{t})\tilde{b}_j(\tilde{t})=\frac{\tilde{\omega}_j^2(0)}{\tilde{b}_j^3(\tilde{t})}
\end{equation}
Note that $\tilde{b}_j(\tilde{t})$ is non-zero at all times~\cite{1991Nunez.etalJournalofPhysicsAMathematicalandGeneral,2006FernandoBarberoG..etalPhys.Rev.D}, 
and in the time-independent limit $\tilde{\omega}(\tilde{t}) \to \tilde{\omega}$, we see that $\tilde{b}_j=1$ and $\dot{\tilde{b}}_j=0$. Thus, in the time-independent limit $\tilde{\tau}_j = t$. Also, $\tilde{b}_j(\tilde{t})$ is related to the classical-time dependent oscillator solution as follows~\cite{2008LoheJournalofPhysicsAMathematicalandTheoretical}:
\begin{equation}
    \ddot{f}+\tilde{\omega}_j^{2}(\tilde{t}) f=0\quad;\quad \tilde{b}_j^{2}(\tilde{t}) =\tilde{\omega}_j(0)\left\{f_{1}^{2}+W^{-2} f_{2}^{2}\right\},
\end{equation}
where $f_1$, $f_2$ are linearly independent solutions of the harmonic oscillator with frequency $\tilde{\omega}_{+}(\tilde{t})$ (or $\tilde{\omega}_-(\tilde{t})$)  and the Wronskian $W=f_1\dot{f}_2-\dot{f}_1f_2$ is a non-zero constant. The solution $\tilde{b}_j(\tilde{t})$ is crucial in constructing the class of invariants (known as Lewis invariants) corresponding to a time-dependent oscillator system~\cite{1967LewisPhys.Rev.Lett.,1968LewisJournalofMathematicalPhysics}, whose eigenvalues are time-independent and evenly spaced~\cite{2008LoheJournalofPhysicsAMathematicalandTheoretical}.

%(\textbf{to delete. clearer discussion in Appendix A})\red{It should be noted that, in general, at an instantaneous time 
%the quantum state of the individual Harmonic oscillator can be decomposed as $\ket{\tilde{\Psi}(\tilde{t})}=\sum_{n}c_{n}(\tilde{t})\ket{n(\tilde{t})}$, i.e., the basis states keep changing with time. As a result, the ground state at $t=0$ does not remain as a ground state in the new basis at a later time, depending on the nature of the time evolution of $\tilde{H}(\tilde{t})$. According to the quantum adiabatic theorem~\cite{2005-Wu.Yang-PRA}, for slow-changing $\tilde{H}(\tilde{t})$, if the system begins close to an eigenstate, it remains close to an eigenstate. In our case, we assume that the system remains in the instantaneous ground state of the time-evolved quantum system. Hence it is sufficient to use the form given in \eqref{GS} to calculate the density matrix at all time slices.}

Recently, the authors have shown that entanglement entropy of various time-independent systems --- CHO, the scalar field in $(1 + 1)-$dimensions, and scalar fields in black-hole space-times --- is invariant under a scaling transformation even though the Hamiltonian is not~\cite{2014-Mallayya.etal-Phys.Rev.D,2019-Chandran.Shankaranarayanan-Phys.Rev.D,2020Chandran.ShankaranarayananPhys.Rev.D,2021Jain.etalPhys.Rev.D}. We generalize the scaling relations to time-dependent systems. Moreover, we explicitly show that the presence of zero-modes corresponds to the divergence entanglement entropy also in time-dependent systems. 

In the rest of this section, we define two quantifying tools for quantum correlations --- entanglement entropy and quantum fidelity --- and use the generalized scaling symmetry to relate the presence of the zero-modes to divergent entanglement entropy for time-dependent systems.
%We also evaluate these four quantifying tools for the two-coupled time-dependent harmonic oscillator Hamiltonian \eqref{eq:CHO-Hamil}.

\subsection{Entanglement entropy and Fidelity}

Like in the case of time-independent CHO~\cite{1993-Srednicki-Phys.Rev.Lett.,2006-Ahmadi.etal-CJP}, to evaluate the entanglement entropy, we must first calculate the reduced density matrix (RDM) of the system~\cite{1993-Srednicki-Phys.Rev.Lett.,2017Ghosh.etalEPLEurophysicsLetters}:
\begin{align}
	\rho_2(\tilde{x}_2,\tilde{x}_2')&=\int d\tilde{x}_1 \tilde{\Psi}_{\rm GS}^*(\tilde{x}_1,\tilde{x}_2')\tilde{\Psi}_{\rm GS}(\tilde{x}_1,\tilde{x}_2)\nonumber\\
	&=\left(\frac{\tilde{\omega}_+(0)\tilde{\omega}_-(0)}{2\pi \tilde{b}_+^2(\tilde{t})\tilde{b}_-^2(\tilde{t})\Re(A)}\right)^{1/2}\exp{-\frac{\gamma}{2}\left(\tilde{x}_2^2+\tilde{x}_2'^2\right)+i\frac{\delta}{2}\left(\tilde{x}_2^2-\tilde{x}_2'^2\right)+\beta \tilde{x}_2\tilde{x}_2'},
\end{align}
where 
\begin{align}
	A&=\frac{1}{4}\left[\frac{\tilde{\omega}_+(0)}{\tilde{b}_+^2(\tilde{t})}+\frac{\tilde{\omega}_-(0)}{\tilde{b}_-^2(\tilde{t})}-i\left(\frac{\dot{\tilde{b}}_+(\tilde{t})}{\tilde{b}_+(\tilde{t})}+\frac{\dot{\tilde{b}}_-(\tilde{t})}{\tilde{b}_-(\tilde{t})}\right)\right]\nonumber\\
	B&=\frac{1}{4}\left[\frac{\tilde{\omega}_+(0)}{\tilde{b}_+^2(\tilde{t})}-\frac{\tilde{\omega}_-(0)}{\tilde{b}_-^2(\tilde{t})}+i\left(\frac{\dot{\tilde{b}}_+(\tilde{t})}{\tilde{b}_+(\tilde{t})}-\frac{\dot{\tilde{b}}_-(\tilde{t})}{\tilde{b}_-(\tilde{t})}\right)\right]\nonumber\\
	\gamma&=2\Re(A)-\left(\frac{\Re(B)^2-\Im(B)^2}{\Re(A)}\right)\nonumber\\
	\beta&=\frac{\abs{B}^2}{\Re(A)}\nonumber\\
	\delta&=2\Im(A)-\frac{2\Re(B)\Im(B)}{\Re(A)} \, .
\end{align}
Since both the harmonic oscillators have the same frequency dependence \eqref{eq:CHO-Hamil}, the functional form of RDM evaluated by integrating over $\tilde{x}_1$ or $\tilde{x}_2$ is the same, and leads to an identical spectrum. However, it should be 
noted that unlike the time-independent case, the RDM is not 
symmetric in $\tilde{x}_2$ and $\tilde{x}_2'$. This is because $\delta$ vanishes for the time-independent CHO. The eigenvalues of the RDM at an instantaneous time can be obtained by solving the following integral equation~\cite{1993-Srednicki-Phys.Rev.Lett.,2006-Ahmadi.etal-CJP}:
\begin{equation}
	\int d\tilde{x}'_2 \, \rho_2(\tilde{x}_2,\tilde{x}_2')f_n(\tilde{x}_2')=p_nf_n(\tilde{x}_2) \, .
\end{equation}
The solution for the above integral equation is~\cite{2017Ghosh.etalEPLEurophysicsLetters}:
\begin{align}
	f_n(x)&=\frac{1}{\sqrt{2^n n!}}\left(\frac{\epsilon}{\pi}\right)^{1/4}H_n(\sqrt{\epsilon}\tilde{x})\exp{-\left(\epsilon+i\delta\right)\frac{\tilde{x}^2}{2}}\nonumber\\
	\epsilon&=\sqrt{\gamma^2-\beta^2}\nonumber\\
	p_n&=\left(1-\tilde{\xi}(\tilde{t})\right)\tilde{\xi}^n(\tilde{t})\\
	\tilde{\xi}(\tilde{t})&=\frac{\beta}{\gamma+\epsilon}=\frac{\sqrt{\left(\frac{\tilde{\omega}_+(0)}{\tilde{b}_+^2(\tilde{t})}+\frac{\tilde{\omega}_-(0)}{\tilde{b}_-^2(\tilde{t})}\right)^2+\left(\frac{\dot{\tilde{b}}_+(\tilde{t})}{\tilde{b}_+(\tilde{t})}-\frac{\dot{\tilde{b}}_-(\tilde{t})}{\tilde{b}_-(\tilde{t})}\right)^2}-2\sqrt{\frac{\tilde{\omega}_+(0)\tilde{\omega}_-(0)}{\tilde{b}_+(\tilde{t})\tilde{b}_-(\tilde{t})}}}{\sqrt{\left(\frac{\tilde{\omega}_+(0)}{\tilde{b}_+^2(\tilde{t})}+\frac{\tilde{\omega}_-(0)}{\tilde{b}_-^2(\tilde{t})}\right)^2+\left(\frac{\dot{\tilde{b}}_+(\tilde{t})}{\tilde{b}_+(\tilde{t})}-\frac{\dot{\tilde{b}}_-(\tilde{t})}{\tilde{b}_-(\tilde{t})}\right)^2}+2\sqrt{\frac{\tilde{\omega}_+(0)\tilde{\omega}_-(0)}{\tilde{b}_+(\tilde{t})\tilde{b}_-(\tilde{t})}}} \nonumber
\end{align}

For the instantaneous GS wave-function subject to an adiabatic evolution of $\tilde{H}(\tilde{t})$~\cite{2005-Wu.Yang-PRA}, the entanglement entropy is calculated as follows: 
\begin{equation}
\label{eq:CHO-ent1}
	\tilde{S}(\tilde{t})=-\sum_n p_n\log{p_n}=-\log{[1-\tilde{\xi}(\tilde{t})]}-\frac{\tilde{\xi}(\tilde{t})}{1-\tilde{\xi}(\tilde{t})}\log{\tilde{\xi}(\tilde{t})},
\end{equation}
%\textcolor{red}{Can you plot entanglement entropy for a simple variation of $\omega(\tilde{t}) = \omega_0 t$?}
It is important to note that the entanglement entropy \emph{only} depends on time as the eigenvalues are time-dependent. The above formalism can in fact be extended to a system of $N$ time-dependent oscillators. See Appendix \eqref{App:A} for details. %One of the features of entanglement in time-dependent quantum systems is that the correlations are asymmetric with respect to the bipartition. In Appendix \eqref{App:B}, we explicitly show this for the three harmonic oscillator case. To our knowledge, this has not been shown in literature, and we discuss its consequences for $N$-oscillators in Section \ref{sec:scalarfielddyn}. 

Fidelity (or overlap function) can be used to determine the extent of the time evolution of a quantum state~\cite{2006-Zanardi.Paunkovic-PRE,2017-Kumar.Shankaranarayanan-SRep}. The overlap between the initial and final states during the evolution is
\begin{equation}
\mathcal{F}_0(\tilde{t})=\abs{\braket{\tilde{\Psi}(0)}{\tilde{\Psi}(\tilde{t})}} \, .
\end{equation}
For the ground state of system, this can be calculated to be:
\begin{equation}
\label{eq:overlapCHO}
	\mathcal{F}_0(\tilde{t})=2\prod_{n=+,-}\sqrt{\frac{2\tilde{\omega}_n(0)}{\tilde{b}_n(\tilde{t})\left[\tilde{\omega}_n^2(0)\left(1+\frac{1}{\tilde{b}_n^2(\tilde{t})}\right)^2+\frac{\dot{\tilde{b}}_n^2(\tilde{t})}{\tilde{b}_n^2(\tilde{t})}\right]}}=\prod_{n}\mathscr{F}_0^{(n)}(t)
\end{equation}

\subsection{Dynamical Scaling Symmetry and its consequences}\label{sec:scaling}

In the time-independent case, the entanglement entropy was shown to have an inherent scaling symmetry that the Hamiltonian of the system did not have~\cite{2014-Mallayya.etal-Phys.Rev.D,2019-Chandran.Shankaranarayanan-Phys.Rev.D,2020Bhattacharyya.etalPhys.Rev.D}. 
Upon rescaling the Hamiltonian by a constant factor, the entropy remained invariant, whereas the Hamiltonian did not. Such a rescaling is convenient as it reduces the number of independent parameters in the Hamiltonian and allows us to probe the cause of entropy divergence. For instance,  the divergence of entanglement entropy can be attributed to the occurrence of zero modes in the scalar field. Here, we generalize the idea to account for a time-dependent system under similar transformations and assess its consequences.

Let us rescale the Hamiltonian \eqref{eq:CHO-Hamil} w.r.t the coupling constant $\tilde{\alpha}$, i. e. 
\begin{equation}
	\label{eq:CHO-Hamil03}
H(t) = \frac{\tilde{H}(\tilde{t})}{\tilde{\alpha}} = \frac{\tilde{p}_1^2}{2 \tilde{\alpha}}+\frac{\tilde{p}_2^2}{2 \tilde{\alpha}}+ \frac{\tilde{\omega}^2(\tilde{t})}{2 \tilde{\alpha}}\left(\tilde{x}_1^2+\tilde{x}_2^2\right)+\frac{\tilde{\alpha}}{2}\left(\tilde{x}_1-\tilde{x}_2\right)^2 \, ,
\end{equation}
On performing the canonical transformations
\[
p_i=\tilde{\alpha}^{-1/2}\tilde{p}_i,\quad x_i=\tilde{\alpha}^{1/2}\tilde{x}_i \,~~\mbox{where}~~i = 1, 2 \, ,
\]
the rescaled Hamiltonian ($H$), {which is now dimensionless}, can be brought to canonical form:
\begin{equation}\label{cho1}
H(t) =\frac{1}{2}\left\{p_1^2+p_2^2+\Lambda(t)\left(x_1^2+x_2^2\right)+\left(x_1-x_2\right)^2\right\}~;\quad \Lambda(t)=\frac{\tilde{\omega}^2(\tilde{t})}{\tilde{\alpha}^2}.
\end{equation}
{It is to be noted that rescaled time $t$ corresponding to the rescaled Hamiltonian is also dimensionless}. Furthermore, the rescaled Hamiltonian is now characterized by a single parameter $\Lambda(t)$. The normal modes of the above rescaled Hamiltonian are:
\begin{equation}
\label{eq:H2-Normalmodes}
{\omega}_-=\sqrt{\Lambda(t)+2};\quad
{\omega}_+=\sqrt{\Lambda(t)}.
\end{equation}
The GS wave-function for $H(t)$ is:
\begin{equation}\label{GStilde}
\Psi_{\rm GS}(x_+, x_-,t)=\prod_{j=\{+,-\}}\left(\frac{\omega_j(0)}{\pi b_j^2(t)}\right)^{1/4}\exp{-\left(\frac{\omega_j(0)}{b_j^2(t)}-i\frac{\dot{b}_j(t)}{b_j(t)}\right)\frac{x_j^2}{2}-\frac{i}{2}\omega_j(0)\tau_j(t)},
\end{equation}
where the scaling parameter $b_j(t)$ for each of these modes satisfies the following Ermakov-Pinney equation:
\begin{equation}
	\ddot{b}_j(t)+\omega_j^2(t)b_j(t)=\frac{\omega_j^2(0)}{b_j^3(t)}\quad;\qquad j=+,-
\end{equation}

The GS entanglement entropy corresponding to $\tilde{H}(t)$ is:
\begin{subequations}
\label{eq:rescaledxiS}
\begin{eqnarray}
\label{eq:CHO-ent2}
\label{eq:rescaledS}
	S(t)&=& -\log{[1-\xi(t)]}-\frac{\xi(t)}{1-\xi(t)}\log{\xi(t)} \\
	\label{eq:rescaledxi}
	\xi(t)&=& \frac{\sqrt{\left(\frac{\omega_+(0)}{b_+^2(t)}+\frac{\omega_-(0)}{b_-^2(t)}\right)^2+\left(\frac{\dot{b}_+(t)}{b_+(t)}-\frac{\dot{b}_-(t)}{b_-(t)}\right)^2}-2\sqrt{\frac{\omega_+(0)\omega_-(0)}{b_+(t)b_-(t)}}}{\sqrt{\left(\frac{\omega_+(0)}{b_+^2(t)}+\frac{\omega_-(0)}{b_-^2(t)}\right)^2+\left(\frac{\dot{b}_+(t)}{b_+(t)}-\frac{\dot{b}_-(t)}{b_-(t)}\right)^2}+2\sqrt{\frac{\omega_+(0)\omega_-(0)}{b_+(t)b_-(t)}}}
\end{eqnarray}
\end{subequations}
Let us now compare the expressions \eqref{eq:CHO-ent1} and \eqref{eq:CHO-ent2}. Using the fact that the rescaled and original variables are related as $\tilde{b}_j(\tilde{t})=b_j(t)$ and $\tilde{\xi}(\tilde{t})=\xi(t)$, we see that $\tilde{S}(\tilde{t})=S(t)$ when $t=\tilde{\alpha} \tilde{t}$. In other words, we see that $\tilde{b}_j$ and $\tilde{S}$ are \emph{invariant under the transformations} $\tilde{H}\to \tilde{H}/\tilde{\alpha}$ and $\tilde{t}\to\tilde{\alpha}  \tilde{t}$. This is valid provided $\tilde{\alpha}$ is a constant. To further explore the consequences of this symmetry, let us look at the following scaling transformations:
\begin{equation}\label{trans}
	\tilde{\omega}\to\eta\tilde{\omega};\quad \tilde{\alpha}\to\eta\tilde{\alpha}
\end{equation}
In the time-independent case, it was shown that these transformations left the entanglement entropy invariant~\cite{2020Chandran.ShankaranarayananPhys.Rev.D}. However, in the time-dependent case, we see that:
\begin{itemize}
	\item For the rescaled Hamiltonian $H$, the entropy remains invariant:
	\begin{equation}
	\label{eq:Scaling01}
		S\left(\eta \tilde{\omega},\eta\tilde{\alpha},t\right)=S\left(\tilde{\omega},\tilde{\alpha},t\right)
	\end{equation}
	\item For the original Hamiltonian $\tilde{H}$, the entropy transforms as:
	\begin{equation}
		\label{eq:Scaling02}
		\tilde{S}\left(\eta \tilde{\omega},\eta\tilde{\alpha},\eta^{-1} \tilde{t}\right)=\tilde{S}\left(\tilde{\omega},\tilde{\alpha}, \tilde{t} \right)
	\end{equation}
\end{itemize}

In the time-independent case, we were able to group all systems with the same $\Lambda$ into a class of systems with the same entropy distinguished only by their energies~\cite{2020Chandran.ShankaranarayananPhys.Rev.D}. The transformations in \eqref{trans} would then take us from one system to another in the same $\Lambda$-class, and the energy gets rescaled appropriately. However, in the time-dependent case, while we can still group the systems into a $\Lambda(t)$-class where they have the same functional form, the transformation \eqref{trans} will rescale not only energy but also the time-scale of evolution.

\begin{figure*}[!ht]
	\begin{center}
		\subfloat[\label{sym1a}][]{%
			\includegraphics[width=0.4\textwidth]{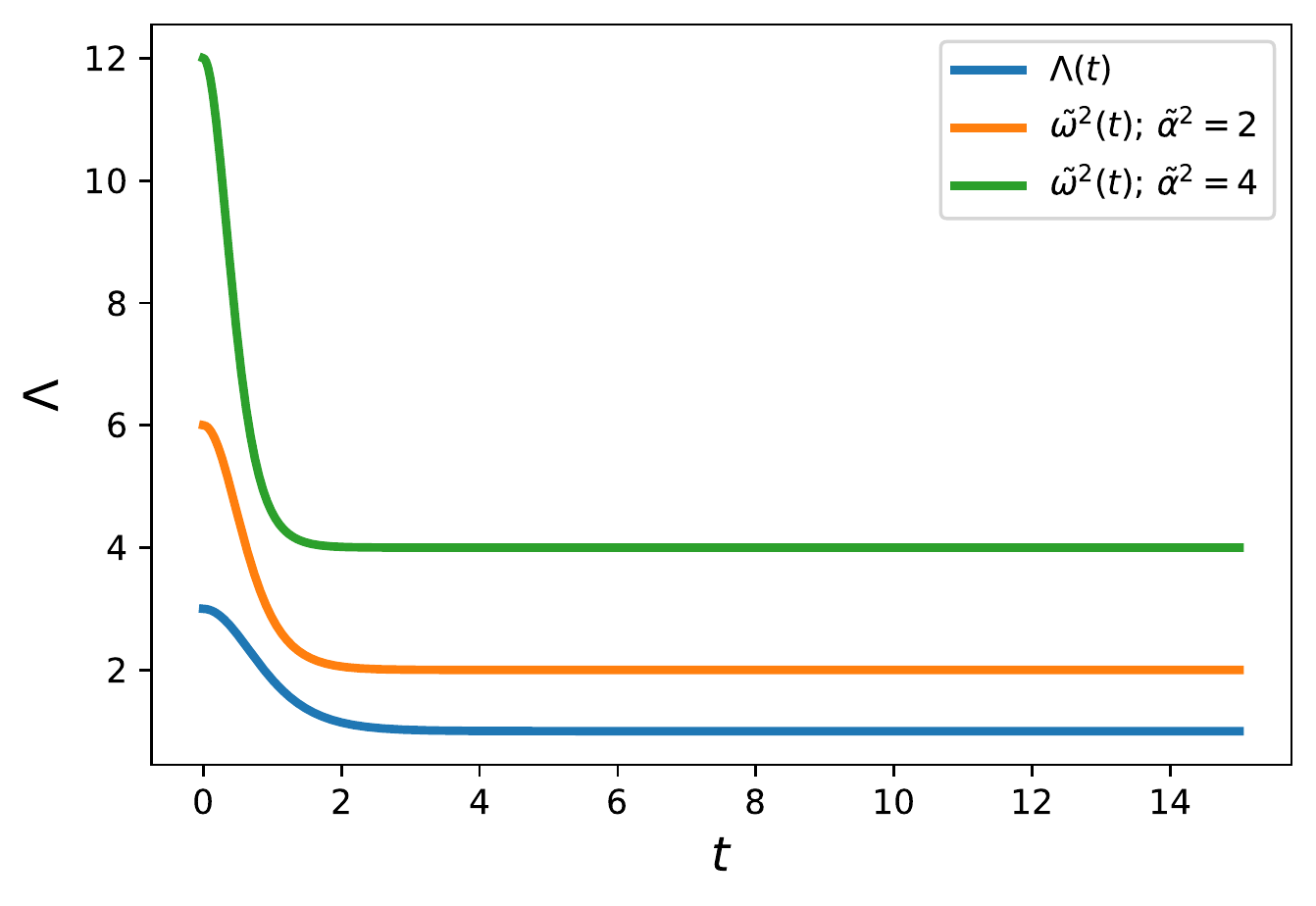}
		}
		\subfloat[\label{sym1b}][]{%
			\includegraphics[width=0.4\textwidth]{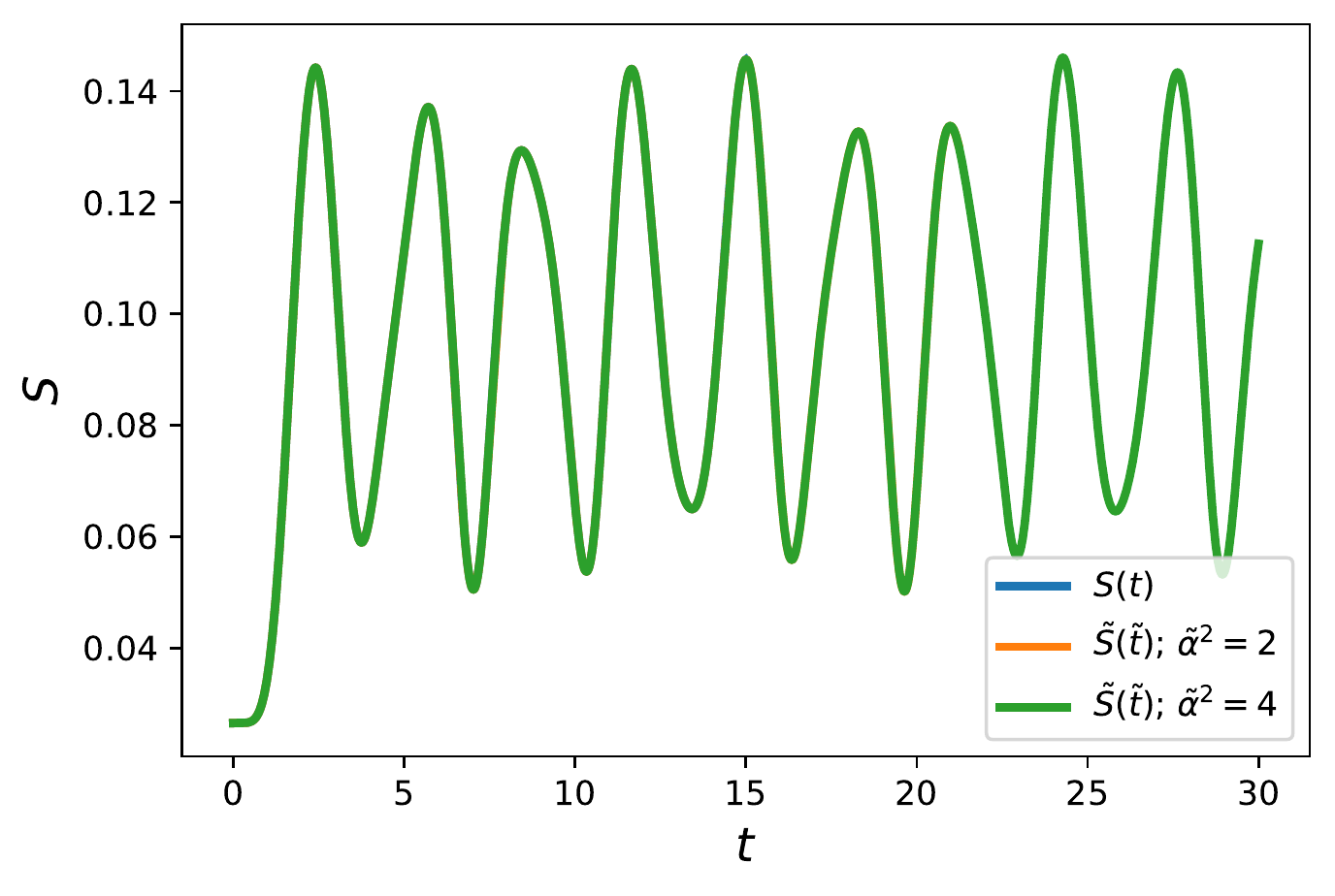}
		}
		
\caption{(a) Time dependence of $\Lambda(t)$ \eqref{eq:LambdaForm3} and $\tilde{\omega}^2(t)$ \eqref{eq:omegaForm3} for $\tilde{\alpha}^2 = 2~\mbox{and}~4$, (b) Dynamics of entanglement entropy in the rescaled system $S(t)$ and the original system $\tilde{S}(\tilde{t})$ for $\tilde{\alpha}^2 = 1,2$. Here, $P=Q=1$.}
		\label{sym}
	\end{center}
\end{figure*}

To illustrate this, we consider the following two different functional forms of $\tilde{\omega}(t)$ for which the exact solution to the Ermakov equation is known:
\begin{enumerate}
    \item The simplest $\tilde{\omega}(\tilde{t})$ for which solutions are well known is:
\begin{equation}
\label{eq:omegaForm1}
    \tilde{\omega}^2(\tilde{t})=\begin{cases}
		C_0\tilde{\alpha}^2& \text{if } \tilde{t}=0\\
		C_1\tilde{\alpha}^2 & \text{if } \tilde{t}>0
	\end{cases}
\end{equation}
In this case, the scaling parameter for the two normal modes are~\cite{2017Ghosh.etalEPLEurophysicsLetters}:
\begin{align}
    \tilde{b}_+(\tilde{t})&= \sqrt{1+\left(\frac{C_0}{C_1}-1\right)\sin^2\left(\tilde{\alpha}\sqrt{C_1}\tilde{t}\right)}\nonumber \\
    \label{eq:bForm1}
    \tilde{b}_-(\tilde{t})&= \sqrt{1+\left(\frac{C_0+2}{C_1+2}-1\right)\sin^2\left(\tilde{\alpha}\sqrt{C_1+2}\tilde{t}\right)}
\end{align}
For the rescaled Hamiltonian $H=\tilde{\alpha}^{-1}\tilde{H}$, we have:
\begin{equation}
    \Lambda(t)=\begin{cases}
		C_0& \text{if } t=0\\
		C_1 & \text{if } t>0
	\end{cases}
\end{equation}
The rescaled scaling parameters in this case are:
\begin{align}
    b_+(t)&= \sqrt{1+\left(\frac{C_0}{C_1}-1\right)\sin^2\left(\sqrt{C_1}t\right)}\nonumber \\
    \label{eq:tildebForm1}
    b_-(t)&= \sqrt{1+\left(\frac{C_0+2}{C_1+2}-1\right)\sin^2\left(\sqrt{C_1+2}t\right)}
\end{align}
Comparing Eqs. (\ref{eq:bForm1}, \ref{eq:tildebForm1}), we see that $\tilde{b}_j(\tilde{t})=b_j(t\to\tilde{\alpha}\tilde{t})$, and in extension, $\tilde{S}(\tilde{t})=S(t\to\tilde{\alpha}\tilde{t})$.
\item Consider the following form of $\tilde{\omega}(t)$:
\begin{equation}
\label{eq:omegaForm3}
	\tilde{\omega}^2(\tilde{t})=\tilde{P}^2+\frac{2}{\tilde{Q}^2}\sech ^{2}\left(\frac{\tilde{t}}{\tilde{Q}}\right)
\end{equation}
Here, the parameter $\tilde{P}=\abs{\tilde{\omega}(\pm \infty)}$ is the asymptotic value of a bell-shaped frequency curve centered at $t=0$, whereas $Q$ captures the ``squeeze" of the bell-curve. The solutions for this form have recently been worked out in Ref.~\cite{2022AndrzejewskiQuantumInformationProcessing}:

\begin{align}
    \tilde{b}_+^2(\tilde{t})&= \left(1+\frac{\tanh^2(\tilde{t}/\tilde{Q})}{\tilde{P}^2\tilde{Q}^2}\right)\left(1-\frac{\sin ^2\left(\tilde{P}\tilde{t}+\tan ^{-1}\left(\frac{\tanh (\tilde{t} / \tilde{Q})}{\tilde{P}\tilde{Q}}\right)\right)}{\left(1+\tilde{P}^2\tilde{Q}^2\right)^2}\right)\nonumber \\
    \tilde{b}_-^2(\tilde{t})&= \left(1+\frac{\tanh^2(\tilde{t}/\tilde{Q})}{(\tilde{P}^2+2\tilde{\alpha}^2)\tilde{Q}^2}\right)\left(1-\frac{\sin ^2\left(\sqrt{\tilde{P}^2+2\tilde{\alpha}^2}\tilde{t}+\tan ^{-1}\left(\frac{\tanh (\tilde{t} / \tilde{Q})}{\sqrt{\tilde{P}^2+2\tilde{\alpha}^2}\tilde{Q}}\right)\right)}{\left(1+\left(\tilde{P}^2+2\tilde{\alpha}^2\right)\tilde{Q}^2\right)^2}\right)
\end{align}
The corresponding functional form in the rescaled Hamiltonian $H=\tilde{\alpha}^{-1}\tilde{H}$ will be:
\begin{equation}
\label{eq:LambdaForm3}
    \Lambda(t)=P^2+\frac{2}{Q^2}\sech ^{2}\left(\frac{t}{Q}\right),
\end{equation}
where $\tilde{P}=\tilde{\alpha} P$ and $\tilde{Q}=\tilde{\alpha}^{-1} Q$. In this case, the solutions are:
\begin{align}
        b_+^2(t)&= \left(1+\frac{\tanh^2(t/Q)}{P^2Q^2}\right)\left(1-\frac{\sin ^2\left(Pt+\tan ^{-1}\left(\frac{\tanh (t / Q)}{PQ}\right)\right)}{\left(1+P^2Q^2\right)^2}\right)\nonumber \\
    b_-^2(t)&= \left(1+\frac{\tanh^2(t/Q)}{(P^2+2)Q^2}\right)\left(1-\frac{\sin ^2\left(\sqrt{P^2+2}t+\tan ^{-1}\left(\frac{\tanh (t / Q)}{\sqrt{P^2+2}Q}\right)\right)}{\left(1+\left(P^2+2\right)Q^2\right)^2}\right)
\end{align}
From here it is easy to see that $\tilde{b}(\tilde{t})=b(t\to\tilde{\alpha}\tilde{t})$, and in extension, $\tilde{S}(\tilde{t})=S(t\to\tilde{\alpha}\tilde{t})$. In this case, we see that to preserve the dynamical scaling symmetry of entanglement, and we must also rescale the parameters $\tilde{P}$ and $\tilde{Q}$ as they are dimensionful in time. In other words, they represent some other time scales in the evolution of the system. From \ref{sym}, we can observe that frequency evolutions with appropriately rescaled time-scales in the original and rescaled Hamiltonians will lead to the same entanglement dynamics. 
\end{enumerate}

In general, it is not possible to obtain the exact solution to the Ermakov equation \eqref{ermakov}. 
However, it is possible to obtain the approximate solution to the Ermakov equation in the asymptotic future provided $\tilde{\omega}(\tilde{t})$ or $\Lambda(t)$ relaxes to a constant value. Moreover, since the dynamical scaling symmetry connects the scale-factors of $H$ and $\tilde{H}$, it is possible to map the dynamics of original Hamiltonian $\tilde{H}(\tilde{t})$ to that of the rescaled Hamiltonian $H(t)$ at all times. In the following subsection, we obtain analytical results by studying the late-time behavior of the scale factors. 
%\item Consider the following form of $\tilde{\omega}(t)$:
%\begin{equation}
%\label{eq:omegaForm2}
%	\omega^2(t)=\tilde{\alpha}^2\left[1-\frac{P}{2}\left(1+\tanh{\left[Q(t-t_q)\right]}\right)\right],
%\end{equation}
%
%\blue{where $P=\abs{\omega^2(\infty)-\omega^2(-\infty)}/\tilde{\alpha}^2$ is the depth, $Q$ is the speed of the variation of $\omega(t)$, and $t_q$ is the time around which the $\omega(t)$ is centered. Unlike \eqref{eq:omegaForm1}, the above frequency has dimensionfull parameters (like $Q$ and $t_q$) and they also need to be rescaled to preserve this symmetry.}
%
%\blue{
%If the quench function involves dimension-full parameters, they may also have to be rescaled to preserve this symmetry. For instance, let us consider the following quench in the original system $H$: In the rescaled system $\tilde{H}=H/\tilde{\alpha}$, according to the generalized scaling symmetry, this corresponds to the following quench:
%\begin{equation}
%	\Lambda(t)=1-\frac{P}{2}\left(1+\tanh{\left[\tilde{Q}(t-\tilde{t}_q)\right]}\right),
%\end{equation}
%where $\tilde{Q}=\tilde{\alpha}^{-1} Q$ and $\tilde{t}_q=\tilde{\alpha} %t_q$. In other words, the parameters that depend on the %dimensions of time ($Q$ and $t_q$) must also be rescaled %to preserve the dynamical scaling symmetry of %entanglement. (Should I add a plot?)
%}

\subsection{Using late-time behavior to understand quantum correlations} \label{sec:Asymptotic}

Upon rescaling the Hamiltonian $H=\tilde{\alpha}^{-1}\tilde{H}$, we were able to simplify the problem by shifting to a Hamiltonian which has only a single, dimensionless time-dependent parameter $\Lambda(t)$ that drives the quench. Since $\Lambda(t)$ also contains the coupling parameter $\tilde{\alpha}$, we refer to $\Lambda(t)$ as the \emph{quench function}. Since the dynamical scaling symmetry of entanglement connects the scale-factors of $H$ and $\tilde{H}$, it is sufficient to work with one to understand the results of the other.

As we will see in Section \ref{sec:scalarfielddyn}, this will have important consequences in field theory. However, first, we fully flesh out the dynamics for a coupled harmonic oscillator. Of special interest is the late-time behavior of entanglement entropy, which not only serves as a diagnostic tool for quantum chaos~\cite{2018Hackl.etalPhys.Rev.A}, but helps identify the presence of zero modes.

To understand the long-term behavior of entanglement due to the quench function, we analyze the solutions to the Ermakov equation. First, let us look at a case where a rescaled frequency $\Lambda(t)$ undergoes a time-evolution and relaxes to a constant value in the asymptotic future:
\begin{equation}
\label{eq:Quench01}
    \Lambda(t)=\begin{cases}
		\Lambda_0& \text{if } t=0\\
		\Lambda_1& \text{if } t\to\infty
	\end{cases}
\end{equation}
Thus, the asymptotic values of the two normal modes --- $\omega_+(t)=\sqrt{\Lambda(t)}$ and $\omega_-(t)=\sqrt{\Lambda(t)+2}$ --- are:
\begin{equation}
    u_j=\lim_{t\to\infty}\omega_j(t)=\begin{cases}
		\sqrt{\Lambda_1}& \text{if } j=+\\
		\sqrt{\Lambda_1+2}& \text{if } j=-
	\end{cases}
\end{equation}
In the asymptotic future ($t\to\infty$), the Ermakov equation takes the following form: 
\begin{equation}
	\ddot{b}_j(t)+u_j^2b_j(t) \sim \frac{\omega_j^2(0)}{b_j^3(t)}\quad;\quad j=+,-
\end{equation}
Since the co-efficient in the second term of the above equation is time-independent, we can obtain the following solutions~\cite{2017Ghosh.etalEPLEurophysicsLetters}:
\begin{equation}
\label{eq:bjlongtime}
	b_j(t)\sim \sqrt{1+\left(\frac{\omega_j^2(0)}{u_j^2}-1\right)\sin^2{u_jt}}\quad;\quad \dot{b}_j(t)\sim \left(\omega_j^2(0)-u_j^2\right)\frac{\sin{2u_jt}}{2u_jb_j(t)} \, .
\end{equation}
From the above expression, we see that the late-time behavior of the scaling parameter crucially depends on the nature of asymptotic normal mode frequencies. Since the asymptotic normal modes can be of three types, the late-time behavior of the scaling parameter can be grouped into the following three categories:
\begin{itemize}
	\item \textbf{Stable mode} $u_j^2>0$ : The solutions $b(t)$ and $\dot{b}(t)$ are finite, bounded oscillations at late times. If both the normal modes are stable, then $b_j(t)$ and $\dot{b}_j(t)$ are bounded at all times, and it clear from 
	Eq. \eqref{eq:rescaledxiS} that the eigenvalues and entropy are  finite at all times. 
	\item \textbf{Zero mode} $u_j^2=0$ : At late times, the solutions \eqref{eq:bjlongtime} further reduce to :
	\begin{equation}\label{b:zero}
		b_j(t)\sim \omega_j(0)t\quad;\quad \dot{b}_j(t)\sim \omega_j(0)
	\end{equation}
	\item \textbf{Inverted mode} $u_j^2<0$ : At late times, the solutions \eqref{eq:bjlongtime} further reduce to :
	\begin{equation}\label{b:inverted}
		b_j(t)\sim \frac{1}{2}\sqrt{1+\frac{\omega_j(0)^2}{v_j^2}}e^{v_jt}\quad;\quad \dot{b}_j(t)\sim\frac{v_j}{2}\sqrt{1+\frac{\omega_j(0)^2}{v_j^2}}e^{v_jt},
	\end{equation}
where we have defined $u_j=iv_j$. 
\end{itemize}
%\red{What do you mean by the following sentence?}
%Since stable modes cannot be further simplified in the long time limit, we focus on the latter cases where $b_j(t)$ is unbounded in time:

Armed with the asymptotic form of the scaling parameters $b_j(t)$, we now obtain the quantum correlations (entanglement entropy and fidelity) for the quench \eqref{eq:Quench01} in the asymptotic limit. As shown in the next subsection, the asymptotic analysis is sufficient for any quench with constant values in the two asymptotic limits ($t \to -\infty$ and $t \to \infty$). 

\subsubsection{Zero mode: $\Lambda_1=0$}

Let us consider that case where one of the normal modes ($\omega_+$) vanish in the asymptotic future. (Note that in the case of CHO only one normal mode can vanish.) The scaling parameter takes the form in Eq. \eqref{b:zero}.
%\begin{equation}
%	b_+\sim\sqrt{\omega_+(0)}t\quad;\quad \dot{b}_+(t)\sim \omega_+(0)
%\end{equation} 
Substituting Eq.~\eqref{b:zero} in Eq. \eqref{eq:rescaledxiS}, the eigenvalues and entropy reduce to:
	\begin{align}
		\xi_{\rm zero}(t)&\sim 1-\frac{4b_-(t)}{t}\sqrt{\frac{\omega_-(0)}{\omega_+(0)\left[\omega_-^2(0)+b_-^2(t)\dot{b}_-^2(t)\right]}}\nonumber\\
		\label{zeromode:ent}
		S_{\rm zero}(t)&\sim \log{\left[\frac{t}{4}\sqrt{\frac{\omega_+(0)}{\omega_-(0)}\left(\frac{\omega_-^2(0)}{b_-^2(t)}+\dot{b}_-^2(t)\right)}\right]}\propto \log(t)
		\end{align}
Similarly, the overlap between initial and final states \eqref{eq:overlapCHO} reduces to:
\begin{equation}
\label{zeromode:overlap}
	\mathcal{F}^{\rm zero}_0(t)\sim 2\sqrt{\frac{\omega_-(0)}{\omega_+(0)b_-(t)t\left[\omega_-^2(0)\left(1+\frac{1}{b_-^2(t)}\right)^2+\frac{\dot{b}_-^2(t)}{b_-^2(t)}\right]}}\propto t^{-1/2}
\end{equation}
This is the first key result of this work, regarding which we would like to stress the following points: First, Eq. \eqref{zeromode:ent} implies that 
if any of the normal modes relaxes to a zero-mode, the entropy of the system increases logarithmically with time, $S\sim\log(t)$. Second, from Eqs. (\ref{zeromode:ent}, \ref{zeromode:overlap}), we obtain the following relation: $S\propto -\log{\mathcal{F}_0^2}$. While this relation might look very speculative at present, as we show later, this is indeed the case.
	
\subsubsection{Inverted modes: $\Lambda_1 < 0$}

This category has three possible scenarios: %Let  
\begin{enumerate}[label=(\Roman*) ]
	\item $-2<\Lambda_1<0$: Consider the scenario, when the system relaxes to one inverted mode ($\omega_+$) and one oscillator mode ($\omega_-$) in the asymptotic future. The solutions for the inverted mode reduce to: 
	\begin{equation}
		b_+(t)\sim \frac{1}{2}\sqrt{1+\frac{\omega_+^2(0)}{v_+^2}}e^{v_+t}\quad;\quad \dot{b}_+(t)\sim\frac{v_+}{2}\sqrt{1+\frac{\omega_+^2(0)}{v_+^2}}e^{v_+t} \, ,
	\end{equation}
	where, $\lim_{t\to\infty}\omega_+(t)=iv_+$. Substituting the above expression in Eq. \eqref{eq:rescaledxiS}, the parameter $\xi$ and entropy $S$ reduce to:
	\begin{align}
		\xi_{\rm Inv}^{\rm (I)}(t)&\sim 1-\frac{8v_+\sqrt{\omega_+(0)\omega_-(0)}\exp{-v_+t}}{\abs{b_-(t)v_+-\dot{b}_-(t)}\sqrt{v_+^2+\omega_+^2(0)}}\nonumber\\
		S_{\rm Inv}^{\rm (I)}(t)&\sim v_+t
	\end{align}
Similarly, the overlap between initial and final states \eqref{eq:overlapCHO} reduces to:
\begin{equation}
\mathcal{F}_0^{\rm (I)}(t)\propto \exp{-\frac{v_+t}{2}} \, .
\end{equation}

\item $\Lambda_1=-2$: Consider the scenario when the system relaxes to one inverted mode ($\omega_+$) and one zero-mode ($\omega_-$). Now, the solutions to $\omega_-$ mode are no longer bound, and the scaling parameter is:
	\begin{equation}
		b_-(t)\sim \omega_-(0)t\quad;\quad \dot{b}_-(t)\sim\omega_-(0)
	\end{equation}
Substituting the above scaling parameter in Eq. \eqref{eq:rescaledxiS}, the parameter $\xi$ and entropy $S$ in the asymptotic future reduce to:
	\begin{align}
		\xi_{\rm Inv}^{\rm (II)}(t)&\sim 1-\frac{8}{t}\sqrt{\frac{\omega_+(0)}{\omega_-(0)\left(v_+^2+\omega_+^2(0)\right)}}\exp{-v_+t}\nonumber\\
		S_{\rm Inv}^{\rm (II)}(t)&\sim v_+t+\log{t}
	\end{align}
Similarly, the overlap between initial and final states \eqref{eq:overlapCHO} reduces to:
	\begin{equation}
		\mathcal{F}_0^{\rm (II)}(t)\propto t^{-1/2}\exp{-\frac{v_+t}{2}}
	\end{equation}
\item $\Lambda_1<-2$: Consider the scenario when the system relaxes to two inverted modes ($\omega_+$, $\omega_-$). The scaling parameter corresponding to $\omega_-$ also grows exponentially:
\begin{equation}
		b_-(t)\sim \frac{1}{2}\sqrt{1+\frac{\omega_-(0)^2}{v_-^2}}e^{v_-t}\quad;\quad \dot{b}_-(t)\sim\frac{v_-}{2}\sqrt{1+\frac{\omega_-(0)^2}{v_-^2}}e^{v_-t} \, ,
	\end{equation}
where $\lim_{t\to\infty}\omega_-(t)=iv_-$. 
Substituting the above scaling parameter in Eq. \eqref{eq:rescaledxiS}, the parameter $\xi$ and entropy $S$ in the asymptotic future reduce to:	
	\begin{align}
		\xi_{\rm Inv}^{\rm (III)}(t)&\sim 1-\frac{16v_+v_-}{v_+-v_-}\sqrt{\frac{\omega_+(0)\omega_-(0)}{\left[v_+^2+\omega_+^2(0)\right]\left[v_-^2+\omega_-^2(0)\right]}}\exp{-(v_++v_-)t}\nonumber\\
		S_{\rm Inv}^{\rm (III)}(t)&\sim (v_++v_-)t
	\end{align}
Similarly, the overlap between initial and final states \eqref{eq:overlapCHO} reduces to: 
	\begin{equation}
		\mathcal{F}_0^{\rm (III)}(t)\propto \exp{-\frac{(v_++v_-)t}{2}}
	\end{equation}
	\end{enumerate}
%	
%General late-time behaviour for oscillator chain:
%\begin{equation}
%	S\sim\left(\sum_{i=1}^nv_i\right)t+\log{t},
%\end{equation}
%where we have $n$ inverted modes with velocities $\{v_i\}$ and one zero mode.

This is the second key result of this work, regarding which we would like to stress the following points: First, from the three scenarios, we see that the eigenvalues $\xi(t)$ and the overlap function have the same functional behavior w.r.t $t$. Moreover, the eigenvalues and the overlap function have an exponential dependence w.r.t the inverted mode frequencies. Second, the asymptotic value of the entanglement entropy scales linearly with time and the normal mode frequency. Third, from the above results, we can conclude that the general late-time behavior of entanglement entropy for CHO is given by:
\begin{equation}
    S\sim\left(\sum_{i=1}^nv_i\right)t+\log{t}+S_0(t),
\end{equation}
where the linear term arises from $n$ (can take $0, 1,$ or $2$) inverted modes, the logarithmic term arises from a zero mode, and $S_0(t)$ is a bounded function arising from stable modes. In the presence of zero modes and inverted modes, the entropy grows unbounded, ultimately diverging as $t\to\infty$. This may be explained by looking at the instantaneous GS wave-function corresponding to the normal mode in question~\cite{2008LoheJournalofPhysicsAMathematicalandTheoretical,2017Ghosh.etalEPLEurophysicsLetters}:
\begin{align}
\Psi_{\rm GS}^{(j)}(x_j,t)&=\left(\frac{\omega_j(0)}{\pi b_j^2(t)}\right)^{1/4}\exp{-\left(\frac{\omega_j(0)}{b_j^2(t)}-i\frac{\dot{b}_j(t)}{b_j(t)}\right)\frac{x_j^2}{2}-\frac{i}{2}\omega_j(0)\tau_j(t)}\nonumber\\
&=\exp{\frac{i\dot{b}_j(t)x_j^2}{2b_j(t)}-\frac{i}{2}\omega_j(0)\tau_j(t)}\psi_0\left(x_j,\omega_j\to\frac{\omega_j(0)}{b_j^2(t)}\right),
\end{align}
where $j\in\{+,-\}$ is the normal mode index, and $\psi_0$ is the ground state solution for the time-independent HO, whose frequency is instantaneously rescaled by scaling parameter $b_j(t)$. Here, we see that the Gaussian has both real and imaginary parts in the exponential when the evolution time-scale is finite, irrespective of whether the mode is stable, zero, or inverted. Due to the real part being non-zero for finite time-scales, the wave-function is normalizable throughout the evolution. However, in the late-time limit, if the normal mode relaxes to zero, using \eqref{b:zero} we get:
\begin{equation}
    \lim_{t\to\infty} \Psi_{\rm GS}^{(j)}(x_j,t)\sim e^{-\frac{i}{2}\omega_j(0)\tau_j(\infty)}\psi_0\left(x_j,\omega_j\to 0\right)\sim\psi_{fp},
\end{equation}
where we see that the oscillator is approximated by a free-particle wave-function $\psi_{fp}$ as $t\to\infty$~\cite{2019-Chandran.Shankaranarayanan-Phys.Rev.D}. Similarly, in the case of an inverted mode in the late-time limit, using \eqref{b:inverted} we get:
\begin{equation}
    \lim_{t\to\infty} \Psi_{\rm GS}^{(j)}(x_j,t)\sim e^{\frac{i}{2}v_jx_j^2-\frac{i}{2}\omega_j(0)\tau_j(\infty)} \psi_{fp} \, .
\end{equation}
Interestingly, this is similar to the functional form obtained in Refs.~\cite{2006Yuce.etalPhysicaScripta,2015Pedrosa.etalCanadianJournalofPhysics}. For the zero-mode and inverted modes, we see that the real part of the exponential in the Gaussian vanishes, and the wave-function is non-normalizable as $t\to\infty$. In the time-independent case, it was shown that entropy divergence is a direct consequence of free particles in the system, as their wave-function is non-normalizable~\cite{2019-Chandran.Shankaranarayanan-Phys.Rev.D,2020Chandran.ShankaranarayananPhys.Rev.D}. This divergence was then truncated by introducing an IR-cutoff in the system. In the time-dependent case, the generation of zero modes and inverted modes leads to non-normalizability exactly as $t\to\infty$. Therefore, entanglement entropy diverges in this limit unless we introduce an IR cut-off. 

We further see that in the presence of zero modes and inverted modes, entropy and fidelity are related as:
\begin{equation}
\label{eq:conjecture01}
    S\sim-\log{\mathcal{F}_0^2}
\end{equation}
Thus the above expression suggests that in the presence of instabilities, subsystem quantities such as entropy may converge to a full system quantity such as logarithmic fidelity at late times. While the above relation is obtained by studying the asymptotic properties of the system, in the next subsection, we show that the asymptotic analysis is sufficient for any quench that has constant values in the two asymptotic limits ($t \to -\infty$ and $t \to \infty$). 

\subsection{Exact results from a quench model}\label{sec:exact}
In this subsection, we will simulate the entanglement dynamics of the CHO subject to a non-trivial evolution of rescaled frequency $\Lambda(t)$. To clearly capture the asymptotic solutions we obtained in the earlier section, we will consider the following functional form:
\begin{equation}
	\Lambda(t)=\begin{cases}
		a+\frac{b-a}{1+\left(\frac{t-t_q}{P}\right)^{2Q}}& \text{if } t\leq t_q\\
		c+\frac{b-c}{1+\left(\frac{t-t_q}{P}\right)^{2Q}} & \text{if } t\geq t_q
	\end{cases}
\end{equation}
Here, $\Lambda(t)$ resembles an asymmetric trough. $P$ represents the width of the trough, $Q$ captures the steepness, and $t_q$ is the center of the trough. Also, we have introduced parameters $a=\Lambda(-\infty)$, $b=\Lambda(t_q)$, and $c=\Lambda(\infty)$ that fix the values of frequency at key points in the evolution. 
\begin{figure*}[!ht]
	\begin{center}
		\subfloat[\label{S1a}][]{%
			\includegraphics[width=0.4\textwidth]{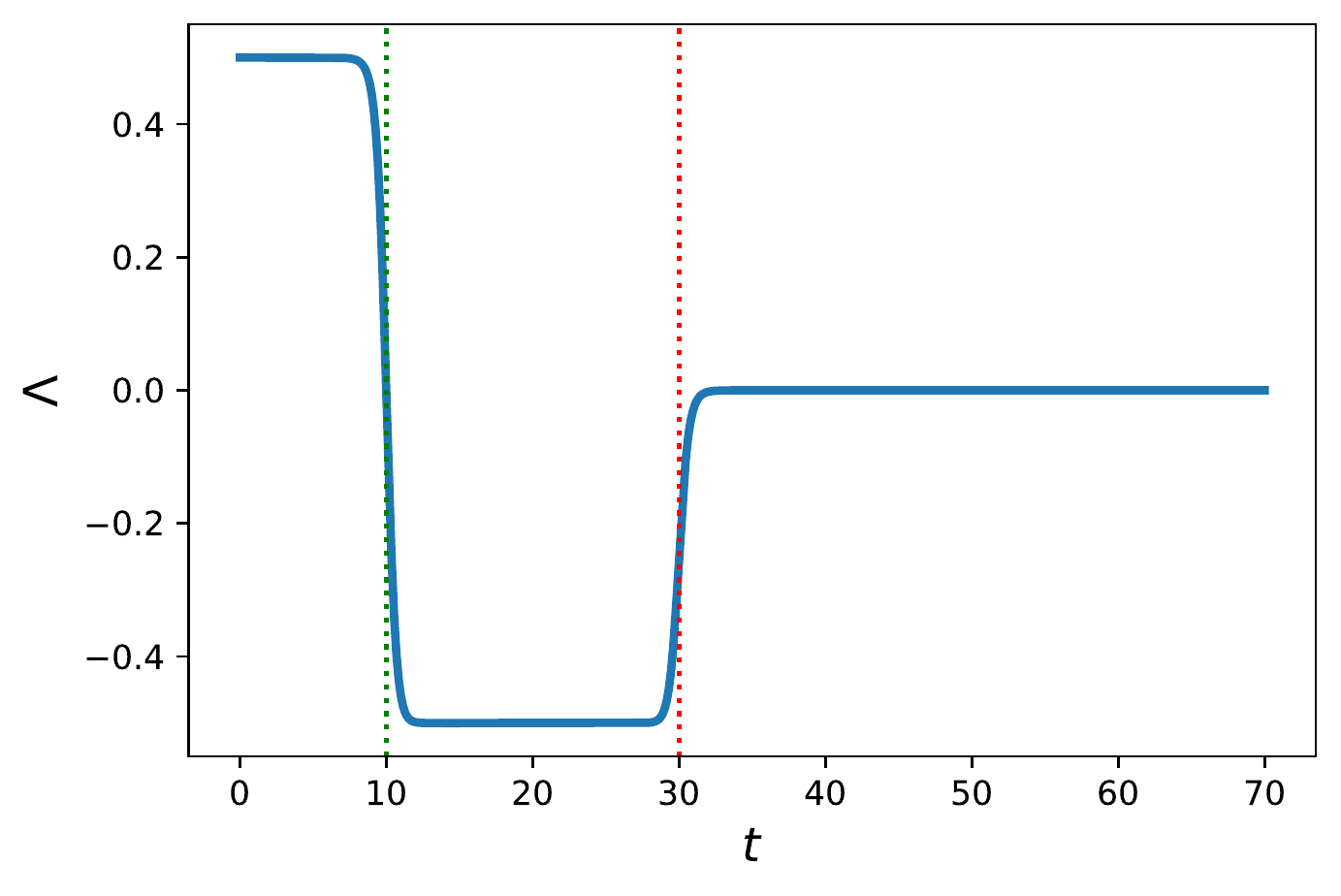}
		}
		\subfloat[\label{S1b}][]{%
			\includegraphics[width=0.4\textwidth]{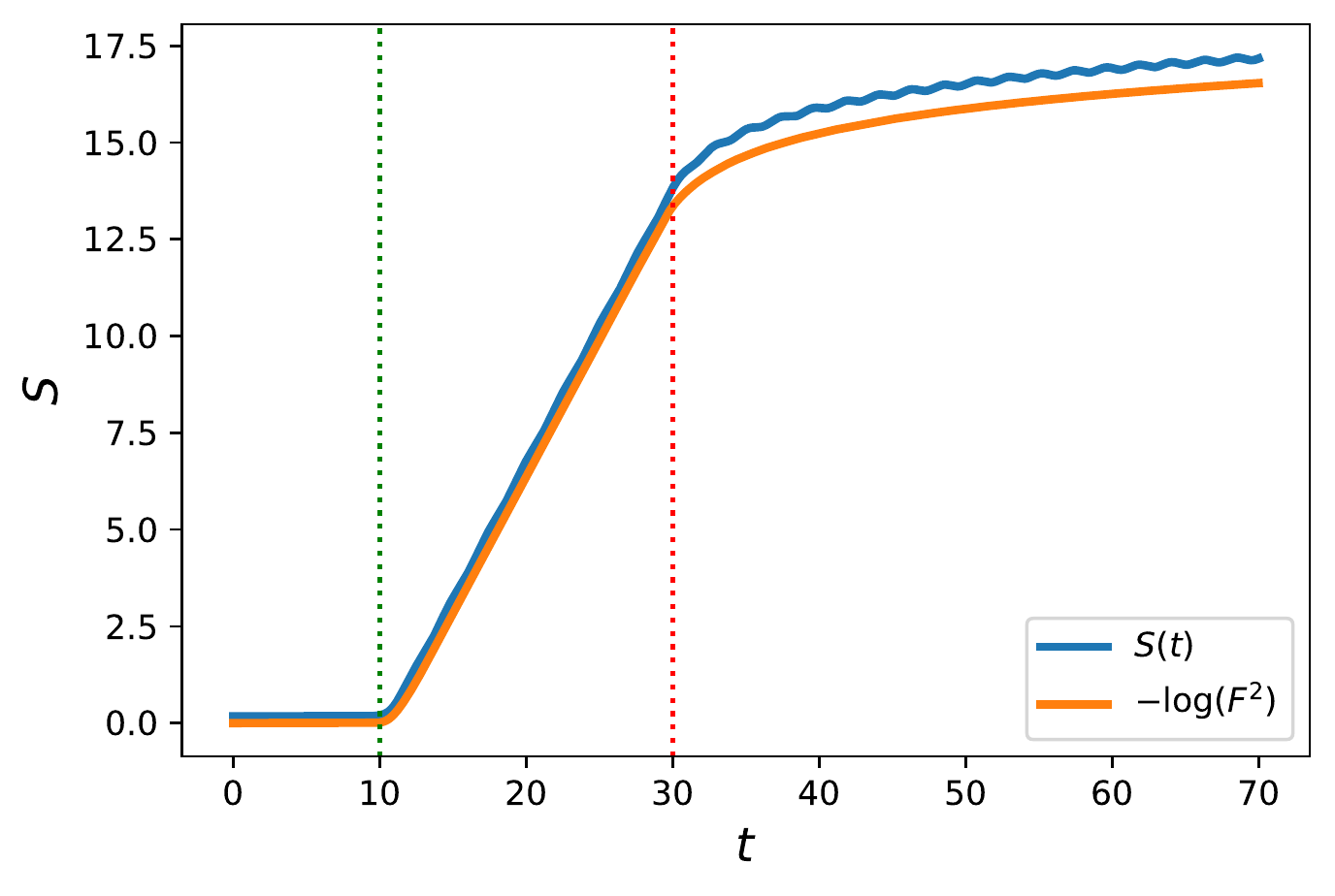}
		}
		
		\caption{(a) Evolution of rescaled frequency $\Lambda(t)$, (b) Dynamics of entanglement entropy $S(t)$ and logarithmic fidelity due to the quench. Here, $a=0.5$, $b=-0.5$, $c=0$, $P=10$, $Q=15$, and $t_q=20$.}
		\label{S1}
	\end{center}
\end{figure*}

In \ref{S1}, we see that during the evolution, the parameters have been tuned such that the time-evolution of rescaled frequency $\Lambda(t)$ covers three regions --- positive (stable), negative (unstable), and zero (metastable). From \ref{S1}, we also observe that in each of these regions, despite the time-intervals being small, the entropy behaves exactly as predicted by the late-time entropy analysis in Section \ref{sec:Asymptotic}. Interestingly, the entanglement entropy and fidelity satisfy the relation \eqref{eq:conjecture01}. Thus, the numerical analysis points to the fact that the asymptotic analysis captures the entanglement dynamics of CHO. This is the \emph{third key result} of this work.

If we were to now shift to the original Hamiltonian $\tilde{H}(\tilde{t})$, we can invoke the scaling symmetry argument by rescaling \eqref{trans} the following parameters:
\begin{equation}
    a\to\tilde{\alpha}^2a\quad;\quad b\to\tilde{\alpha}^2b \quad;\quad c\to\tilde{\alpha}^2c \quad;\quad t_q\to\tilde{\alpha} t_q \quad;\quad P\to\tilde{\alpha} P
\end{equation}
We may then use the relations $\tilde{S}(\tilde{t})=S(t\to\tilde{\alpha} \tilde{t})$ and $\mathcal{F}_0(t)=\tilde{\mathcal{F}}_0(t\to\tilde{\alpha} \tilde{t})$ to reproduce the dynamics of original system $\tilde{H}(\tilde{t})$ from $H(t)$.

In this section, we have shown how the scaling symmetry of entanglement entropy for the time-independent case \cite{2020Chandran.ShankaranarayananPhys.Rev.D} can be extended to the time-dependent case. In the time-independent case, we were able to group all systems in the same $\Lambda$-class, which are found to be generated by the transformations \eqref{trans}. Since entropy was found to depend on $\Lambda$ alone monotonically, all such systems with the same $\Lambda$ have the same entropy and vice-versa. Under the transformations \eqref{trans}, entropy also, therefore, remains invariant, while the ground state energies differ. We group all systems with the same functional form of $\Lambda(t)$ when considering a time-dependent Hamiltonian. In this case, the dynamical scaling symmetry \eqref{eq:Scaling01} implies that while the entropy is invariant under transformations \eqref{trans}, the evolution time-scales must be rescaled accordingly. Consequently, all systems in the same $\Lambda(t)$-class will have the same functional form for entropy but with different instantaneous ground state energies and time-scales of evolution. 

We also studied the late-time behavior of entanglement and the fidelity function using the dynamical scaling symmetry. The linear and logarithmic contributions to entropy from inverted (unstable) and zero (metastable) modes are in agreement with results in Ref.~\cite{2018Hackl.etalPhys.Rev.A}, and the corresponding coefficients have been exactly derived for the CHO. These contributions grow unbounded in time, and entropy divergence as $t\to\infty$ is explained by the non-normalizability of the normal mode wave-function in this limit. Here, while we see that the late-time behavior of entropy can serve as a diagnostic test for quantum chaos, it may be insufficient to measure the scrambling time and Lyapunov exponents associated with the chaotic behavior. For this, we must explore measures such as Loschmidt echo and circuit complexity that can be used to quantify such instabilities. Therefore, in the next section, we focus on understanding these measures and seeing how they may relate to entropy and fidelity in the late-time limit.

\section{Quantifying instabilities using dynamical scaling symmetry}\label{qchaos}
%\subsection{Loschmidt Echo and Circuit complexity}

In Section \ref{model}, we derived the dynamical scaling symmetry of entanglement entropy, and studied its late-time behaviour. We saw that there were linear and logarithmic contributions that grow unbounded with time, arising from instabilities in the system. We also saw that in this limit, entanglement entropy which is a subsystem quantity can be related to GS fidelity corresponding to the global wave-function. To quantify the instabilities, we may use well-established measures such as Loschmidt echo and circuit complexity, {which are codified by the global wave-function}. Here, we use the dynamical scaling symmetry of entanglement to explore the following --- i) The late-time behaviour of Loschmidt echo and circuit complexity, and ii) their relation to entanglement entropy in this limit. This would help us see how measures characterizing chaos such as Lyapunov exponents may also be derived from entanglement entropy in the late-time limit, in spite of it being a subsystem quantity.

\subsection{Loschmidt Echo}
\label{sec:LoschEcho}

A standard way of measuring the sensitivity to perturbations in a quantum system is by looking at the overlap between states that follow slightly different evolutions of the Hamiltonian. This can be defined by what is known as the fidelity function~\cite{1984PeresPhys.Rev.A,2006Gorin.etalPhysicsReports}:
\begin{equation}
	\mathcal{F}(t)=\abs{\braket{ \Psi(t)}{\Psi_1(t)}},
\end{equation}
where the Hamiltonian $H'(t)$ that evolves $\Psi_1(t)$ is slightly different from the Hamiltonian $H(t)$ that evolves $\Psi(t)$. Numerically, this is also the same as Loschmidt Echo~\cite{2001Jalabert.PastawskiPhys.Rev.Lett.,2006Gorin.etalPhysicsReports}, which similarly measures the sensitivity of the system to perturbations in time-evolved quantum systems, but instead is carried out by performing a forward evolution followed by a backward evolution on the initial state $\Psi_0$. In the case of CHO, this can be done by taking slightly different Hamiltonians $H(t)$ and $H'(t)$ respectively, resulting in a new state $\Psi_2$:
\begin{equation}
	\mathcal{M}(t)=\abs{\langle\Psi_0|e^{i\int H'dt}e^{-i\int Hdt}|\Psi_0\rangle}={\abs{\langle\Psi_0|\Psi_2\rangle}}
\end{equation} 
While $\mathcal{F}(t)$ and $\mathcal{M}(t)$ have the same value, the states prepared for both overlaps are different. A quantity that can help distinguish these two different scenarios is the circuit complexity calculated from the wave-function~\cite{2019Ali.etalJournalofHighEnergyPhysics}, which we will discuss in the latter part of this section.

In the case of CHO, we can calculate the ground state Loschmidt echo corresponding to Hamiltonians $H(\Lambda,t)$ and $H'=H(\Lambda+\delta\Lambda,t)$ as follows:
\begin{equation}\label{def:Loschmidt}
\mathscr{\mathscr{M}}(t)= \prod_{j=+,-}\!\!\mathscr{M}_j 
= 2^{N/2}\prod_{j=+,-}\left[\frac{\omega_j(0)\omega_j'(0)}{b_j^2(t)b_j'^2(t)\kappa_j}\right]^{1/4} \!\! ;
~\kappa_j=\left[\frac{\omega_j(0)}{b_j^2(t)}+\frac{\omega_j'(0)}{b_j'^2(t)}\right]^2+\left[\frac{\dot{b}_j(t)}{b_j(t)}-\frac{\dot{b}'_j(t)}{b_j'(t)}\right]^2
\end{equation}
Here, the modification is brought about by an infinitesimal change $\delta\Lambda$ in the evolution of physical parameter $\Lambda(t)$. Suppose the rescaled mass relaxes to a constant $\Lambda_1$ at late times; we may consider the asymptotic solutions of the scaling parameter as discussed in Section \ref{sec:Asymptotic}. In the asymptotic limit, we isolate the contribution $\mathscr{M}_+(t)$ from the $\omega_+$ mode for three different categories discussed in Sec. \eqref{sec:Asymptotic}, by expanding $\mathscr{M}$ about $\delta\Lambda$:
\begin{enumerate}
	\item $\Lambda_1>0$: In this category, $\omega_+$ mode is stable and the Loschmidt echo contribution takes the form:
	\begin{equation}
	    \mathscr{M}_+(t)\sim 1-\mathscr{O}(\delta\Lambda^2)
	\end{equation}
	We can see that the effect of infinitesimal change $\delta\Lambda$ on stable modes is negligible even at late times.
	%%%
	\item $\Lambda_1=0$: In this category, $\omega_+$ mode is a zero mode and we have:
	\begin{equation}
	    \mathscr{M}_+(t)\sim \begin{cases}
		1-\mathscr{O}(\delta\Lambda^2)& \text{if } 0\ll t\ll t_{\rm zero}\\
		\sqrt{\frac{2}{\omega_+(0)t}}\left(1-\frac{\delta\Lambda}{8\omega_+^2(0)}\right) & \text{if } t\gg t_{\rm zero}
	\end{cases}
	\end{equation}
	Thus, the Loschmidt echo behaves differently above and below the new time-scale $t_{\rm zero} ={2\omega_+(0)}/{\delta\Lambda}$. In the asymptotic limit, $\mathscr{M}_+$ has the same behavior as the ground state overlap function \eqref{zeromode:overlap}. Thus, one can interpret $t_{\rm zero}$ as the delay experienced by the system for the Loschmidt echo to switch to a power-law decay after quench. {We also see that when $t\gg t_{\rm zero}$, Loschmidt echo is related to entropy as:
	\begin{equation}
	    S\sim -\log{\mathscr{M}^2}\sim -\log{\mathscr{F}_0^2}
	\end{equation}}

	\item $\Lambda_1<0$: In this category, $\omega_+$ mode is inverted and  the Loschmidt echo is:
\begin{equation}
	    \mathscr{M}_+(t)\sim \begin{cases}
		1-\frac{\delta\Lambda}{8\omega_+^2(0)} & \text{if } 0\ll t\ll t_{\rm scram} \\\left(1-\frac{\delta\Lambda}{8\omega_+^2(0)}\right)\exp{-v_+(t-t_{\rm scram})} & \text{if } t\gg  t_{\rm scram}
	\end{cases}
	\end{equation}
	where we obtain a new timescale $t_{\rm scram}$ arising due to the infinitesimal change ($\delta \Lambda$):
	\begin{equation}\label{scram}
	    t_{\rm scram}\sim -\frac{1}{2v_+}\log{\frac{\delta\Lambda}{v_+\omega_+(0)}} 
	\end{equation}
	Here, we see that the Loschmidt echo undergoes an exponential decay, but there is sufficient delay $t_{\rm scram}$ from the quench time for this decay to kick in. Since $t_{\rm scram}$ captures how quickly the information about the original wave-function is lost upon introducing a slight change in the initial conditions, it is natural to identify $t_{\rm scram}$ as the \emph{scrambling time}~\cite{Qu2021}. 
	Another way to quantify the exponential decay of the Loschmidt echo is the \emph{maximal Lyapunov exponent}~\cite{Qu2021}. The maximal Lyapunov exponent is:
	\begin{equation}\label{lyap}
	    \lambda_L=-\lim_{t\to\infty}\frac{1}{t}\log{\mathscr{M}}\sim v_+
	\end{equation}
We see that the Lyapunov exponent is independent of the perturbation $\delta\Lambda$. It is to be noted that when we have two inverted modes with different timescales $\{t_{\rm scram}^{(j)}\}$, the scrambling time corresponds to the earliest onset of exponential decay. This is determined by the mode with the largest $v_j$ as can be seen in \eqref{scram}. For CHO, this always corresponds to the smallest (more negative) normal mode $\omega_+^2(t)$. The Lyapunov exponent, in this case, becomes $\lambda_L=v_++v_-$. {We also see that for inverted modes, we get a slightly different asymptotic relation between the entropy and fidelity as compared to the zero-mode case:
\begin{equation}
    S\sim -\log{\mathscr{M}}\sim -\log{\mathscr{F}_0^2}
\end{equation}}
\end{enumerate}
{To our knowledge, the relevance of the timescale $t_{\rm zero}$ has not been discussed earlier in the literature. However, for sufficiently small $\delta\Lambda$, we see that the scale $t_{\rm zero}$ is much larger as compared to $t_{s}$, i.e., a zero-mode instability retains information about the original wave-function for a considerable amount of time when subject to a small change in initial conditions. For instance, setting $\delta\Lambda\sim \mathscr{O}(10^{-10})$ and other constants to be $\mathscr{O}(1)$, we see that:}
\begin{equation}
	    t_{\rm zero}\sim \mathscr{O}(10^{10})\quad;\quad t_{\rm scram}\sim \mathscr{O}(10) \, .
\end{equation}
This has important implications for quantum perturbations generated during cosmological inflation and the black hole physics. We will discuss the implications in Sec. \ref{sec:conc}.

\begin{figure*}[!ht]
	\begin{center}
		\subfloat[\label{le1a}][]{%
			\includegraphics[width=0.4\textwidth]{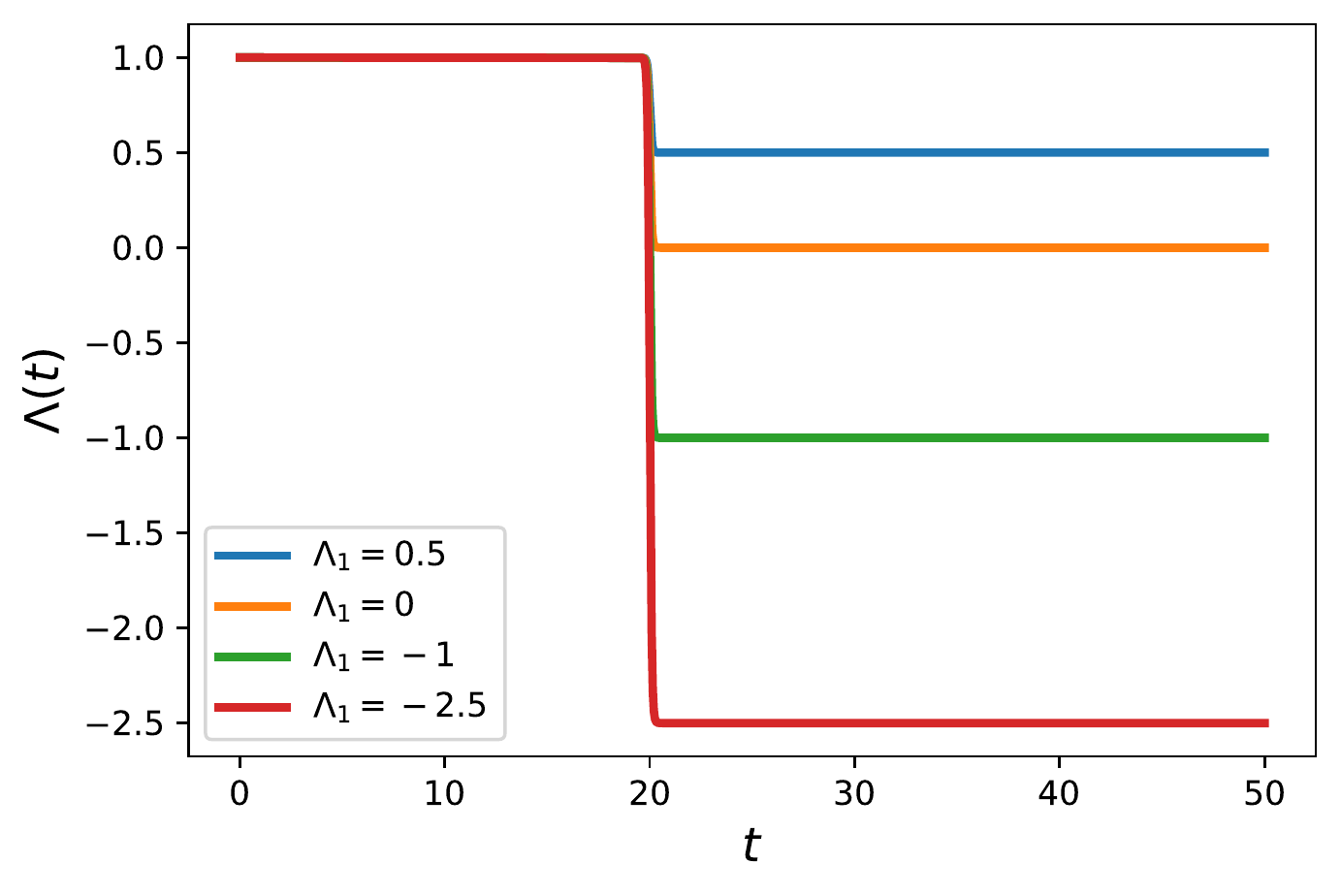}
		}
		\subfloat[\label{le1b}][]{%
			\includegraphics[width=0.4\textwidth]{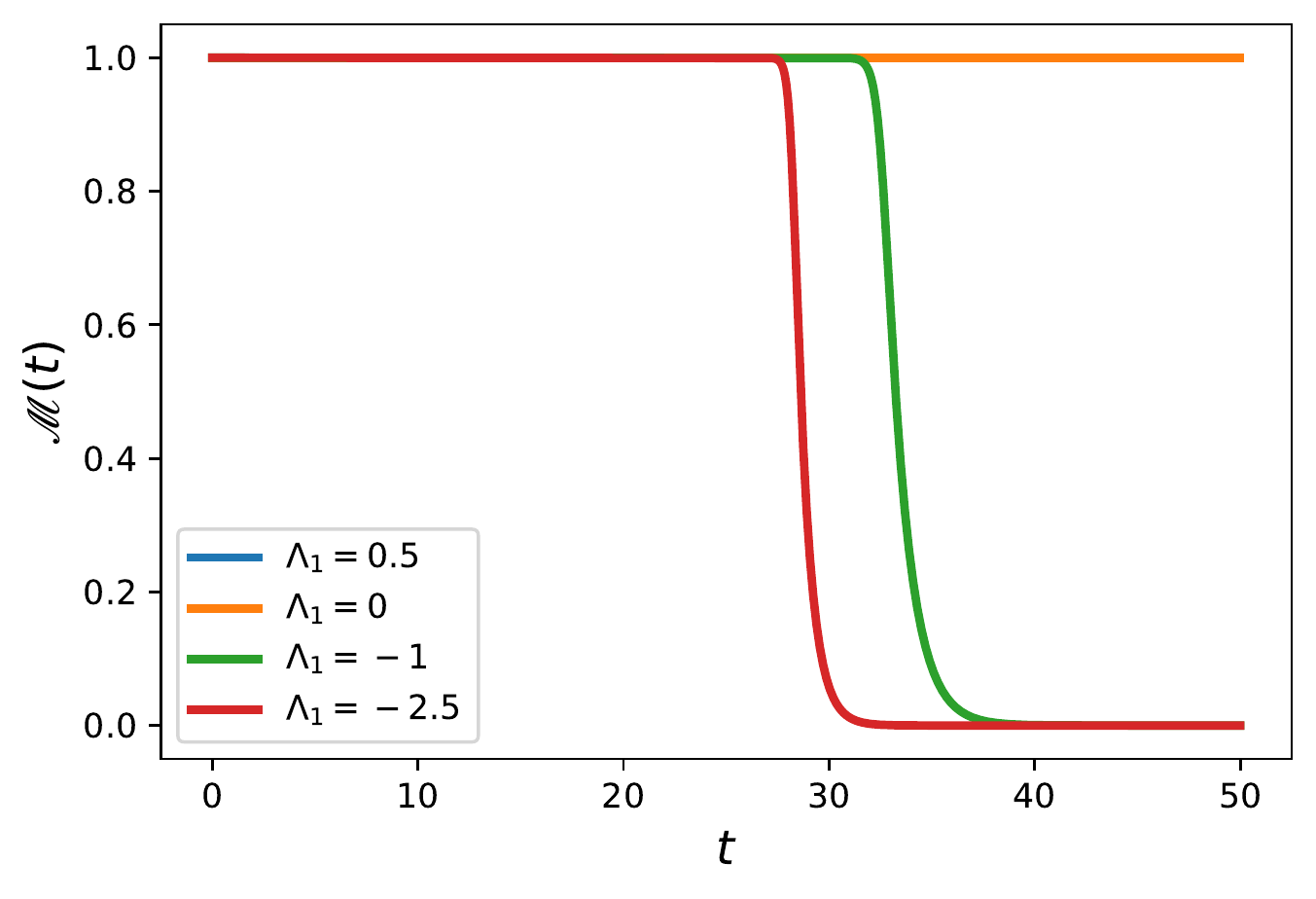}
		}
		
		\caption{(a) Quench functions $\Lambda(t)$ and (b) Loschmidt Echo $\mathscr{M}(t)$ for the corresponding quenches. Here, $Q=10$, $t_q=20$ and $\delta\Lambda=10^{-10}$.}
		\label{LE1}
	\end{center}
\end{figure*}

\begin{figure*}[!ht]
	\begin{center}
		\subfloat[\label{tlag1a}][]{%
			\includegraphics[width=0.4\textwidth]{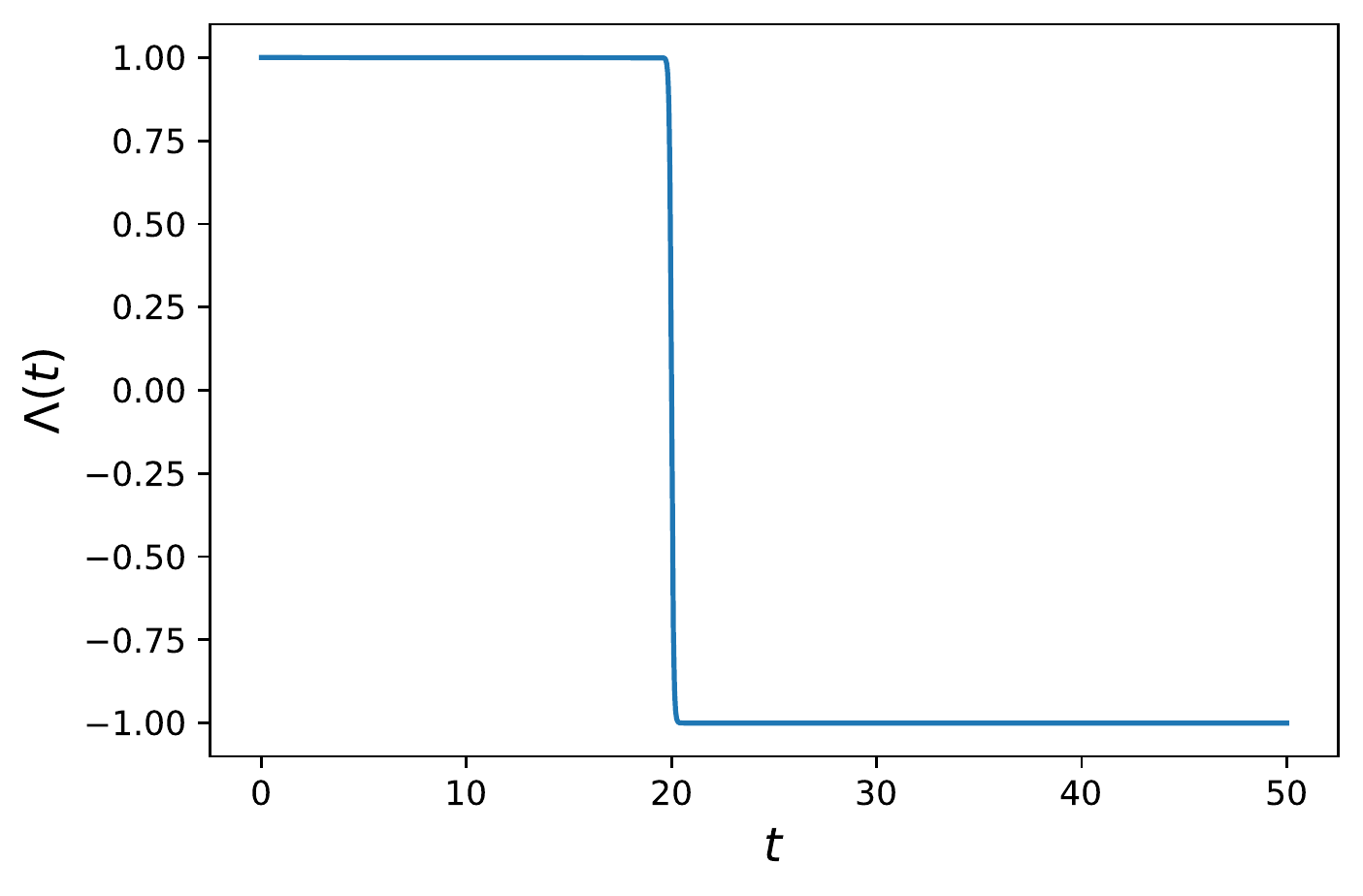}
		}
		\subfloat[\label{tlagb}][]{%
			\includegraphics[width=0.4\textwidth]{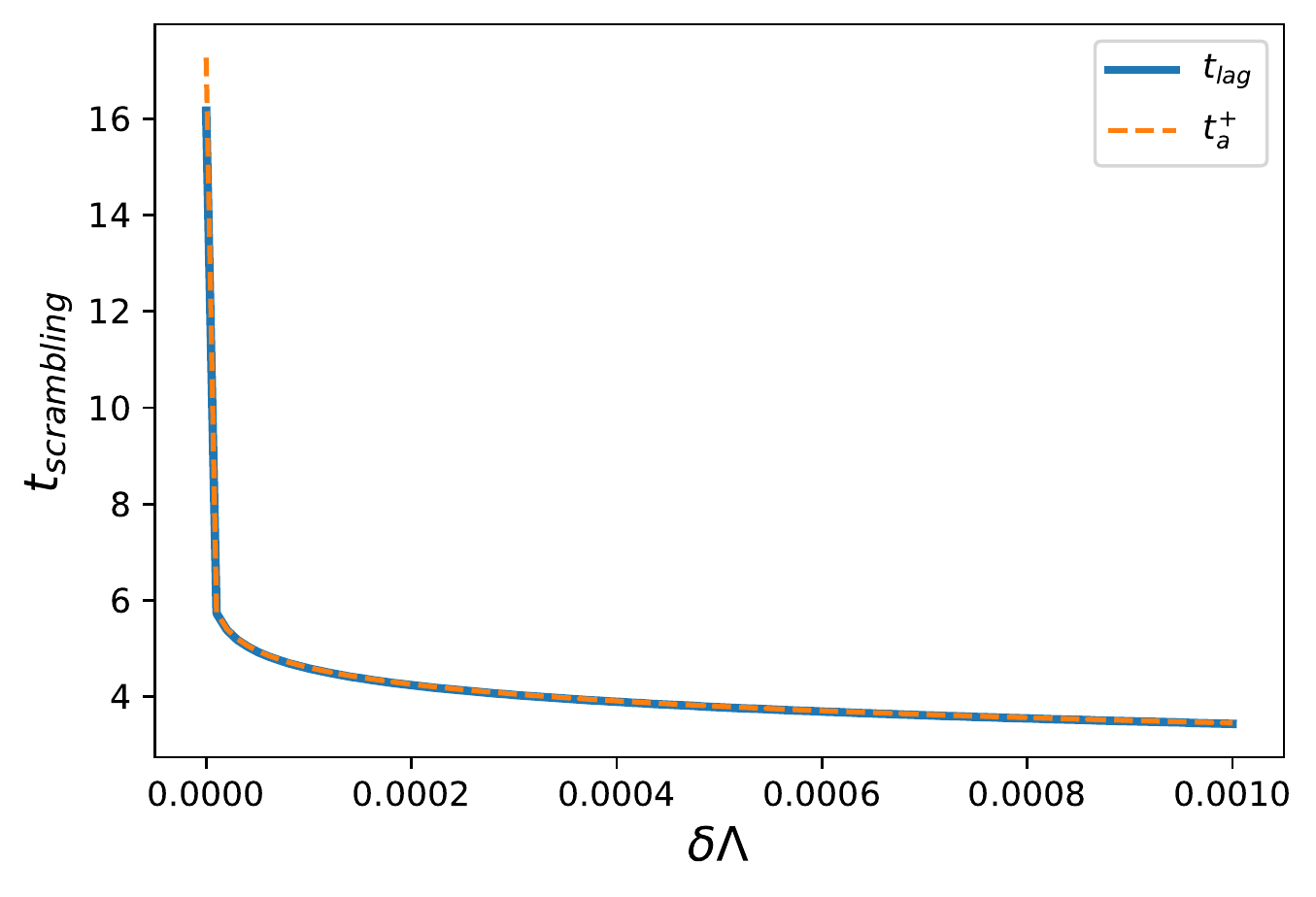}
		}
		
		\caption{(a) Quench function and (b) scrambling time for exponential decay. Here, $t_q=20$, $P=2$ and $Q=10$.}
		\label{tlag1}
	\end{center}
\end{figure*}

\begin{figure*}[!ht]
	\begin{center}
		\subfloat[\label{tlag1c}][]{%
			\includegraphics[width=0.4\textwidth]{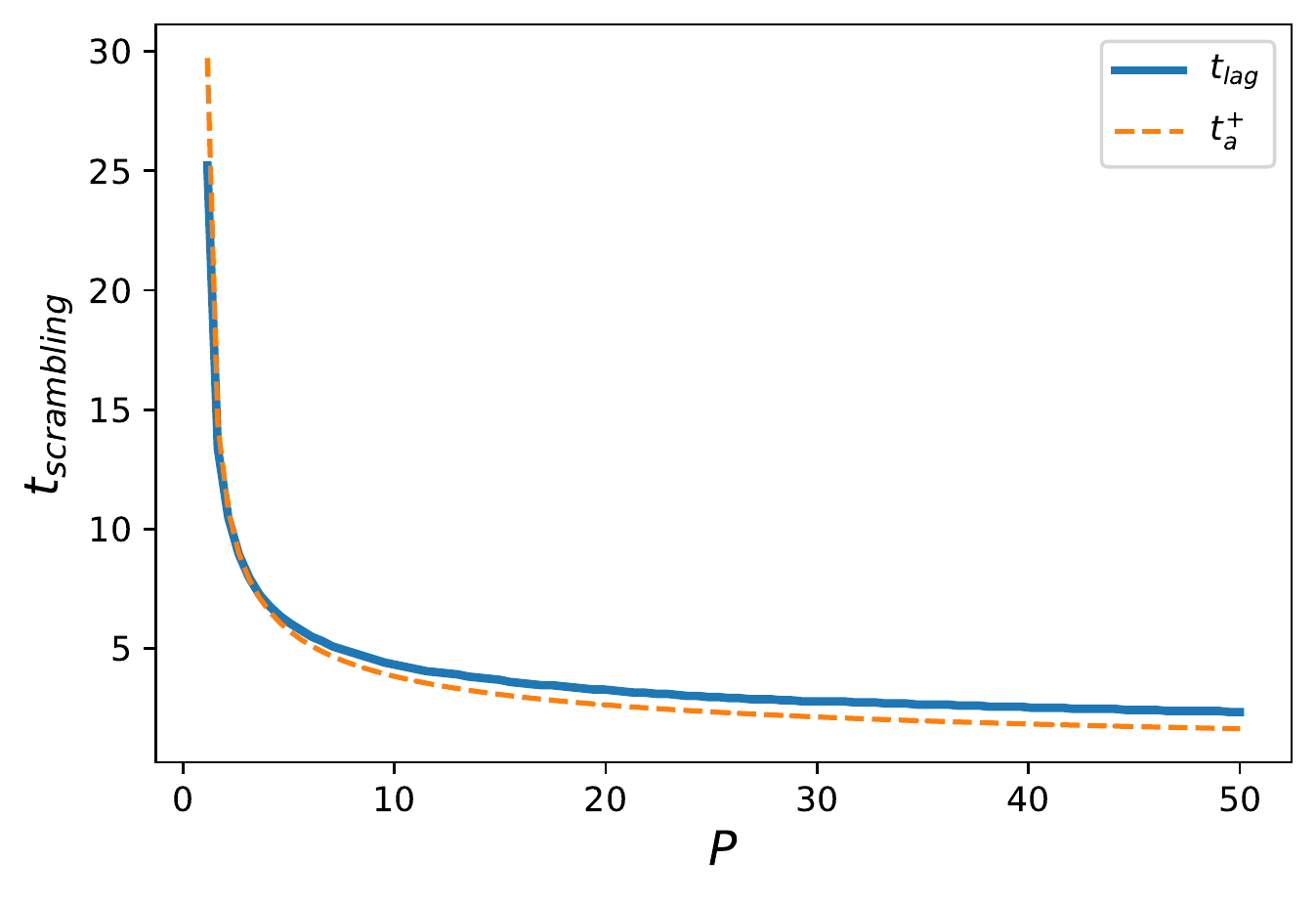}
		}
		\subfloat[\label{tlagd}][]{%
			\includegraphics[width=0.4\textwidth]{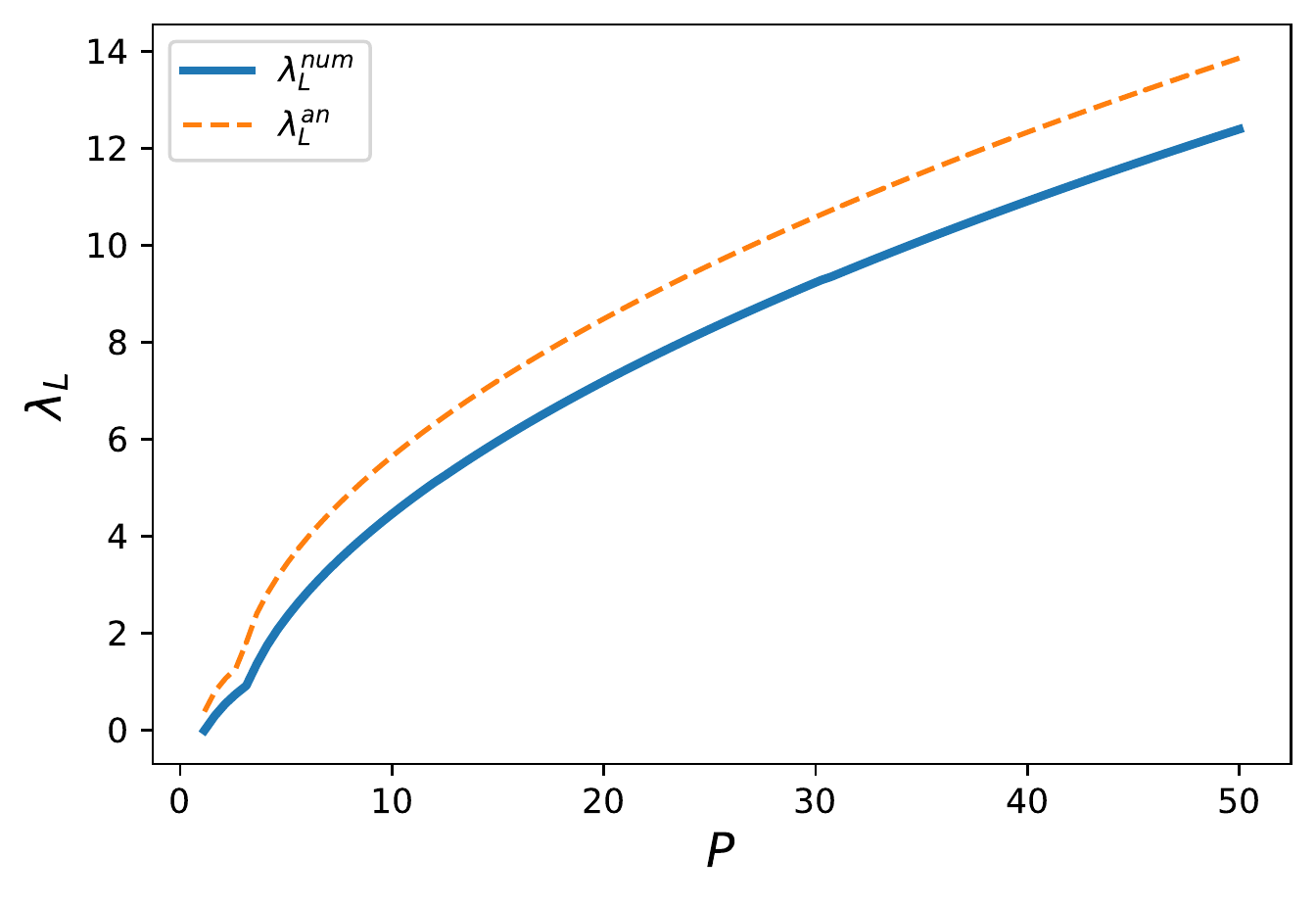}
		}
		
		\caption{(a) Scrambling time and (b) Lyapunov exponents of the system with respect to quench depth $P$. Here, $t_q=20$, $Q=10$, $\delta\Lambda=10^{-10}$ and $t_{max}=45$. {The deviation in the right plot is found to reduce for a longer time $t_{\rm max}$, but it is constrained by the exponentially growing scaling parameters ($b_k(t)$) that demand higher computational capacity.}}
		\label{tlag2}
	\end{center}
\end{figure*}

Similar to previous sections, we now show that the analytic results obtained in the late-time limit match with numerics for a generic quench function. We consider the following evolution of the rescaled frequency (quench function):
\begin{equation}\label{quench}
	\Lambda(t)=1-\frac{P}{2}\left(1+\tanh{\left[Q(t-t_q)\right]}\right),
\end{equation}
where $P=\abs{\Lambda(\infty)-\Lambda(-\infty)}$ is the depth of quench, $Q$ is the speed of quench, and $t_q$ is the time about which the quench function is centered. The results of the numerics are plotted in \ref{LE1} - \ref{tlag2} which we discuss below:  \ref{LE1} shows that: 
\begin{itemize}
\item The Loschmidt echo remains very close to $1$ for a stable/zero-mode, whereas for an inverted mode, the Loschmidt echo begins to exponentially decay as predicted.
\item There is a time lag for the exponential decay to kick in from when the quench occurs. 
\end{itemize}

To further investigate this, in \ref{tlag1} and \ref{tlag2}, we plot the lag with respect to quench parameters. The lag is calculated as the time difference $t_{lag}$ between the beginning of quench $t_{quench}$ when $\Lambda(t+\delta t)-\Lambda(t)>10^{-5}$, and $t_{exp}$, the beginning of exponential decay when $\mathscr{M}(t+\delta t)-\mathscr{M}(t)>10^{-5}$. The time steps here are fixed to be $\delta t\sim 0.005$. From the two figures, we see that the lag matches with the scrambling time that were obtained for late-times in Eq. \eqref{scram}. Also, as can be seen in the \ref{tlag2}, the numerically evaluated Lyapunov exponent ($\lambda_L^{num}=-(t_{max}-t_{quench})^{-1}\log{\mathscr{M}(t_{max})}$) matches with the late-time predictions in \eqref{lyap}.

The above results are obtained for the rescaled Hamiltonian, and, as mentioned, the above results are identical to a group of systems with the same $\Lambda-$class. 
Using the dynamical scaling symmetry in Sec. \eqref{sec:scaling}, we can infer some key properties of 
Loschmidt echo for the original Hamiltonian [$\tilde{H}(\tilde{t})$]. This is the third key result of this work, regarding which we would like to discuss the following points: First, under the dynamical scaling symmetry, we have:
\begin{equation}
    \tilde{\mathscr{M}}(\tilde{t},\delta\tilde{\omega}^2)=\mathscr{M}(t\to\tilde{\alpha}\tilde{t},\delta\Lambda\to\tilde{\alpha}^{-2}\delta\tilde{\omega}^2)\quad;\quad \tilde{\omega}_j(\tilde{t})=\tilde{\alpha}\omega_j(t\to\tilde{\alpha}\tilde{t}) \quad;\quad \tilde{v}_j=\tilde{\alpha}v_j
\end{equation}
Therefore, the scrambling time and Lyapunov exponents for the original system are given by:
\begin{equation}
    \tilde{t}_{\rm scram}=\tilde{\alpha}^{-1}t_{\rm scram} \quad;\quad \tilde{\lambda}_L=\tilde{\alpha} \lambda_L \, .
\end{equation}
We thus see that a larger coupling leads to quicker scrambling and increased exponential sensitivity to initial conditions. Second, while scrambling time tells us how quickly information about the original wave-function is lost when the system is subject to small change $\delta\Lambda$ in the initial conditions, the Lyapunov exponent is independent of $\delta\Lambda$ and is instead a characteristic feature of inverted modes in the system. For the CHO, we may therefore relate it to entanglement entropy as follows:
\begin{equation}
    \lambda_L=\lim_{t\to\infty}\frac{1}{t}S(t)=\sum_{i} v_i,
\end{equation}
where $v_i$ is summed over all the inverted modes in the system at late times. {The late-time linear growth of entanglement entropy for CHO is, therefore, a quantum version of a well-known relation for statistical entropy in classical chaotic systems~\cite{1999Latora.BarangerPhys.Rev.Lett.,2018Bianchi.etalJournalofHighEnergyPhysics}:
\begin{equation}
    S=h_{\rm KS} t,
\end{equation}
where $h_{\rm KS}$ is the Kolmogorov-Sinai rate defined as the sum of positive Lyapunov exponents~\cite{1977PesinRussianMathematicalSurveys}. Lyapunov exponents typically characterize the exponential divergence of nearby phase-space trajectories of the entire dynamical system. However, 
the entanglement entropy exactly mirrors its classical, full system counterpart, whose production rate is determined by these exponents. Nevertheless, such a mapping between classical and quantum measures of entropy in dynamical systems is not always trivial~\cite{2018Bianchi.etalJournalofHighEnergyPhysics}, particularly when we consider arbitrary subsystem sizes as discussed in Sec. \ref{sec:scalarfielddyn}.}

\subsection{Circuit Complexity}\label{sec:complexity}

The unitary time evolution operator \eqref{eq:UnitaryOper} can, in principle, be implemented as a quantum circuit. Circuit complexity~\cite{2017Jefferson.MyersJournalofHighEnergyPhysics,2019Ali.etalJournalofHighEnergyPhysics,2020Ali.etalPhys.Rev.D} measures the computational cost associated with each such circuit and essentially tells us with what ease a certain target state can be prepared. To calculate circuit complexity from the wave-function, let us first consider a unitary transformation $U(t)$ that represents a quantum circuit that inputs a reference state $\ket{\Psi^R}$ at time $t=0$ and outputs a target state $\ket{\Psi^T}$ at a later time $t$:
\begin{equation}
	\ket{\Psi^T(t)}=U(t)\ket{\Psi^R(0)}
\end{equation}
The unitary operator can be written as a path-ordered exponential as follows~\cite{2017Jefferson.MyersJournalofHighEnergyPhysics,2019Ali.etalJournalofHighEnergyPhysics,2020Ali.etalPhys.Rev.D}:
\begin{equation}
	U(t)=\overleftarrow{\mathcal{P}} \exp \left(i \int_{0}^{t} d \tau H(t)\right),
\end{equation}
where $H(t)$ is Hermitian. We can decompose this operator as follows:
\begin{equation}
	H(t)=\sum_IY^I(t)M_I
\end{equation}
Here, the basis $\{M_I\}$ is the set of fundamental gates, and the control functions $\{Y^I(t)\}$ determine the contribution of each gate to the circuit. For instance, the scaling/entangling gate is one such fundamental gate \cite{2017Jefferson.MyersJournalofHighEnergyPhysics}:
\begin{equation}
	Q_{ab}=\exp{\epsilon M_{ab}}\quad;\quad M_{ab}=ix_ap_b+\frac{\delta_{ab}}{2},
\end{equation}
where $\epsilon$ is an infinitesimal parameter. The functions $\{Y^I(t)\}$ can be obtained using the identity:
\begin{equation}
	Y^{I}(t) M_{I}=\partial_{t} U(t) U^{-1}(t)
\end{equation}
 To proceed further, let us rewrite the ground state for the CHO as follows:
\begin{equation}
	\Psi(t)=\mathscr{N}(t)\exp{-\frac{1}{2}\tilde{X}^TW(t)\tilde{X}},
\end{equation}
where $W(t)$ is a diagonal matrix and $\mathscr{N}(t)$ is the normalization factor:
\begin{equation}
	W_{nn}=\frac{\omega_n(0)}{b_n^2(t)}-i\frac{\dot{b}_n(t)}{b_n(t)}\quad;\quad \mathscr{N}(t)=\left(\prod_{n}\frac{\omega_n(0)}{\pi b_n^2(t)}\right)^{1/4}\exp{-\frac{i}{2}\sum_{n=1}^N\omega_n(0)\tau_n}\quad;\quad n=+,-
\end{equation}
We now choose a matrix representation for the fundamental gates which act on diagonal matrix $W$ as follows:
\begin{equation}
	W(t)=U(t).W(0).U^T(t)\quad;\quad U=\exp{\sum_k\alpha_k(t)M_k^{D}},
\end{equation}
where $\{\alpha_k\}$ are complex, and the $\{M_k^D\}$ are diagonal generators of $GL(N,C)$ (since $W$ is complex) with only one identity in the $(k,k)$ position. We now apply the boundary conditions:
\begin{equation}
	W(t)=U(t).W(0).U^T(t)\quad;\quad W(0)=U(0).W(0).U^T(0)
\end{equation}
We may now parameterize the unitary operator as follows:
\begin{equation}
	U(t)=\overleftarrow{\mathcal{P}} \exp \left(\int_{0}^{t}Y^I(t)M_I dt\right)
\end{equation}
The control functions can therefore be obtained from:
\begin{equation}
	Y^{I}=\operatorname{tr}\left(\partial_{t} \tilde{U}(t) \cdot \tilde{U}(t)^{-1} \cdot\left(M^{I}\right)^{T}\right),
\end{equation}
While complexity measures the cost of the optimal circuit that implements unitary, there can be multiple ways in which the contributions from each gate can be accumulated. Here, we choose the following definition:
\begin{equation}
	C(U)=\sum_{+,-}\abs{\alpha_k}
\end{equation}

For the CHO, the above choice of calculating complexity from the wave-function leads to~\cite{2019Ali.etalJournalofHighEnergyPhysics}:
\begin{equation}
\label{def:CWF}
	C_{WF}=\frac{1}{2}\sum_{k=+,-}\left[\left\{\frac{1}{2}\ln\left(\frac{\omega_k^2(0)}{b_k^4(t)}+\frac{\dot{b}_k^2(t)}{b_k^2(t)}\right)\right\}^2+\left\{\tan^{-1}\left(\frac{\dot{b}_k(t)b_k(t)}{\omega_k(0)}\right)\right\}^2\right]^{1/2}=\sum_k C_{WF}^{(k)}
\end{equation}
Another popular definition involves computing the circuit complexity from covariance matrix~\cite{2020Ali.etalPhys.Rev.D} as opposed to the wave-function, leading to the following expression~\cite{2019Ali.etalJournalofHighEnergyPhysics}:
\begin{equation}
\label{def:CCM}
	C_{CM}=\frac{1}{2}\sum_{k=+,-}\cosh^{-1}\left\{\frac{\omega_{k}^{2}(0)\left(1+b_k^2(t)\right)+\dot{b}_k^2(t)}{2 \omega_{k}^2(0)}\right\}=\sum_k C_{CM}^{(k)}
\end{equation}
Unlike $C_{CM}$, $C_{WF}$ is able to distinguish between the evolution routes of fidelity and Loschmidt echo~\cite{2019Ali.etalJournalofHighEnergyPhysics}. In this work, we  further elaborate on the features that distinguish the two measures, along with their connection with entanglement entropy. 

With the help of asymptotic solutions to the Ermakov equation, we can also calculate the long term behavior of complexity in the two categories:
\begin{itemize}
	\item Zero mode: Substituting $b_k(t)\sim\omega_k(0)t$ in Eqs. (\ref{def:CWF}, \ref{def:CCM}), we have:
	\begin{equation}
		C_{WF}\sim \frac{1}{2}\log{t} \qquad;\qquad C_{CM}\sim\log{t}
	\end{equation}
	
	\item Unstable mode: Let us consider the mode contributions individually. Substituting $b_k(t)\sim\frac{1}{2}\sqrt{1+\frac{\omega_k^2(0)}{v_k^2}}e^{v_kt}$ in Eqs. (\ref{def:CWF}), we have: 
	\begin{equation}
		C_{WF}^{(k)}\sim \begin{cases}
		v_kt & \text{if } 0\ll t\ll t_{\rm sat}^{(k)} \\\frac{1}{2}\sqrt{\frac{\pi^2}{16}+\left(\log{v_k}\right)^2} & \text{if } t\gg  t_{\rm sat}^{(k)}
	\end{cases}
	\end{equation}
	We see that the wave-function complexity saturates after a time $t_{sat}$ given by:
	\begin{equation}
	    t_{sat}= \max \{t_{sat}^{(k)}\} \sim \frac{1}{2v_k}\log{\frac{4\omega_k(0)v_k}{\omega_k^2(0)+v_k^2}},
	\end{equation}
	where $v_k$ corresponds to the inverted mode that takes the longest time to saturate. Adding them up, we see that the complexity at late-times ($t\gg t_{sat}$) becomes constant:
	\begin{equation}
	    C_{WF}= \frac{1}{2}\sum_k\sqrt{\frac{\pi^2}{16}+\left(\log{v_k}\right)^2}
	\end{equation}
	As for $C_{CM}$, we obtain a late-time behaviour that is similar to that of entanglement entropy:
	\begin{equation}
	    C_{CM}\sim \left(\sum_k v_k\right)t=h_{\rm KS}t
	\end{equation}
   {As a result, we demonstrate that, similar to entropy, the rate of development of complexity for unstable quantum systems is identical to the classical Kolmogorov-Sinai rate.}

\end{itemize}

\begin{figure*}[!ht]
	\begin{center}
		\subfloat[\label{chocomp1a}][]{%
			\includegraphics[width=0.4\textwidth]{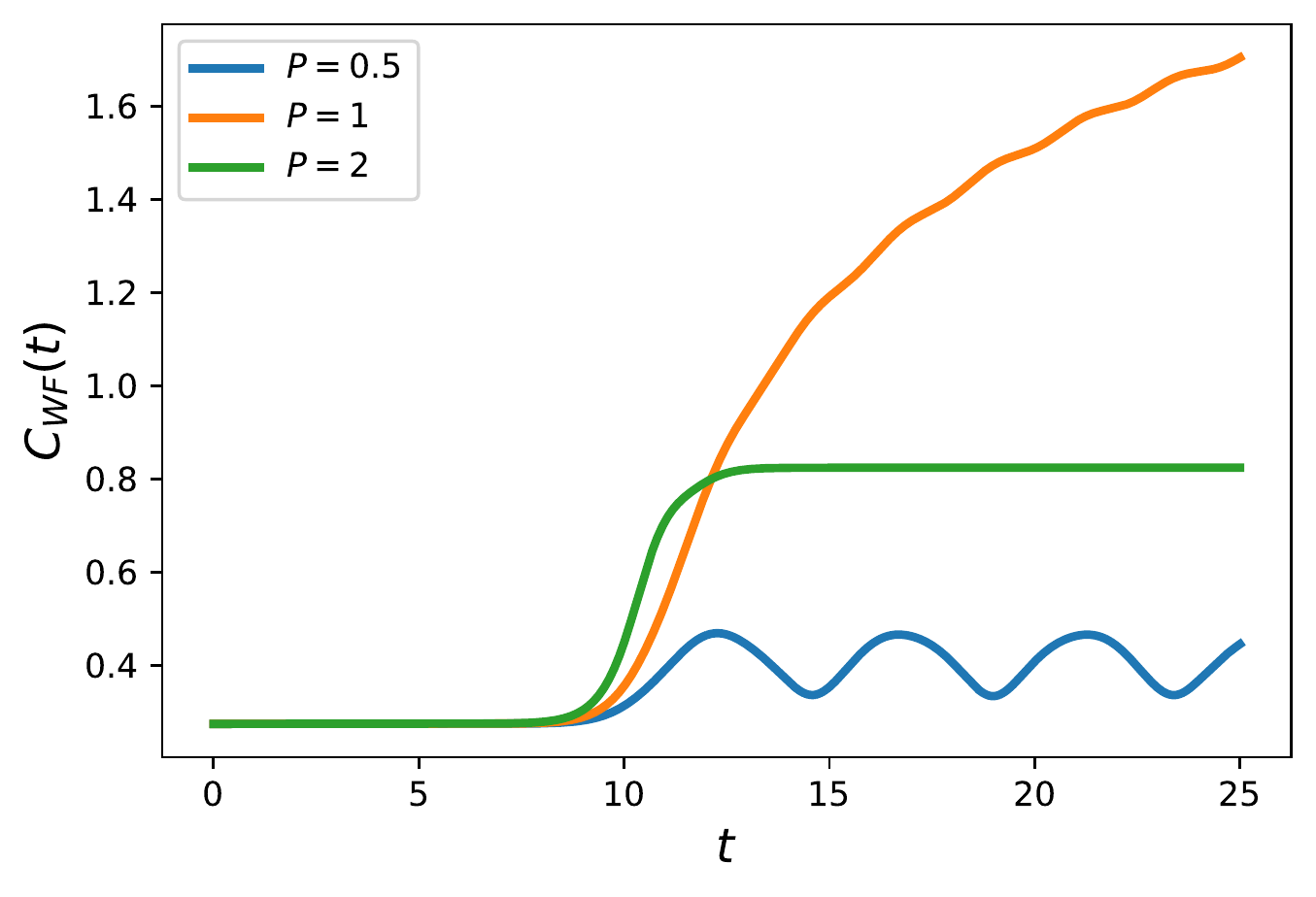}
		}
		\subfloat[\label{chocomp1b}][]{%
			\includegraphics[width=0.4\textwidth]{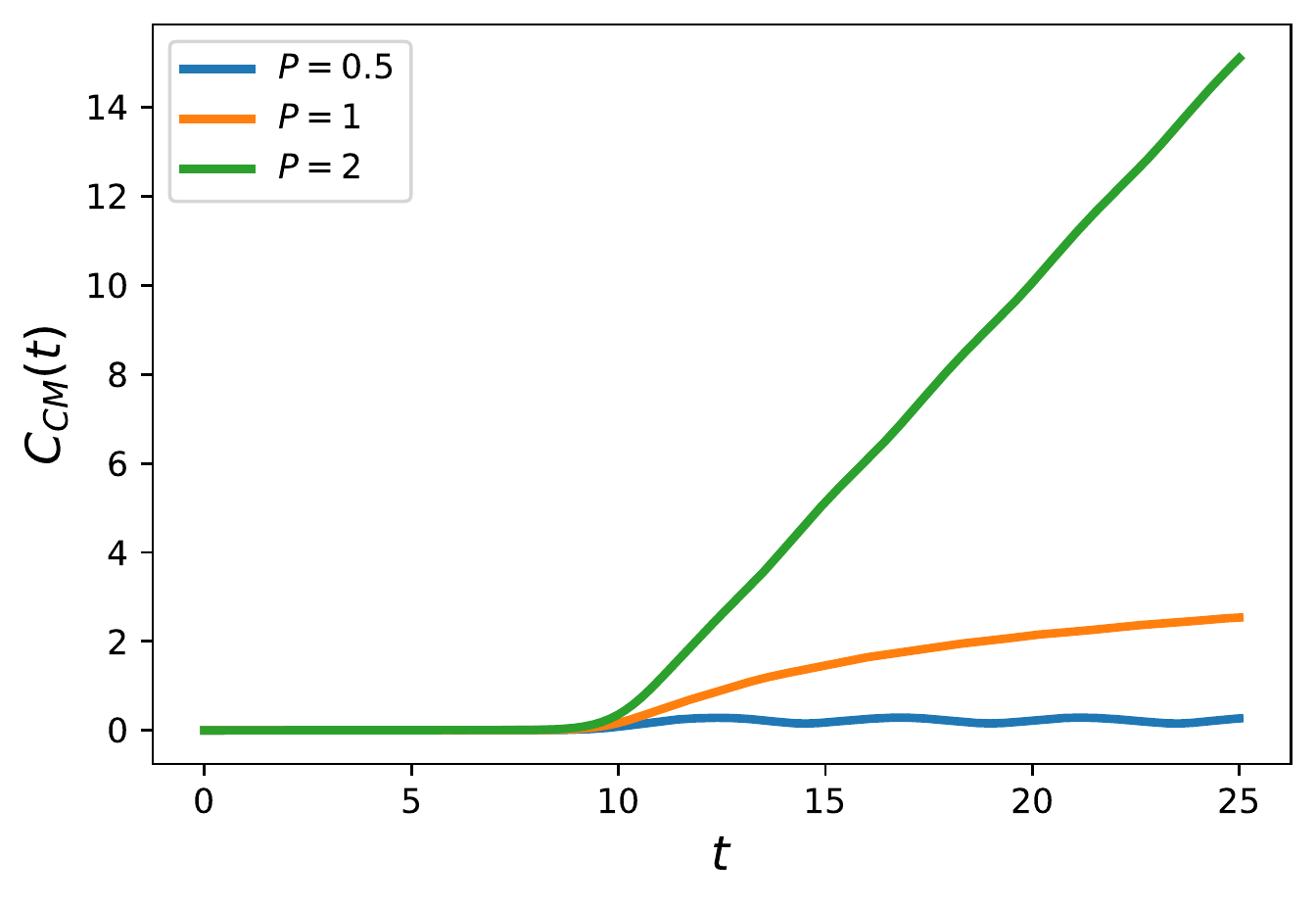}
		}
		
		\caption{Circuit complexity as calculated from (a) wave-function and (b) covariance matrix for a quenched coupled harmonic oscillator. Here, $t_q=10$ and $Q=10$.}
		\label{fig:chocomp}
	\end{center}
\end{figure*}
Like in the case of entropy, Fidelity, and Loschmidt echo, we now show that the analytic results obtained in the asymptotic limit match with the numerically evaluated complexity for a generic quench function.
For the quench model (\ref{quench}), we obtain the dynamics of circuit complexity for the three categories --- stable, zero, and unstable modes. \ref{fig:chocomp} is the plot of the numerical results from which we infer the following: The stable oscillator modes result in a periodic behavior. In contrast, the onset of metastable (zero) and unstable (inverted) modes increase complexity. For both $C_{WF}$ and $C_{CM}$, zero modes result in a logarithmic increase with time, similar to that of entanglement entropy. However, for inverted modes, $C_{WF}$ exhibits an initial increase until it attains saturation, whereas $C_{CM}$ exhibits a linear, unbounded increase with time, similar to entanglement entropy. 

These measures can therefore be used depending on what modes we are interested in --- $C_{WF}$ picks up zero-mode contributions at late times, whereas $C_{CM}$ picks up inverted mode contributions at late times. Interestingly, $C_{WF}$ has also been shown to exhibit a linear increase with time for unstable modes in cosmological models~\cite{2020Bhattacharyya.etalPhys.Rev.D}.

Since the results derived in this section hold for all systems in the same $\Lambda(t)$-class, we can obtain the complexity of the original Hamiltonian $[\tilde{H}(\tilde{t})]$ using the dynamical scaling symmetry in Sec. \eqref{sec:scaling}. We see that like entropy and fidelity, complexity also remains invariant:
\begin{equation}
    \tilde{C}_{WF}(\tilde{t})=C_{WF}(t\to \tilde{\alpha}\tilde{t})\quad;\quad \tilde{C}_{CM}(\tilde{t})=C_{CM}(t\to \tilde{\alpha}\tilde{t})
\end{equation}
As a result, the saturation time-scale for wave-function complexity gets rescaled as $\tilde{t}_{sat}=\tilde{\alpha}^{-1}t_{sat}$. Similar to scrambling time, the saturation time decreases with an increasing coupling constant.

\subsection{Connection between correlation measures}\label{sec:connection}

\begin{figure*}[!ht]
	\begin{center}
		\subfloat[\label{}][P=1]{%
			\includegraphics[width=0.4\textwidth]{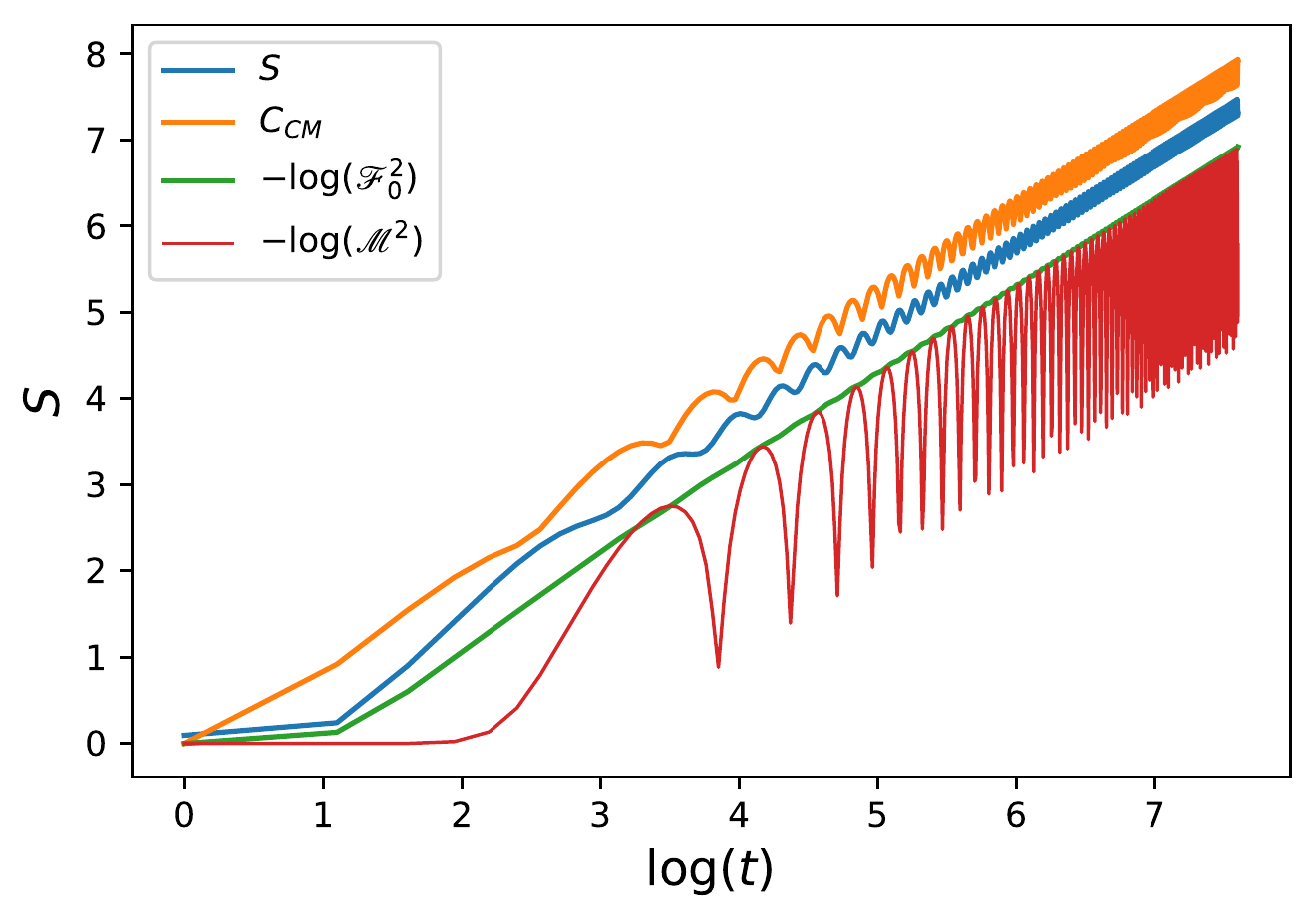}
		}
		\subfloat[\label{}][P=1.5]{%
			\includegraphics[width=0.4\textwidth]{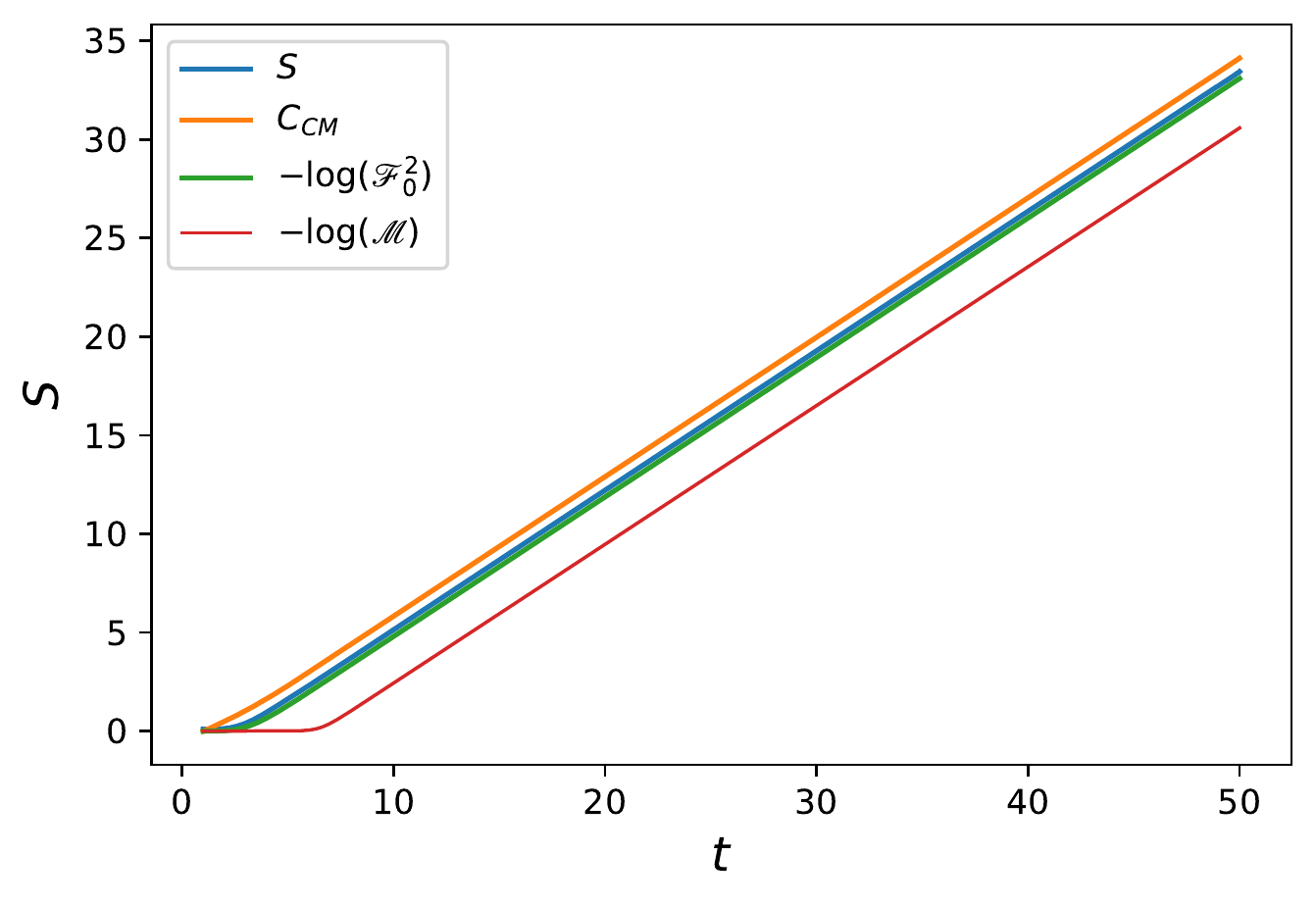}
		}
		
		\caption{Asymptotic behavior of various correlation measures in CHO in the presence of (a) zero mode and (b) inverted mode. Here, we use the quench function in \eqref{quench} where $Q=1$ and $t_q=2$. For calculating Loschmidt echo $\mathscr{M}(t)$, we have fixed $\delta\Lambda=0.01$.}
		\label{fig:asymptotic}
	\end{center}
\end{figure*}

Having studied the dynamics of CHO using dynamical scaling symmetry, we now arrive at the connection between these four correlation measures in the asymptotic limit. In the presence of a zero-mode/inverted modes at late-times, we see that:
\begin{equation}
    S \sim -\log{\mathscr{F}_0^2}\sim C_{CM} \sim \begin{cases}
		-\log{\mathscr{M}^2}\sim\log{t}& \text{zero mode}\\ %\lim_{t\to\infty}\omega_{min}^2=0\\
		-\log{\mathscr{M}}\sim h_{\rm KS}t & \text{inverted mode} %\lim_{t\to\infty}\omega_{min}^2<0
	\end{cases}
\end{equation}
The above relations for correlation measures in the presence of a zero-mode/inverted mode instability can be observed in \ref{fig:asymptotic}. {In classical chaotic systems, the Lyapunov exponents characterize the exponential divergence of nearby phase-space trajectories of the full dynamical system. In quantum systems, while the earlier notion of chaos was initially attributed to systems whose classical counterparts were already known to be chaotic, its more recent definitions account for exponential sensitivity of the wave-function to nearby paths in the parameter space via Loschmidt echo \cite{Qu2021}, or in the space of all unitary transformations via circuit complexity~\cite{2019Ali.etalJournalofHighEnergyPhysics}. The quantum Lyapunov exponents that characterize this sensitivity are the inverted modes in the system, which in turn determine entanglement entropy analogously to its classical counterpart, the Kolmogorov-Sinai entropy. Despite being a subsystem measure, entanglement entropy is thermal in the presence of instabilities~\cite{2016Kaufman.etalScience}, and converges with the global measures such as fidelity, Loschmidt echo, and complexity at late times. Such a late-time convergence is still intact even when the Kolmogorov-Sinai rate $h_{\rm KS}$ approaches zero, and the resultant zero-mode in the system leads to a logarithmic growth typical of metastable~\cite{2018Hackl.etalPhys.Rev.A} or MBL (Many-Body Localized) phases~\cite{2013Serbyn.etalPhys.Rev.Lett.,2020Anza.etalQuantum}.}

\section{Massive Scalar Field in $1+1$-dimensions}
\label{sec:scalarfielddyn}

In the earlier two sections, we used dynamical scaling symmetry in time-dependent CHO to study the late-time behavior of entanglement and its relation to measures that quantify instabilities in the system, such as Loschmidt echo and circuit complexity. This section studies the implications of scaling symmetry in lattice-regularized time-dependent scalar field theories. The Hamiltonian of a time-dependent massive scalar field ($\tilde{\varphi}$) in $(1+1)-$dimensions is~\cite{2020Chandran.ShankaranarayananPhys.Rev.D,2021Jain.etalPhys.Rev.D}:
\begin{eqnarray}
	\tilde{H}=\frac{1}{2}\int d\tilde{x} \left[\tilde{\pi}^2+(\nabla\tilde{\varphi})^2+\tilde{m}_f^2(\tilde{t}) \, \tilde{\varphi}^2\right]
\end{eqnarray}
where $\tilde{m}_f(\tilde{t})$ is the mass of the scalar field with an explicit time-dependence. To evaluate the real-space entanglement entropy, we discretize the above Hamiltonian into a chain of harmonic oscillators by imposing a UV cutoff $\tilde{a}$ and an IR cutoff $\tilde{L}=(N+1)\tilde{a}$. On employing a mid-point discretization procedure, the resultant Hamiltonian takes the following form~\cite{Das2010,2020Chandran.ShankaranarayananPhys.Rev.D}: 
\begin{equation}
	\label{eq:1DHami-Mink}
	\tilde{H}=\frac{1}{2\tilde{a}}\sum_j\left[\pi_j^2+\Lambda(t)\varphi_j^2+(\varphi_j-\varphi_{j+1})^2\right]=\frac{1}{a}H\quad;\quad H=\frac{1}{2}\left[\sum_{j=1}^{N}\pi_j^2+\sum_{i,j=1}^{N}K_{ij}(t)\varphi_i \varphi_j\right]
\end{equation}
where $\Lambda(t)=\tilde{a}^2\tilde{m}_f^2(\tilde{t})$. The time-dependent Hamiltonian \eqref{eq:1DHami-Mink} is crucial for understanding quantum correlations of scalar fields in dynamical backgrounds such as cosmological inflation and understanding the stability of horizons. Appendix \eqref{app:BH-Ham} shows that the Hamiltonian of a massive scalar field propagating in a time-dependent spherically symmetric space-time, when discretized, effectively reduces to \eqref{eq:1DHami-Mink}. In Sec. \eqref{sec:TD-Metrics} we discuss the implications of the results for this model Hamiltonian for the time-dependent space-times. 

Here, we have mapped the degrees of freedom of a scalar field to a lattice of harmonic oscillators with nearest-neighbor coupling. All information about correlations is therefore encoded in the coupling matrix $K(t)$ with time evolution (See Appendix \ref{App:A} for details about the $N-$CHO system). The form of $K(t)$ also depends on the boundary conditions employed~\cite{2020Chandran.ShankaranarayananPhys.Rev.D}:
\begin{itemize}
	\item \textbf{Dirichlet Condition} (DBC) : Here, we impose the condition $\varphi_0=\varphi_{N+1}=0$. The coupling matrix $K_{ij}$ becomes a symmetric Toeplitz matrix with the following non-zero elements:
	\begin{align}
		K_{jj}&=\Lambda(t)+2\nonumber\\
		K_{j,j+1}=K_{j+1,j}&=-1
	\end{align}
	The normal modes are~\cite{2008-Willms-SIAMJournalonMatrixAnalysisandApplications}:
	\begin{eqnarray}\label{eq:dirmodes}
		\tilde{\omega}_k^2(t)=\Lambda(t)+4\sin^2{\frac{k\pi}{2(N+1)}}\quad k=1,..N
	\end{eqnarray}

	\item \textbf{Neumann Condition} (NBC): We impose the condition $\partial_x \varphi=0$ at the 
	two ends of the chain by setting $\varphi_0=\varphi_1$ and $\varphi_{N+1}=\varphi_N$. 
	The resultant coupling matrix is, therefore, a perturbed symmetric Toeplitz matrix whose non-zero elements are given below:
	\begin{align}
		K_{jj\neq1,N}&=\Lambda(t)+2\nonumber\\
		K_{11}=K_{NN}&=\Lambda(t)+1\nonumber\\
		K_{j,j+1}=K_{j+1,j}&=-1
	\end{align}
	The normal modes for this boundary condition are~\cite{2008-Willms-SIAMJournalonMatrixAnalysisandApplications}:
	\begin{eqnarray}\label{eq:neumodes}
		\tilde{\omega}_k^2(t)=\Lambda(t)+4\sin^2{\frac{(k-1)\pi}{2N}};\quad k=1,...N
	\end{eqnarray}
\end{itemize}
In both the above cases, it is to be noted that the smallest mode corresponds to $k=1$. 
On invoking the dynamical scaling symmetry of entanglement that was developed in Section \ref{model}, we can therefore obtain the entanglement entropy of the field ($\tilde{H}$) from the rescaled Hamiltonian $H$ as follows: 
\begin{equation}
    \tilde{S}(\tilde{t})=S\left(t\to \tilde{a}^{-1}\tilde{t}\right) \, .
\end{equation}
From its definition, it is also clear that $\Lambda(t)$ is invariant under the scaling $(\eta)$ transformations:
\begin{equation}
	\tilde{a}\to \eta \tilde{a};\quad \tilde{m}_f\to\eta^{-1}\tilde{m}_f \, .
\end{equation}  
Under these scaling transformations, the entanglement entropy of the original system varies as:
\begin{equation}
    \tilde{S}\left(\eta^{-1} \tilde{m}_f,\eta \tilde{a},\eta \tilde{t}\right)=\tilde{S}\left(\tilde{m}_f,\tilde{a}, \tilde{t} \right)
\end{equation}

In the rest of this section, we will explore the entanglement dynamics of the lattice-regularized field subject to two different quenches.

\subsection{Obtaining late-time behavior using covariance matrix approach}

The covariance matrix is given by $\sigma=\frac{1}{2}M\mathscr{S}(t)M^T$, where $\mathscr{S}=\oplus \mathscr{S}_i$, whose elements are~\cite{2021HabArrih.etalInternationalJournalofGeometricMethodsinModernPhysics}:
\begin{equation}
	\mathscr{S}_i=\begin{bmatrix}
		B_i(t)&-C_i(t)\\-C_i(t)&A_i(t)
	\end{bmatrix},
\end{equation}
where,
\begin{equation}
A_i=\frac{\omega_i(0)}{b_i^2(t)}+\frac{\dot{b}_i^2(t)}{\omega_i(0)}\, ;\quad
B_i=\frac{b_i^2(t)}{\omega_i(0)}\, ; \quad
C_i=\frac{b_i(t)\dot{b}_i(t)}{\omega_i(0)}
\end{equation}
It is also useful to note that:
\begin{equation}
    A_iB_i-C_i^2=1
\end{equation}
For the single oscillator reduced state, the reduced covariance matrix has the following elements~\cite{2014-Mallayya.etal-Phys.Rev.D,2021Jain.etalPhys.Rev.D}:
\begin{equation}
	\sigma_1=\frac{1}{2}\begin{bmatrix}
		\sum_i M_{1i}^2B_i&-\sum_i M_{1i}^2C_i\\-\sum_i M_{1i}^2C_i& \sum_i M_{1i}^2A_i
	\end{bmatrix}
\end{equation}
The entanglement entropy of the system is then given by~\cite{2002Audenaert.etalPhysicalReviewA,2014-Mallayya.etal-Phys.Rev.D,2021Jain.etalPhys.Rev.D}:
\begin{equation}
	S_1=\left(\alpha+\frac{1}{2}\right)\log{\left(\alpha+\frac{1}{2}\right)}-\left(\alpha-\frac{1}{2}\right)\log{\left(\alpha-\frac{1}{2}\right)},
\end{equation}
where $\alpha=\sqrt{\det{\sigma_{1}}}$. If the determinant (and hence $\alpha$) is very large, we may simplify the expression as follows:
\begin{equation}
	S_1\approx\log{\alpha}=\frac{1}{2}\log{\left(\det{\sigma_{red}}\right)}
\end{equation}

First, we note that the diagonalizing matrix $M$ is independent of the mass $\Lambda(t)$~\cite{2021Jain.etalPhys.Rev.D}. Therefore, global mass quenches will not affect the matrix elements of $M$ at any time. It is therefore sufficient to look at the asymptotic behavior of the elements of $\mathscr{S}_i$. Like in the case of CHO, we now consider two categories: %
\begin{itemize}
\item Suppose, at late times, the lowest mode $\omega_k(t)$ corresponds to a zero-mode, we have:
	\begin{equation}
		A_k\sim\frac{1}{\omega_k(0)}\, ;\quad B_k\sim \omega_k(0)t^2\, ;\quad C_k\sim t
	\end{equation}
	The determinant on tracing out the first oscillator can then be rewritten as:
	\begin{equation}
		\det{\sigma_{red}}\sim\frac{1}{4}\left[\left\{M_{1k}^2B_k+\sum_{j\neq k}M_{1j}^2B_j\right\}\left\{M_{1k}^2A_k+\sum_{j\neq k}M_{1j}^2A_j\right\}-\left\{M_{1k}^2C_k+\sum_{j\neq k}M_{1j}^2C_j\right\}^2\right]
	\end{equation}

	At late times, we see that the $B_k\propto t^2$ and $C_k^2\propto t^2$ terms dominate:
	\begin{equation}
		\det{\sigma_{red}}\sim\frac{t^2}{4}\left\{\omega_{k}(0)M_{1k}^2\sum_{j}M_{1j}^2A_j-M_{1k}^4\right\}=\frac{\omega_k(0)M_{1k}^2t^2}{4}\sum_{j\neq k}M_{1j}^2A_j
	\end{equation}
    The entanglement entropy at late times in the presence of a zero-mode reduces to:
	\begin{equation}
	 	S_1\sim\frac{1}{2}\log{\left(\det{\sigma_{red}}\right)}\sim \log{t}
	\end{equation}

	\item Suppose, at late times, a single mode $\omega_k(t)$ becomes inverted mode ($\Lim{t\to\infty}\omega_k\to i v_k$), we see that:
	\begin{equation}\label{corr-elements}
		A_k\sim v_k^2B_k\quad;\quad B_k\sim \frac{1}{4\omega_k(0)}\left\{1+\frac{\omega_k^2(0)}{v_k}\right\}\exp{2v_kt} \quad;\quad C_k\sim v_k B_k
	\end{equation}
	Similar to what was done for zero modes, the determinant can be simplified as follows:
	\begin{equation}\label{determinant}
		\det{\sigma_{red}}\sim \frac{M_{1k}^2B_k}{4}\left[v_k^2\sum_{j\neq k} M_{1j}^2A_j-2v_k\sum_{j\neq k} M_{1j}^2C_j+\sum_{j\neq k}M_{1j}^2B_j\right]
	\end{equation}
	 The entanglement entropy at late times in the presence of an inverted mode, therefore, reduces to:
	\begin{equation}
		S_1\sim\frac{1}{2}\log{\left(\det{\sigma_{red}}\right)}\sim v_k t
	\end{equation}
\end{itemize}
Generalizing these results for an arbitrary subsystem size using the same approach is non-trivial. 
{For inverted modes, however, the entanglement entropy of a subsystem has a
known form and is bounded by its classical counterpart, the Kolmogorov-Sinai entropy that arises from Lyapunov exponents of exponentially diverging phase-space trajectories in chaotic systems~\cite{2018Bianchi.etalJournalofHighEnergyPhysics}:
\begin{equation}\label{eq:KSbound}
    S_n\sim\left(\sum_{k=1}^{2n}\lambda_k\right) t\leq h_{\rm KS}t,
\end{equation}
where $S_n$ is the entanglement entropy of the $n$-oscillator subsystem,  $\{\lambda_k\}$ are the $2n$ largest positive Lyapunov exponents, and $h_{\rm KS}$ is the sum of all positive Lyapunov exponents. It can be noted from here that the half chain entropy $S_{N/2}$ always saturates the bound}. From the Lyapunov exponents derived in Sec. \ref{sec:LoschEcho}, we can deduce that for a massive scalar field:
\begin{equation}\label{eq:Kolmogorov}
    S_n\sim \left(\sum_{k=1}^{2n}v_k\right)t\quad;\quad \lim_{t\to\infty}\omega_k(t)\to iv_k,
\end{equation}
where $\{v_k\}$ are automatically indexed from largest inverted mode to the smallest based on Eqs. \eqref{eq:dirmodes} and \eqref{eq:neumodes}. 

\subsection{Late-time dynamics after quenching}
In this section, we analyze the entanglement dynamics of a lattice-regularized massive scalar field that undergoes two different kinds of quench --- i) Quench of the rescaled field mass by considering a global evolution of $\Lambda(t)$, and ii) Quench of boundary conditions from Dirichlet to Neumann, which is a localized event implemented at the edges of the harmonic chain. We see distinct characteristics in the dynamics that follow, particularly how entanglement peaks travel throughout the system.

\subsubsection{Quench of the scalar field mass}
Similar to previous sections, we will consider the generic quench function \eqref{quench}. From Eqs. \eqref{eq:dirmodes} and \eqref{eq:neumodes}, it is clear that a global evolution in $\Lambda(t)$ will drive the evolution of all the normal modes. Like in the case of CHO, we will consider three categories of evolution: \\[2pt]

\begin{figure*}[!ht]
	\begin{center}
		\subfloat[\label{mq1a}][]{%
			\includegraphics[width=0.4\textwidth]{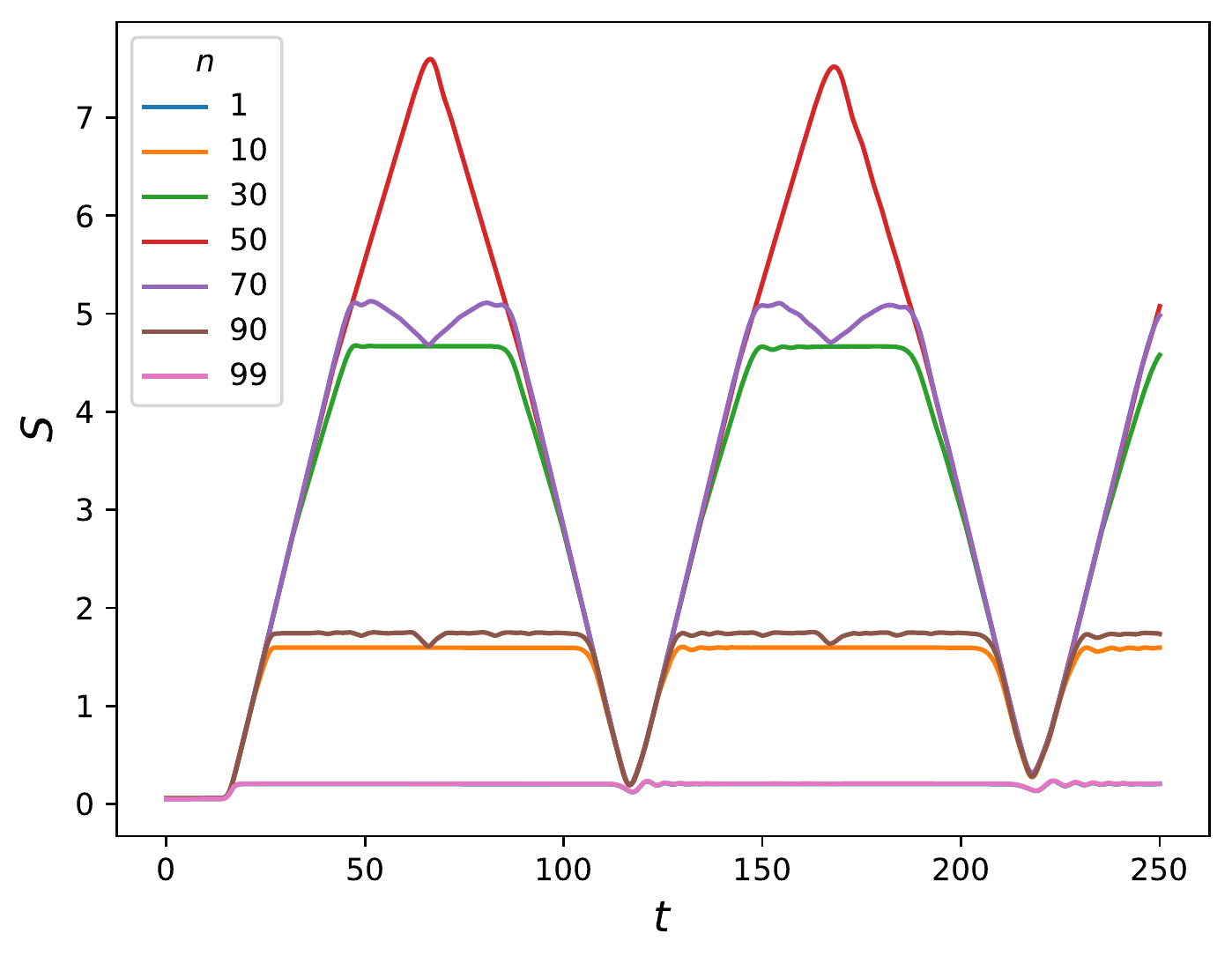}
		}
		\subfloat[\label{mq1b}][]{%
			\includegraphics[width=0.4\textwidth]{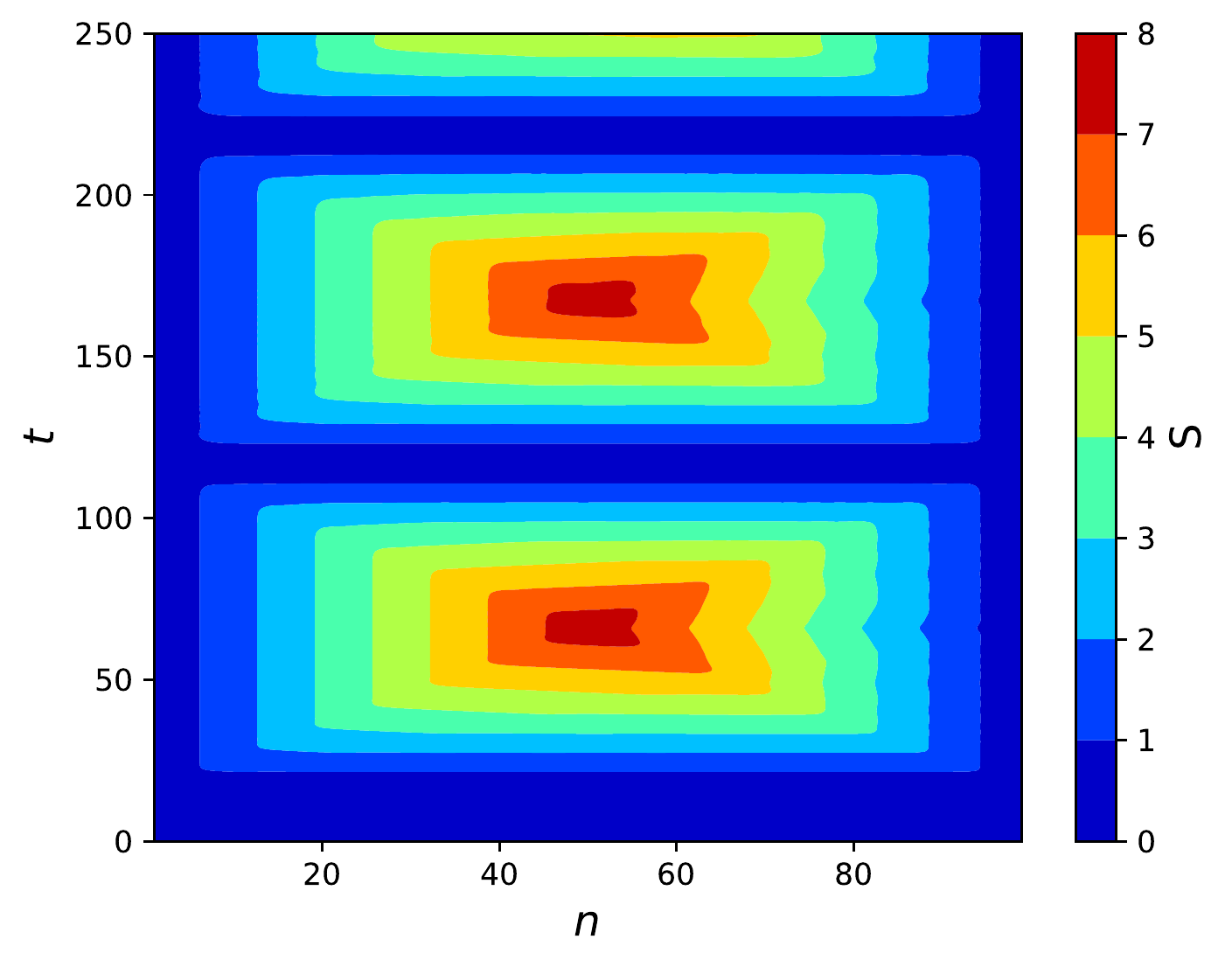}
		}
		
		\caption{Entanglement dynamics after a mass quench ($\Lambda(t)$) for DBC resulting in late-time stable modes --- (a) Time-evolution for various subsystem sizes where we see a periodic linear growth, plateau and descent, and (b) Sub-system scaling of entanglement entropy at each time slice in the evolution. Here, $N=100$, $P=1$, $Q=1$ and $t_q=15$.}
		\label{fig:mq1}
	\end{center}
\end{figure*}

\begin{figure*}[!ht]
	\begin{center}
		\subfloat[\label{mqf1}][]{%
			\includegraphics[width=0.4\textwidth]{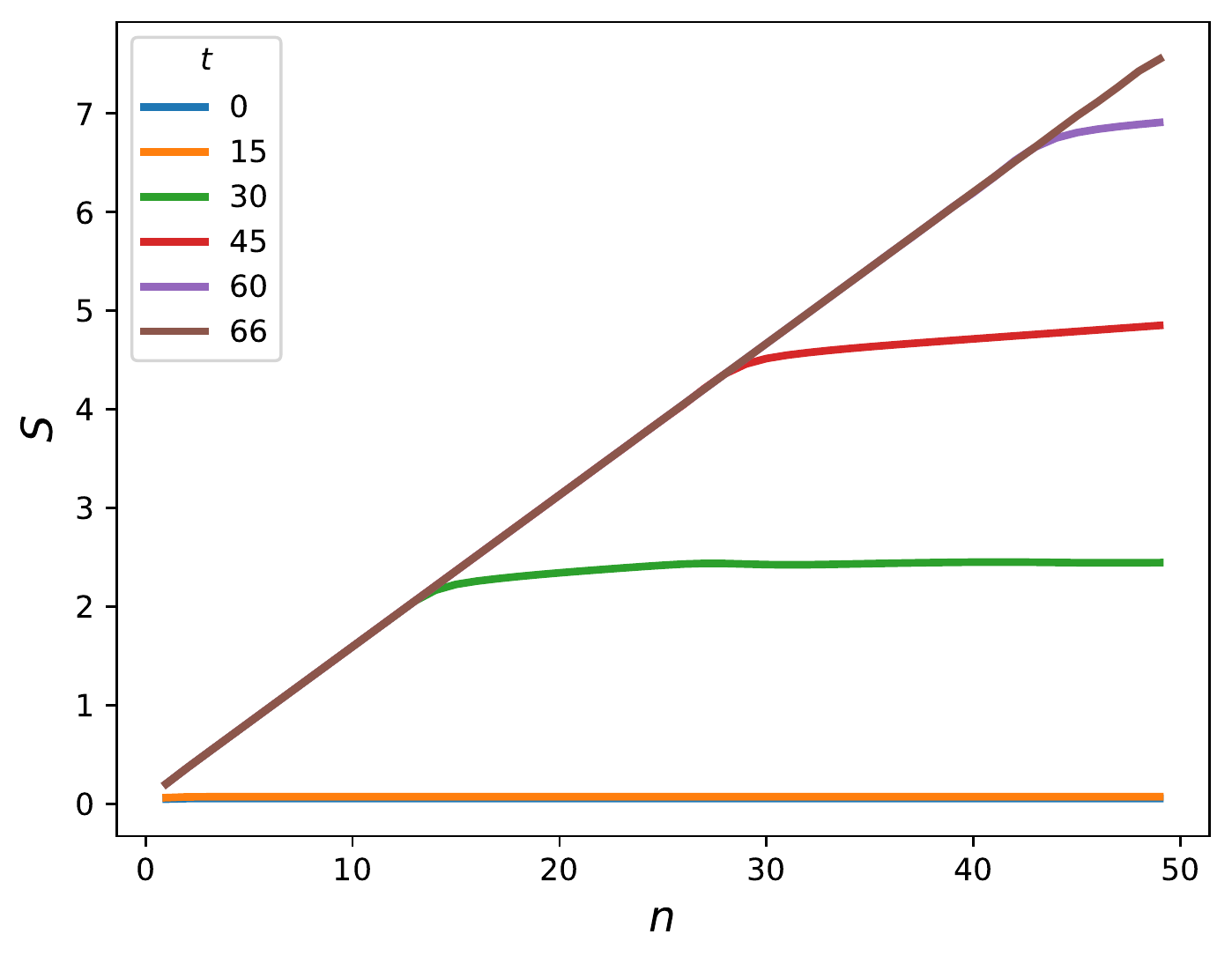}
		}
		\subfloat[\label{mqf2}][]{%
			\includegraphics[width=0.45\textwidth]{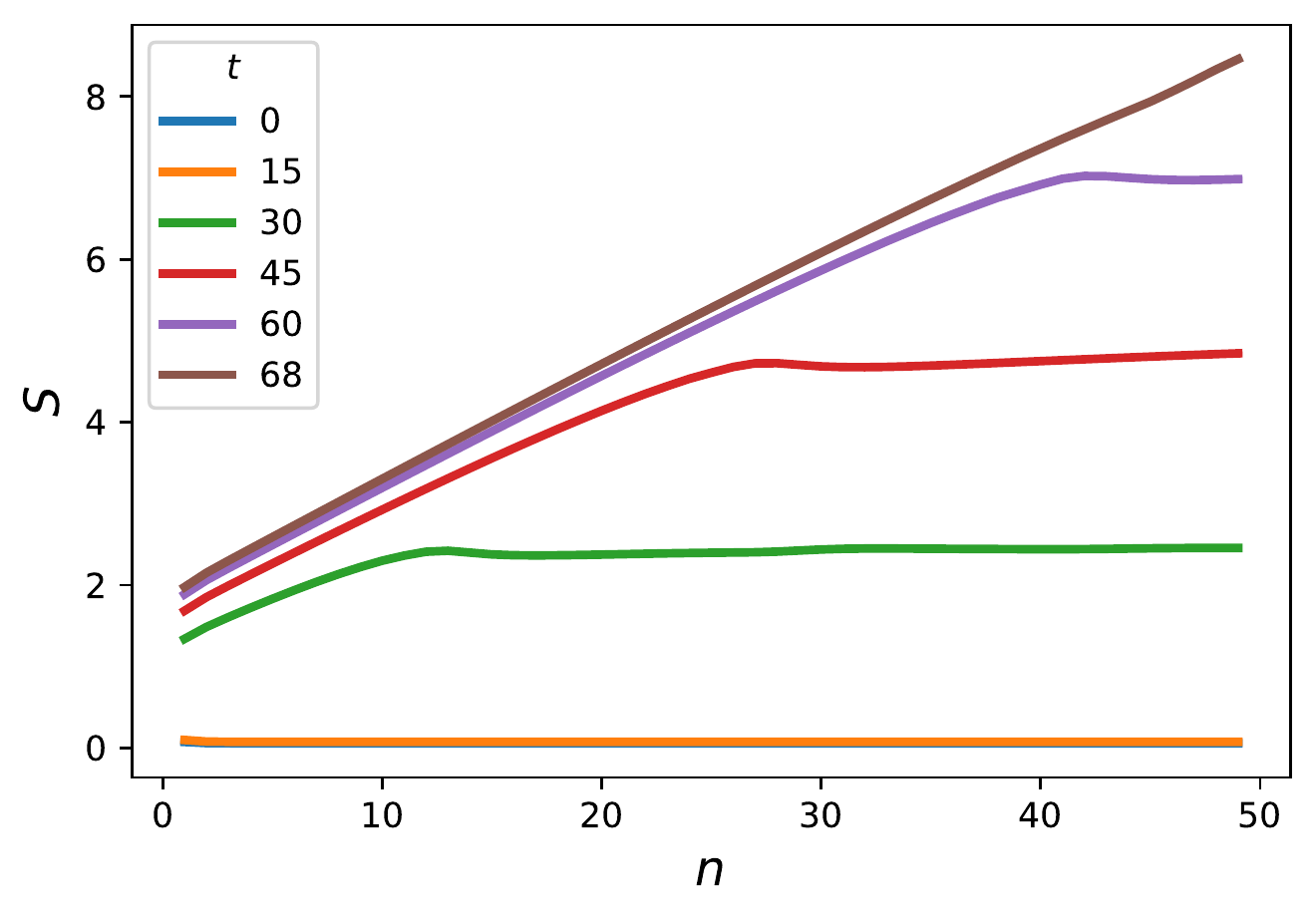}
		}
		
		\caption{Area-law to volume-law transition of entanglement entropy for a massive scalar field with (a) a stable mode spectrum (DBC), and (b) a zero-mode instability (NBC). Here, we use quench function \eqref{quench}, where $P=1$, $Q=1$ and $t_q=15$ for a system of $N=100$ oscillators. At $t=0$, when the rescaled field mass is $\Lambda=1$, the entanglement entropy follows a typical area-law for 1-D systems ($S_n\sim n^0$). At the peak of linear growth of half-chain entropy $S_{N/2}$, the entropy assumes a volume-law ($S_n\sim n$) for 1-D systems. Further evolution indicates that the scaling oscillates between a volume-law (at the peaks of $S_{N/2}$) characteristic of thermal behavior, and an area-law (at the troughs of $S_{N/2}$).}
		\label{fig:area1}
	\end{center}
\end{figure*}

\noindent {\bf Stable modes}: Let us look at a global evolution of $\Lambda(t)$ which only results in stable modes at late times. For this, we may consider $\Lim{t\to\infty}\Lambda(t)\to0$ in the DBC chain, where Eq. \eqref{eq:dirmodes} ensures that all modes remain stable at late-times for a finite $N$. \ref{fig:mq1} is the plot of the numerical results from which we infer the following: First, the entanglement dynamics is multi-oscillatory across various subsystem sizes. Hence, unlike time-independent systems,  we do not observe a fixed subsystem scaling behavior of entropy. Second, there is a periodic linear growth, plateau, and descent of entropy with time. This tells us that entropy periodically mimics inverted-mode dynamics (linear growth) despite all modes being stable. We also see that the time-scale for this linear growth increases with subsystem size, reaching a maximum when the bipartition is at the middle of the chain. The entanglement dynamics is therefore always bounded by the half-chain entropy $S_{N/2}$ at late-times. Lastly, whenever the half-chain entropy peaks, we see from \ref{fig:area1} that the entanglement entropy satisfies the volume law.

\begin{figure*}[!ht]
	\begin{center}
		\subfloat[\label{mq2a}][]{%
			\includegraphics[width=0.4\textwidth]{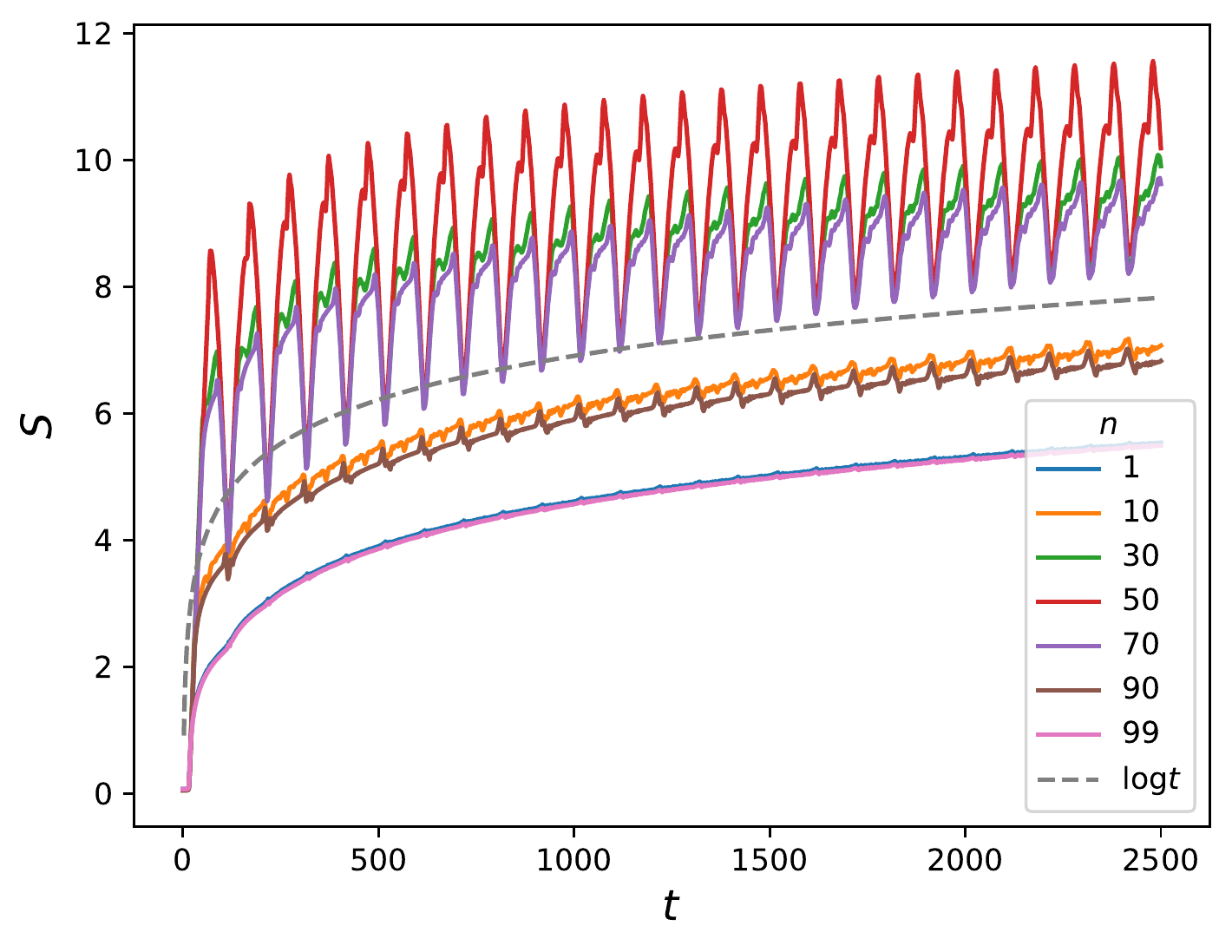}
		}
		\subfloat[\label{mq2b}][]{%
			\includegraphics[width=0.4\textwidth]{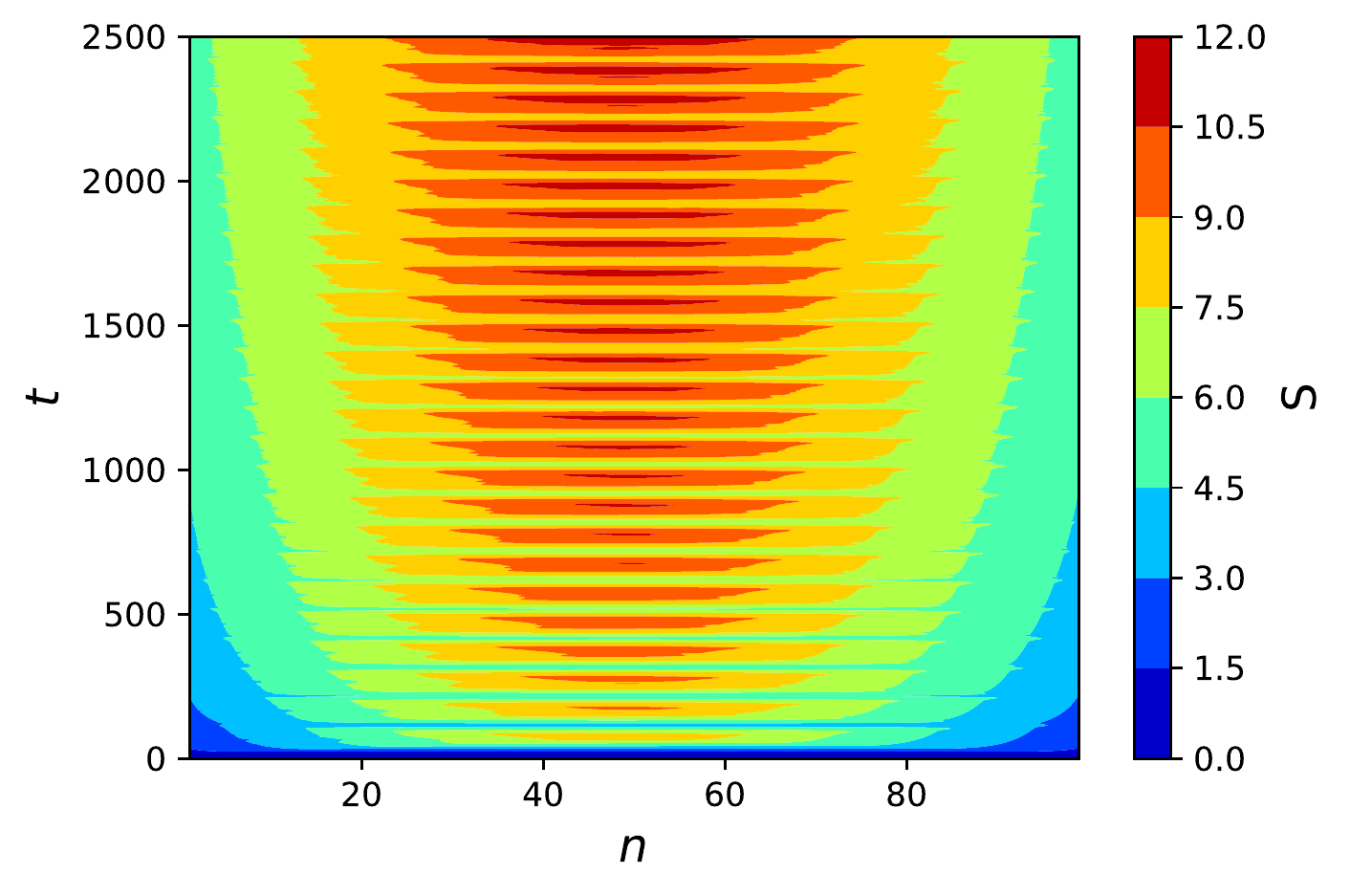}
		}
		
		\caption{Entanglement dynamics after a mass quench ($\Lambda(t)$) for NBC resulting in a late-time zero mode --- (a) Time-evolution for various subsystem sizes where we see an overall logarithmic growth, and (b) Sub-system scaling of entanglement entropy at each time slice in the evolution. Here, $N=100$, $P=1$, $Q=1$ and $t_q=15$.}
		\label{fig:mq2}
	\end{center}
\end{figure*}

\noindent {\bf Zero-mode:} Let us now consider a situation where, at late times, the system has a zero-mode. For this, we may consider the same evolution $\Lambda(t)$ \eqref{quench}, but now for an NBC chain. Here, Eq. \eqref{eq:neumodes} ensures that there will be exactly one zero-mode at late times, even for a finite $N$, in clear contrast with DBC. From \ref{fig:mq2}, we see that there is a logarithmic production of entropy with time that dominates the oscillatory behavior across all subsystem sizes. Just like in DBC, entanglement entropy of any subsystem size at late-times is bounded from above by $S_{N/2}$, corresponding to the half-chain entropy. From \ref{fig:area1}, similar to the case of stable-modes, we see that the volume-law of entropy features during the peaks of $S_{N/2}$. 

\begin{figure*}[!ht]
	\begin{center}
		\subfloat[\label{mq2a}][]{%
			\includegraphics[width=0.4\textwidth]{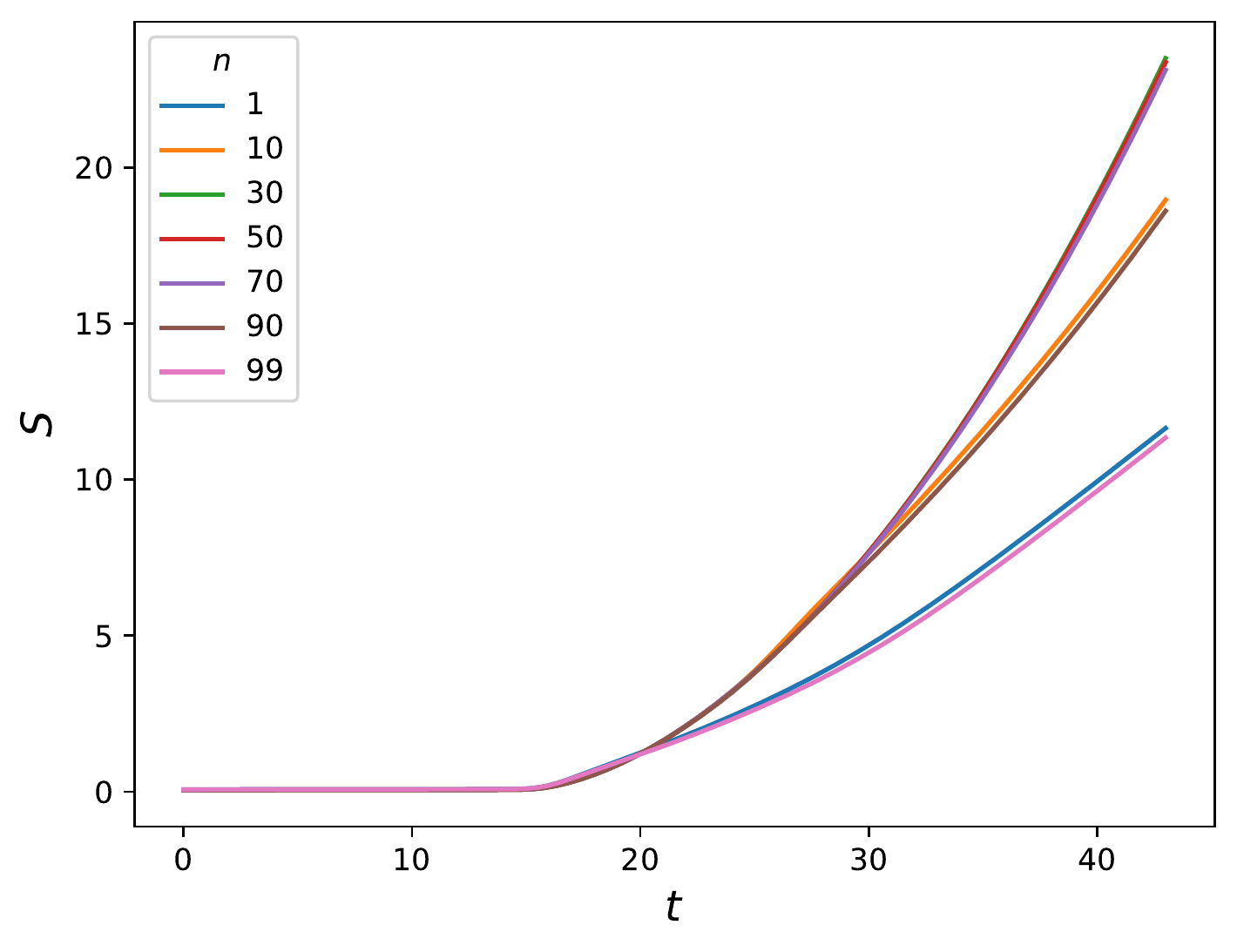}
		}
		\subfloat[\label{mq2b}][]{%
\includegraphics[width=0.4\textwidth]{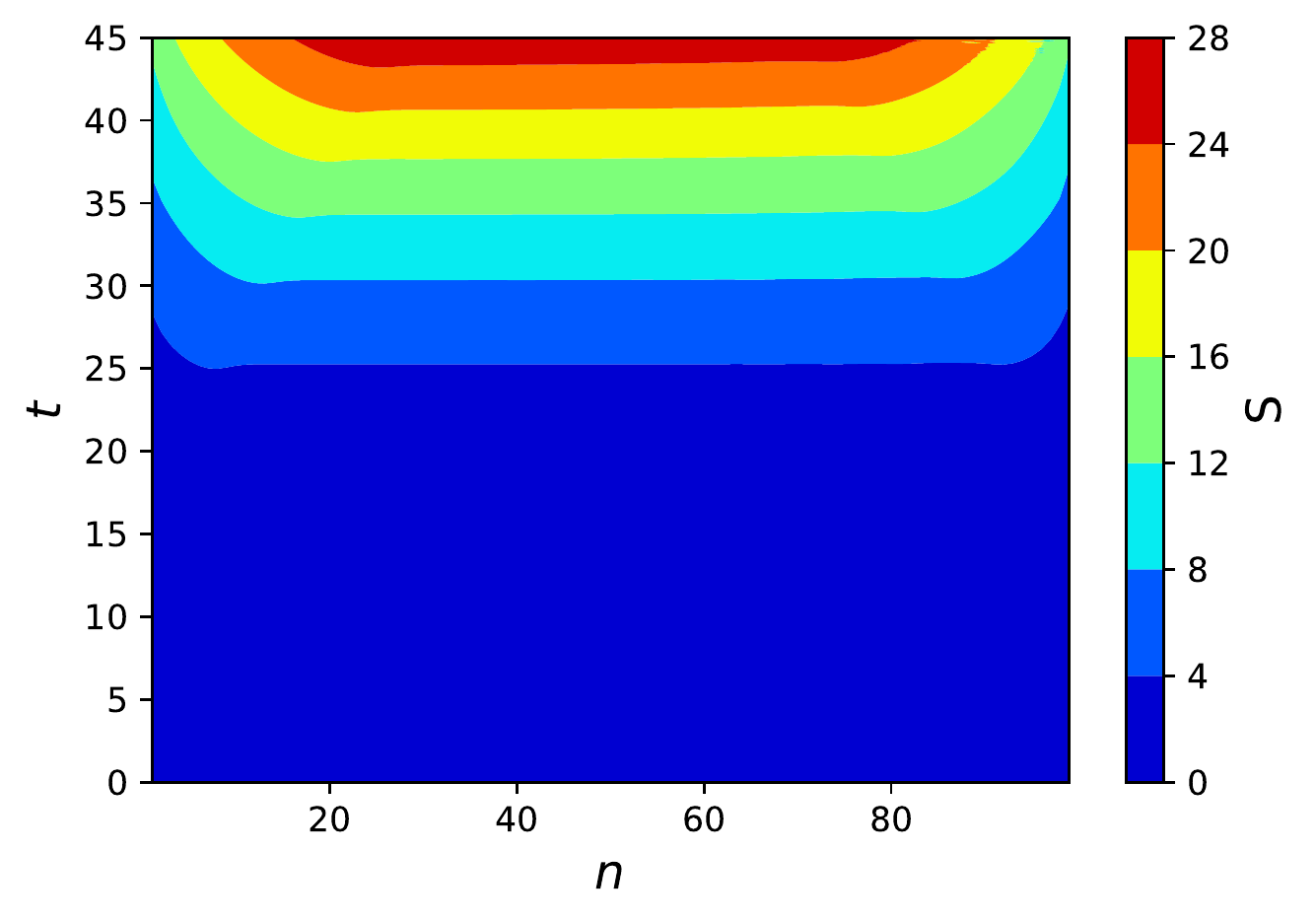}
		}
		\caption{Entanglement dynamics after a mass quench ($\Lambda(t)$) for NBC resulting in late-time inverted modes --- (a) Time-evolution for various subsystem sizes where we see an overall linear growth, and (b) Sub-system scaling of entanglement entropy at each time slice in the evolution. Here, $N=100$, $P=1.1$, $Q=1$ and $t_q=15$.}
		\label{fig:mq3}
	\end{center}
\end{figure*}

\begin{figure*}[!hbt]
	\centering
	\includegraphics[scale=0.5]{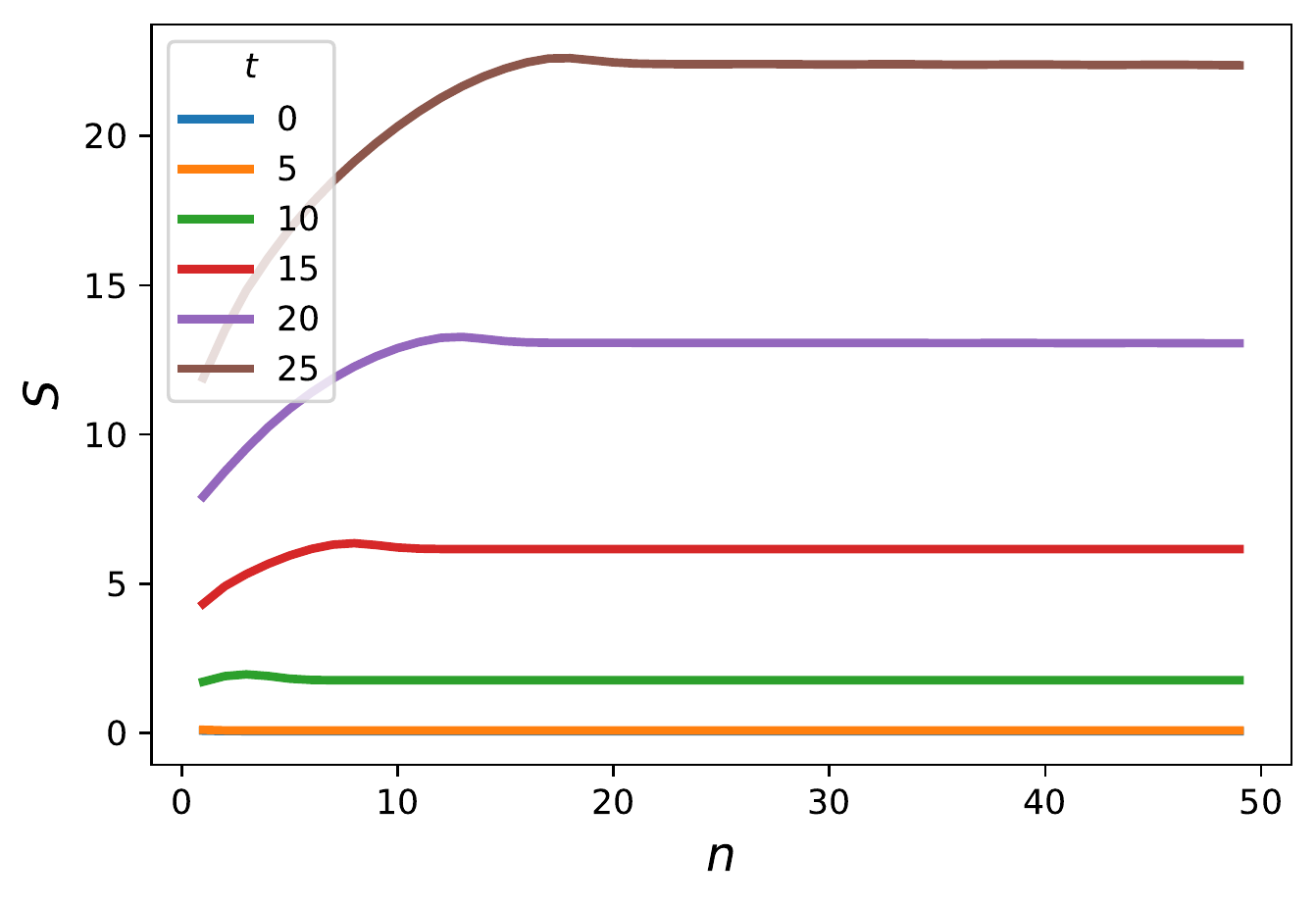}
	\caption{Area-law violation of entanglement entropy for a massive scalar field with an inverted mode instability. Here, we use the quench function in \eqref{quench}, where $P=1.2$, $Q=1$ and $t_q=5$ for a system of 100 oscillators. At $t=0$, the entanglement follows a typical area-law, and the entropy scaling is found to deviate from it with time. However, due to the exponentially growing solutions $b_k(t)$ arising from inverted mode instability, we are unable to obtain the late-time scaling relation of entropy.}
		\label{fig:area2}

\end{figure*}

\noindent {\bf Inverted modes:} Lastly, we consider a quench that results in late-time inverted modes. For this, we may consider an evolution such that $\Lim{t\to\infty}\Lambda(t)<0$ in an NBC chain. Here, Eq. \eqref{eq:neumodes} ensures that the system generates a finite number of inverted modes depending on the system size $N$. From \ref{fig:mq3}, we see that there is an overall linear growth of entropy with time, whose slope varies with subsystem size as predicted in Eq \eqref{eq:Kolmogorov}. Similar to previous cases, we expect the late-time entropy of any subsystem size to be bounded by $S_{N/2}$ corresponding to the half-chain entropy, however we are unable to simulate the dynamics for longer times due to the exponentially growing solutions $b_k(t)$ of the Ermakov equation. From \ref{fig:area2}, we see that subsystem scaling of entropy deviates from the area-law with time, but we are unable to obtain the late-time scaling relation the system exhibits due to computational constraints.

\subsubsection{Boundary quench}

\begin{figure*}[!ht]
	\begin{center}
		\subfloat[\label{bcq1c}][]{%
			\includegraphics[width=0.4\textwidth]{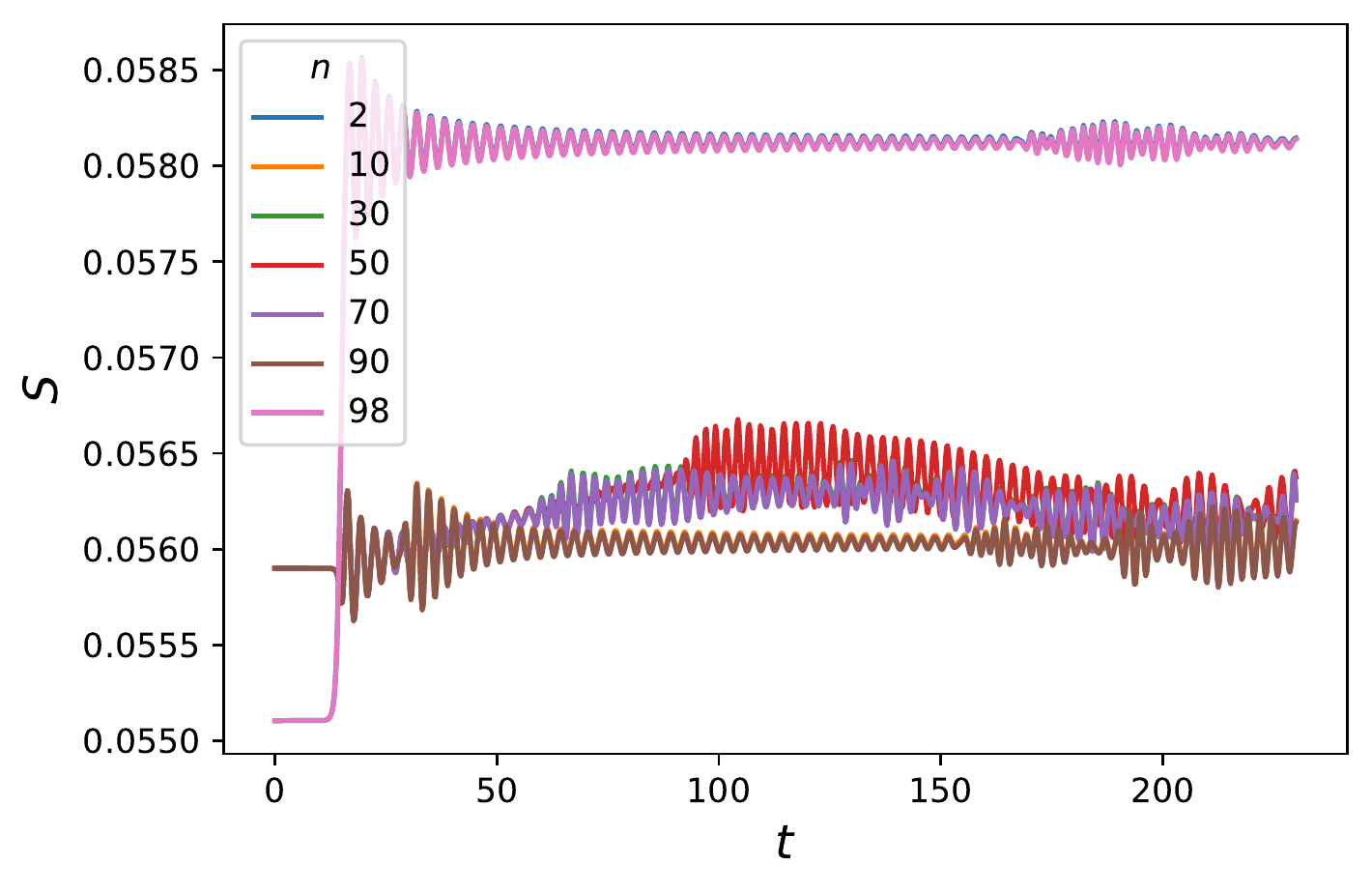}
		}
		\subfloat[\label{bcq1d}][]{%
			\includegraphics[width=0.4\textwidth]{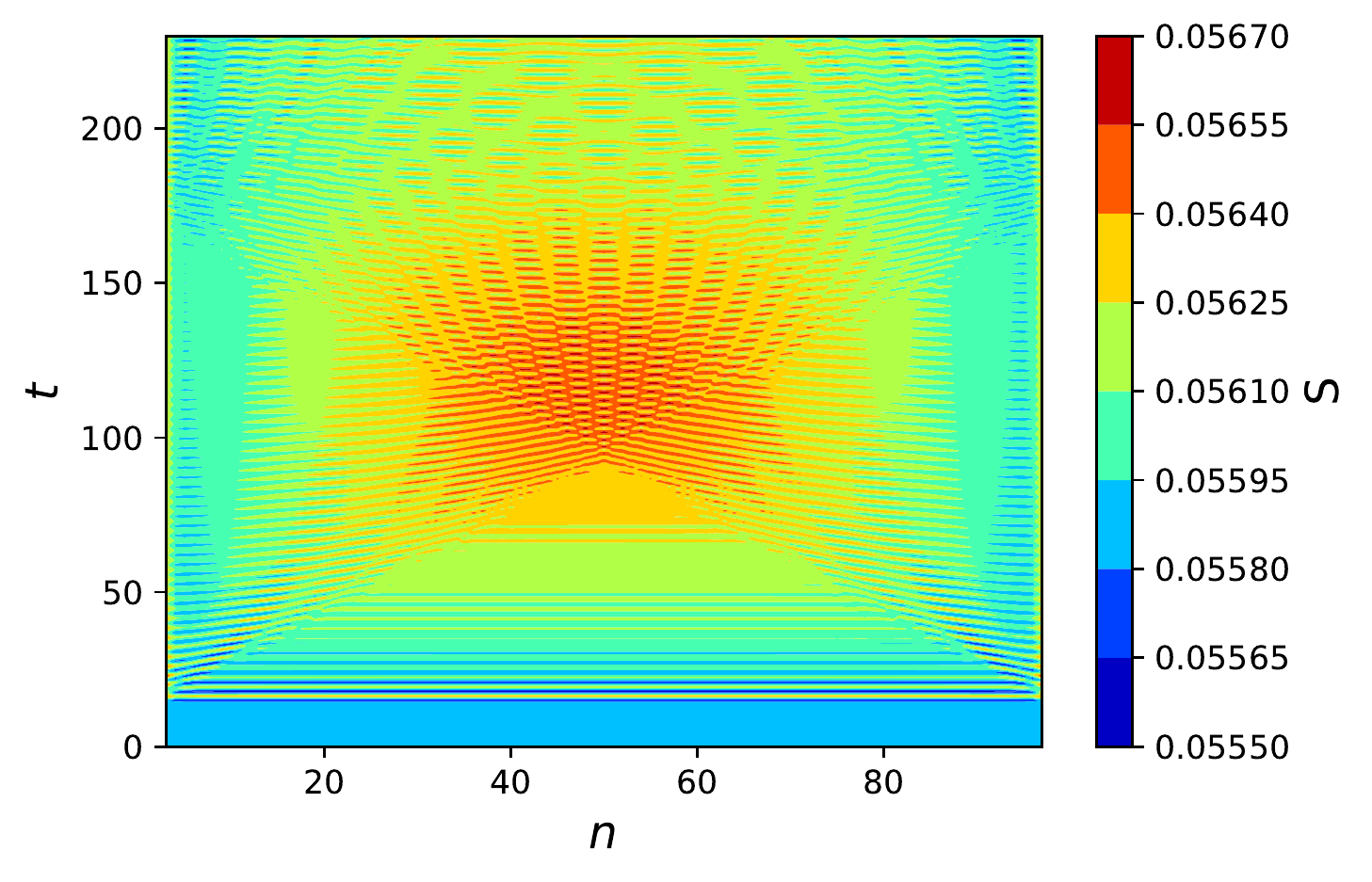}
		}
		
		\caption{Entanglement Dynamics after a boundary condition quench from Dirichlet to Neumann --- (a) Time-evolution for various subsystem sizes, and (b) Sub-system scaling of entanglement entropy at each time slice in the evolution. In (b) we have suppressed the edge entropies, namely for the first two and last two oscillator subsystems, for better visual clarity. Here, $N=100$, $Q=1$ and $t_q=15$.}
		\label{fig:bcq}
	\end{center}
\end{figure*}

In contrast with a global quench of the rescaled field mass $\Lambda(t)$, we may also consider a local quench in the lattice and analyze the subsequent entropy evolution. For this, we consider the lattice-regularized field that initially obeys DBC and at late times evolves to NBC. Physically, this corresponds to a system in constant contact with an infinite ``bath" and is insulated from the ``bath" at late times. Such a setup can be used to simulate the Dynamical Casimir Effect (DCE), which leads to particle creation due to time-dependent properties of the material. Here, this can be simulated by imposing time-dependent Robin boundary conditions at the boundaries that undergo a quantum quench~\cite{2011Silva.FarinaPhys.Rev.D,2012FARINA.etalInternationalJournalofModernPhysicsConferenceSeries}. The Robin boundary condition at each time-slice is imposed as follows:
\begin{equation}
	\varphi+\zeta(t)\partial_x\varphi=0,
\end{equation}
where $\zeta_i(t)$ is time-dependent. When $\zeta(t)\to 0$ it takes the Dirichlet form and when $\zeta(t)\to\infty$ it takes the Neumann form. A boundary quench can also be brought about by constructing the non-zero elements of the coupling matrix as follows:
\begin{align}
K_{11}&=K_{NN}=\Lambda +1+g^2(t)\nonumber\\
K_{jj}&=\Lambda +2 \qquad j \neq 1, N\nonumber\\
K_{j,j+1}&=K_{j+1,j}=-1\nonumber\\
g^2(t)&=\frac{1}{2}\left[1-\tanh{\left\{Q\left(t-t_q\right)\right\}}\right] 
\end{align}
From the above construction, it can be seen that $K(t\to -\infty)= K_{DBC}$ and $K(t\to\infty)= K_{NBC}$. Simulating this in \ref{fig:bcq}, we see that there are entanglement peaks or ``ripples" that travel from the edges to the middle of the chain, where they meet to form overtones that later spread outwards. This contrasts with the global quench of $\Lambda(t)$ where entanglement always traveled from the middle of the chain towards the edges, with the half-chain entropy consistently serving as an upper bound to subsystem entropy. Furthermore, we see that in the case of boundary quench these peaks initially propagate with a constant velocity through the chain, resembling a light-cone like structure. This may serve as an indication of particle-creation at the boundaries as a consequence of DCE, the details of which will be addressed in a separate work.

\subsection{Scaling symmetry and connection between correlation measures}

\begin{figure*}[!ht]
	\begin{center}
		\subfloat[\label{}][P=1]{%
			\includegraphics[width=0.4\textwidth]{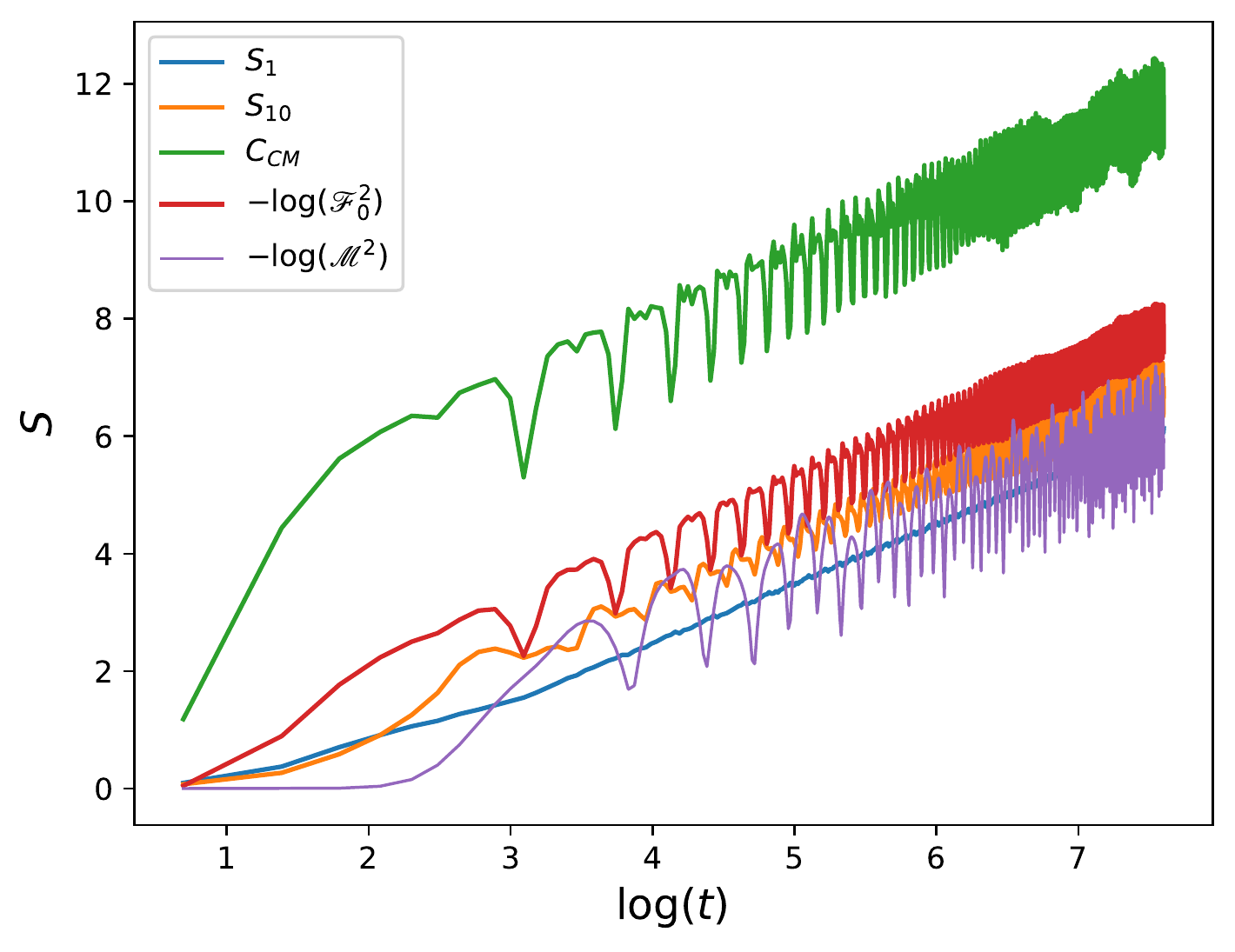}
		}
		\subfloat[\label{}][P=1.5]{%
			\includegraphics[width=0.4\textwidth]{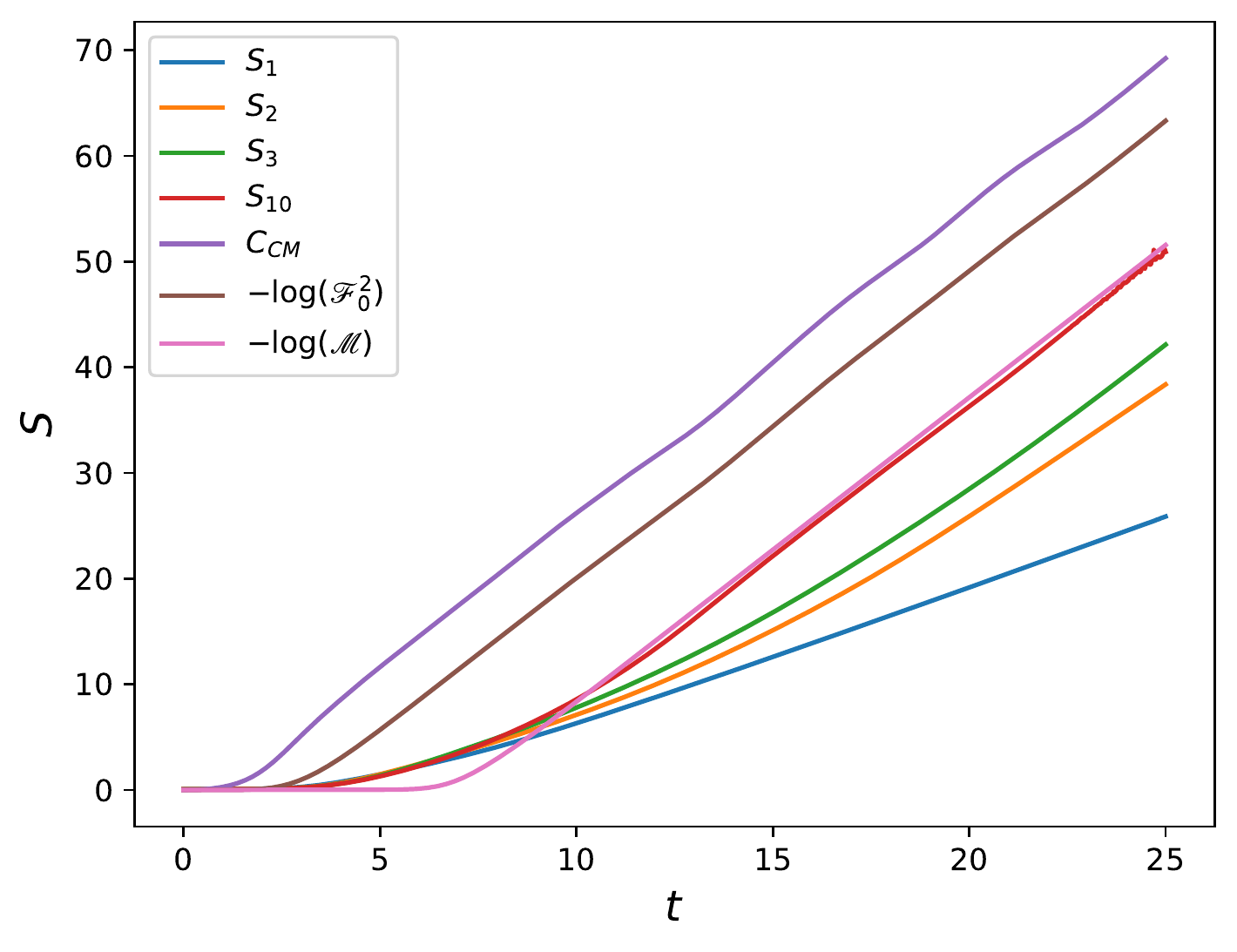}
		}
		
		\caption{Asymptotic behaviour of various correlation measures in a lattice-regularized massive scalar field with $N=20$ oscillators satisfying NBC in the presence of (a) zero mode and (b) inverted modes. Here, we use the quench function in \eqref{quench} where $Q=1$ and $t_q=2$. For calculating Loschmidt echo $\mathscr{M}(t)$, we have fixed $\delta\Lambda=0.01$. In (b), setting $P=1.5$ results in $5$ inverted modes at late-times, as a result of which entropy shares the same leading-order behavior as other correlation measures when Kolmogorov-Sinai bound is saturated as per Eq. \eqref{eq:entNHO}, i.e., when subsystem size $n\geq 3$.}
		\label{fig:asymptoticNCHO}
	\end{center}
\end{figure*}

The calculation of correlation measures associated with the global wave-function, such as GS fidelity, Loschmidt echo and complexity, can be easily generalized for the lattice as follows:
\begin{equation}
    \mathscr{F}_0(t)= \prod_{j=1}^N \mathscr{F}_0^{(j)}(t)\quad;\quad \mathscr{M}(t)=\prod_{j=1}^N \mathscr{M}_j(t) \quad ;\quad  C_{CM}(t)=\sum_{j=1}^N C_{CM}^{(j)}
\end{equation}
where the individual normal mode contributions have been defined in Eqs. \eqref{eq:overlapCHO}, \eqref{def:Loschmidt} and \eqref{def:CCM} respectively. 
In the presence of a zero mode, entanglement entropy for any subsystem size has a dominant logarithmic behavior at late-times as seen in \ref{fig:mq2}. Therefore, similar to the CHO, entropy for a subsystem size $n$ converges with other correlation measures at late-times as was established in Sec. \ref{sec:connection}, and confirmed by \ref{fig:asymptoticNCHO}:
\begin{equation}
    S_n^{\rm zero}\sim -\log{\mathscr{F}_0^2} \sim -\log{\mathscr{M}^2} \sim C_{CM} \sim \log{t}
\end{equation}

However, entropy growth in the presence of late-time inverted modes depends drastically on the subsystem size as can be seen in Eq. \eqref{eq:Kolmogorov}. More specifically, 
the entanglement entropy $S_n$ \emph{only} 
includes the number of inverted modes up to $2 \, n$. 
However, we expect entropy to converge with other measures only when the Kolmogorov-Sinai bound in Eq. \eqref{eq:KSbound} is saturated. To see this, let us suppose that there are $m$ late-time inverted modes $\{v_k\}$ in the system. For a massive scalar field, these are automatically indexed with decreasing magnitude as per Eqs. \eqref{eq:dirmodes} and \eqref{eq:neumodes}. Let us now consider the entanglement entropy of subsystem size $n$. If $m>2n$, the Kolmogorov-Sinai bound is not saturated, and we see that the entropy does not converge with other correlation measures. However, if $m\leq 2n$, the Kolmogorov-Sinai bound is saturated, and we see that the entropy converges with other correlation measures in the asymptotic limit. If all the modes are inverted, i.e., if $m=N$, then only the half-chain entropy $S_{N/2}$ converges with other correlation measures. From the above arguments, and from the results in \ref{fig:asymptoticNCHO}, we conclude that:
\begin{equation}\label{eq:entNHO}
    S_{n\geq \frac{m}{2}}^{\rm inv} \sim -\log{\mathscr{F}_0^2}\sim -\log{\mathscr{M}} \sim C_{\rm CM}  \sim h_{\rm KS} t
\end{equation}
The scrambling time and Lyapunov exponents for the massive scalar field are therefore:
\begin{equation}
    t_{\rm scram}\sim-\frac{1}{2v_1}\log{\frac{\delta\Lambda}{v_1\omega_1(0)}}\quad;\quad \lambda_L=\sum_{k=1}^m v_k=h_{\rm KS}
\end{equation}

Let us now go back to the original Hamiltonian $\tilde{H}(\tilde{t})$. Using the dynamical scaling symmetry developed in Sec. \ref{sec:scaling}, we obtain the following relations for a massive scalar field:
\begin{equation}
    \tilde{S}(\tilde{t})=S(\tilde{a}^{-1}\tilde{t})\quad;\quad \tilde{\mathscr{F}}_0(\tilde{t})=\mathscr{F}_0(\tilde{a}^{-1}\tilde{t}) \quad;\quad \tilde{\mathscr{M}}(\tilde{t})=\mathscr{M}(\tilde{a}^{-1}\tilde{t}) \quad;\quad \tilde{C}_{CM}(\tilde{t})=C_{CM}(\tilde{a}^{-1}\tilde{t}) 
\end{equation}
From the above relation, we see that as we take the continuum limit $\tilde{a}\to0$, the field correlations even at early times correspond to the late-time behavior of correlations in the rescaled system $H(t)$. Exactly at $\tilde{a}=0$, however, entropy and complexity diverge, whereas fidelity and Loschmidt echo vanish. We also see that the symmetry implies:
\begin{equation}
    \tilde{t}_{\rm scram}
    =\tilde{a}t_{\rm scram}\quad;\quad \tilde{\lambda}_L=\tilde{a}^{-1}\lambda_L
\end{equation}
As we take the continuum limit $\tilde{a}\to0$, we see that there is instant scrambling, and the exponential sensitivity to initial conditions is divergent.

%It is in fact a simplified version of the models that have been previously studied in Refs.~\cite{1985Guth.PiPhys.Rev.D,1993PadmanabhanStructureFormationUniverse,1995Mueller.LoustoPhys.Rev.D}.

\section{Conclusions and Discussions}\label{sec:conc}

Unitary evolution ensures that the state remains pure for an isolated quantum system that is initially in a pure state. As a result, the entropy for the total state is trivially zero. However, a natural, non-trivial description of entropy arises from the quantum entanglement between its constituent subsystems, wherein we obtain a thermal-like mixture of states upon integrating out some of the degrees of freedom. Entanglement entropy and other measures of quantum correlations provide a framework in which we can study the less-understood thermal properties of quantum systems compared to their well-understood classical counterparts. Our results show that an explicit connection between the quantum and classical regimes can indeed be established in the presence of instabilities, and we see that it has direct consequences in simple systems such as the CHO all the way to quantum fields propagating in time-dependent backgrounds.

In Section \ref{model}, we looked at the simple case of a CHO with time-dependent frequencies and reviewed the prescriptions for calculating entanglement entropy ($S$) and GS fidelity ($\mathscr{F}_0$) assuming pure-state adiabaticity. Then, in Section \ref{sec:scaling}, we showed that the dynamical evolution of quadratic Hamiltonians such as that of a CHO exhibits an inherent scaling symmetry that proves to be useful in more ways than one. For instance, with the help of some special transformations, we shifted back and forth between different Hamiltonian descriptions that belonged to the same $\Lambda(t)-$ class and therefore correspond to the same dynamical evolution of quantum correlations. In Section \ref{sec:Asymptotic}, we employed this scaling symmetry to analytically obtain the late-time behavior of correlations for $\Lambda(t)-$classes that result in inverted mode or zero-mode instabilities. Such instabilities resulted in an unbounded entropy growth (logarithmic for zero-modes and linear for inverted modes) as $t\to\infty$, culminating in a quantum state that is non-normalizable. Finally, in Section \ref{sec:exact}, we simulated a realistic quench scenario to show that the analytic predictions of correlations at late-times also manifest over shorter time-scales of instability in the system. Quantum correlations, therefore, serve as an effective diagnostic tool for instabilities in the system.

In Section \ref{qchaos}, we used the dynamical scaling symmetry to formulate a general prescription for quantifying zero-mode and inverted mode instabilities. First, in Section \ref{sec:LoschEcho}, we used Loschmidt echo ($\mathscr{M}$) arising from an infinitesimal change in the initial conditions to identify the scrambling time ($t_{\rm scram}$) and Lyapunov exponent ($\lambda_L$) corresponding to a mode that undergoes inversion. The largest inverted mode is responsible for the earliest onset of exponential decay of the echo, whereas the smallest inverted mode was responsible for the last bout of exponential decay. The overall rate of exponential decay of Loschmidt echo at late-times was found to coincide exactly with the classical Kolmogorov-Sinai rate $h_{\rm KS}$, which is the sum of all positive Lyapunov exponents. In addition to this, we obtained a much longer time-scale $t_{\rm zero}$ that corresponds to the onset of a power-law decay of the echo arising from a zero-mode instability. Therefore, a CHO with zero-mode instability remains stable to small changes in the initial conditions for a much longer time than one with an inverted mode instability. 

In Section \ref{sec:complexity}, we compared two different measures of circuit complexity, namely, the wave-function complexity $C_{WF}$ and correlation matrix complexity $C_{CM}$. While these measures exhibited a logarithmic growth similar to entanglement entropy in the case of a zero-mode instability, they responded differently to an inverted mode instability. While $C_{CM}$ exhibited a linear growth similar to entanglement entropy, $C_{WF}$ grew linearly until it saturated after a time $t_{\rm sat}$. Therefore, in systems that are dominated by inverted mode effects, we may employ $C_{WF}$ to pick up signatures of zero-mode instabilities that are otherwise suppressed. From the results of Sections \ref{model} and \ref{qchaos}, we conclude that for a CHO, leading-order terms of quantum correlations converge at late-times in the presence of instabilities:
\begin{equation}
    S_{\rm CHO} \sim -\log{\mathscr{F}_0^2}\sim C_{CM} \sim \begin{cases}
		-\log{\mathscr{M}^2}\sim\log{t}& \text{zero mode}\\ %\lim_{t\to\infty}\omega_{min}^2=0\\
		-\log{\mathscr{M}}\sim h_{\rm KS}t & \text{inverted mode} %\lim_{t\to\infty}\omega_{min}^2<0
	\end{cases}
\end{equation}

In Section \ref{sec:scalarfielddyn}, we extended the dynamical scaling symmetry to a massive scalar field propagating in a $(1+1)-$dimensional flat space-time. Using the dynamical scaling symmetry, we explored the late-time dynamics of entanglement when the field is subjected to two different types of quench scenarios. First, we considered the mass quench of a scalar field and analyzed the stable modes, zero-modes, and inverted modes at late-times on a case-by-case basis. We observed the following --- i) Like the CHO, a logarithmic/linear entropy growth manifested in scalar fields for zero/inverted mode instabilities for arbitrary subsystem sizes, ii) Entanglement evolution for stable modes was found to periodically mimic the inverted-mode instability by way of a linear entropy growth (followed by a plateau and descent), iii) Unlike the CHO, however, the slope of linear entropy growth for inverted mode instability is only bounded by the Kolmogorov-Sinai entropy rate as opposed to being equal to it, and iv) The convergence of entanglement entropy with other asymptotic quantum correlations occur only for subsystem sizes for which this bound is saturated, i.e., wherein the mapping to its classical counterpart holds exactly. For an inverted mode instability at late-times, we see that
\begin{equation}
    S_{n\geq \frac{m}{2}}^{\rm inv} \sim -\log{\mathscr{F}_0^2}\sim -\log{\mathscr{M}} \sim C_{\rm CM}  \sim h_{\rm KS} t
\end{equation} 

A similar convergence of leading order terms of quantum correlations holds for zero-mode instability at late-times, even as the Lyapunov exponents vanish, corresponding to a meta-stable phase:
\begin{equation}
    S_n^{\rm zero}\sim -\log{\mathscr{F}_0^2} \sim -\log{\mathscr{M}^2} \sim C_{CM} \sim \log{t}
\end{equation}

For a global quench of the scalar field mass, the mapping between entanglement entropy and thermal entropy also manifests in the subsystem scaling of entropy, regardless of mode-stability. For stable modes and zero-modes, when the half-chain entropy $S_{N/2}$ peaks periodically (following a linear growth that mimics inverted mode dynamics), the entropy was found to scale linearly with subsystem size. Hence, the entropy scaling oscillates between an area-law and a volume law with time. Similarly, for inverted modes, entanglement entropy consistently violates area law. Since we have shown the area-law to volume-law transition to occur for stable and zero-modes, it is expected that the same transition will also occur for the inverted modes. Therefore, our analysis potentially points to the possibility that the entanglement entropy of the scalar field assumes thermal characteristics in the presence of instabilities. This is currently under investigation.

Secondly, we looked at a scenario wherein the boundary condition that the scalar field satisfied transitioned from Dirichlet to Neumann. Physically, this models the Dynamical Casimir Effect (DCE). We saw that after the quench, there were entanglement peaks or ``ripples" that originated at the boundaries and traveled at almost a constant velocity to the center of the chain (resembling a light-cone). Tbis is possibly an indication of particle-creation at the boundaries due to DCE, which we wish to address in a later work. Furthermore, for a global quench of scalar field mass, we observed that the half-chain entropy bounded the subsystem entropy. In contrast, a local quench at the boundaries often violated the bound, as the peaks traveled from the edges to the middle.

Additionally, we explored the consequences of dynamical scaling symmetry on a massive scalar field regularized by a UV cut-off $\tilde{a}$. We showed that the time-scales for scrambling ($\tilde{t}_{\rm scram}\propto \tilde{a}$) and complexity saturation ($\tilde{t}_{\rm sat}\propto \tilde{a}$) approached zero on taking the continuum limit $\tilde{a}\to0$, whereas the Kolmogorov-Sinai rate diverged as a power law ($\tilde{h}_{\rm KS}\propto \tilde{a}^{-1}$). On taking this limit, we also showed that the early-time behavior of quantum correlations of the scalar field (described by $\tilde{H}$) corresponded to the late-time behavior of these measures in the rescaled system (described by $H$). 

The results reported here have implications for cosmological perturbations and black-hole quasi-normal modes (QNMs). For instance, the action for the first-order scalar perturbations for a canonical 
scalar-field ($\varphi$) driven inflation is~\cite{1999-Garriga.Mukhanov-PLB}:
\begin{equation}
	S=\frac{1}{2} \int 
	\left[v^{\prime 2}
	+ v \nabla^2 v 
	- \frac{z''}{z} v^2 \right] d \eta d^{3} x \quad 
	z = \frac{a \varphi'}{\cal H}
\end{equation}
where $\eta$ is the conformal time and prime denotes the derivative with respect to $\eta$ and ${\cal H} = a'/a$. For any background evolution $a(\eta)$, the quantum scalar fluctuations satisfy the following time-dependent equation: 
\begin{equation}
	v_k^{\prime\prime} + \left[k ^2 - \frac{z''}{z} \right] v_k = 0  
	\label{eq:MSequation}
\end{equation}
Like in the CHO, as in Sec.~\ref{sec:Asymptotic}, we have three categories: 
\begin{enumerate}
	\item {\bf Sub-Hubble scales}: In this category $k^2 \gg z''/z$ and  the above differential equation reduces to:
	\begin{equation}
		v_k^{\prime\prime} + k ^2 v_k \simeq 0   
	\end{equation}
	and the effective frequencies are always positive corresponding to plane-wave solutions in Minkowski background.
	\item {\bf Horizon crossing}: This corresponds to the category $k^2 = z''/z$ and the differential equation \eqref{eq:MSequation} has a zero-mode. 
	\item {\bf Super-Hubble scales}: In this category $k^2 \ll z''/z$ and  the  differential equation \eqref{eq:MSequation} reduces to:
	\begin{equation}
		v_k^{\prime\prime} - \frac{z''}{z} 
		v_k \simeq 0 
	\end{equation}
	Two points to note: First the solution to the differential equation is identical for all the $k-$modes. Second, the effective frequency for all $k-$mode is negative. 
\end{enumerate}
For any background evolution $a(\eta)$, the quantum tensor fluctuations satisfy the following time-dependent equation: 
\begin{equation}
	\mu_T^{\prime\prime} + \left[k ^2 - \frac{a''}{a} \right] \mu_T = 0  
	\label{eq:tensorequation}
\end{equation}
The same analysis can be extended to the tensor perturbations, however, the only difference is that the zero mode occurs at a different horizon crossing --- $k^2 = a''/a$.
Since the horizon crossing occurs differently for the scalar and tensor, there will be a slight difference in the evolution of the two perturbations. Studying the quantum correlations in inflation and bounce can hence be potentially be useful in systematically distinguishing between these two early Universe scenarios.

{For black holes, in Ref. \cite{2015-Hintz.Vasy-JMP}, the authors proved that the strength of the Cauchy horizon instability is determined by the half-life of the most slowly decaying quasinormal modes (QNM). In Ref. \cite{2018-Cardoso.etal-PRL}, the authors studied this for Riessner-Nordstrom-de Sitter space-time by numerically evaluating the QNM frequencies. In this case, the authors excluded the zero mode in their analysis. Our analysis show that system with zero modes decays slowly compared to the unstable modes. Thus, the analysis of Cauchy horizon instability is incomplete without a complete understanding of the zero-modes. Since QNMs are dissipative systems, the analysis reported here can be translated to black-holes. This is currently under investigation. 
}

\begin{acknowledgements}
SMC is supported by Prime Minister's Research Fellowship offered by the Ministry of Education, Govt. of India. The work is supported by the MATRICS SERB grant.
\end{acknowledgements}
\appendix
\section{Solution to the time-dependent Schrodinger equation}\label{App:GaussianState}
Let us consider the Schrodinger equation for a harmonic oscillator with time-dependent frequency:
\begin{equation}
	i \frac{\partial \Psi(x,t)}{\partial t}=-\frac{1}{2} \frac{\partial^2 \Psi(x, t)}{\partial x^2}+\frac{1}{2} \omega^2(t) x^2 \Psi(x, t)
\end{equation}
We are interested in a particular class of solutions, known as form-invariant Gaussian states (GS). This approach is fundamental to understanding how particle creation manifests when an initial quantum vacuum state evolves with respect to a time-dependent Hamiltonian~\cite{2008Mahajan.PadmanabhanGeneralRelativityandGravitation,2018Rajeev.etalGeneralRelativityandGravitation}:
\begin{equation}
	\Psi_{GS}=\mathscr{N}(t)\exp{-R(t)x^2},
\end{equation}
where $\mathscr{N}$ is the time-dependent normalization factor. Upon solving the time-dependent Schrodinger equation, we obtain the following equations:

\begin{equation}\label{Riccati}
	i\frac{\dot{\mathscr{N}}}{\mathscr{N}}=R\quad;\quad i\dot{R}=2R^2-\frac{\omega^2(t)}{2}
\end{equation}
%The normalization condition further gives us :
%\begin{equation}
	%\abs{\mathscr{N}(t)}=\left(\frac{R+R^*}{\pi}\right)^{1/4}
%\end{equation}
The non-linear Riccati-type equation in $R(t)$ given above is equivalent to the non-linear Ermakov equation~\cite{1967LewisPhys.Rev.Lett.,1968LewisJournalofMathematicalPhysics} in $b(t)$, which can be seen by making the following substitution:
\begin{equation}
	R(t)=\frac{1}{2}\left[\frac{\omega(0)}{b^2(t)}-i\frac{\dot{b}(t)}{b(t)}\right]\quad\Rightarrow \quad \ddot{b}(t)+\omega^2(t)b(t)=\frac{\omega^2(0)}{b^3(t)}
\end{equation}
The general solution to the Schrodinger equation is therefore~\cite{2008LoheJournalofPhysicsAMathematicalandTheoretical,2008Mahajan.PadmanabhanGeneralRelativityandGravitation}:
\begin{equation}
	\Psi_{GS}(x,t)=\left(\frac{\omega(0)}{\pi b^2(t)}\right)^{1/4}\exp{-\left(\frac{\omega(0)}{b^2(t)}-i\frac{\dot{b}(t)}{b(t)}\right)\frac{x^2}{2}-\frac{i}{2}\omega(0)\tau(t)},
\end{equation}
where $\tau=\int b^{-2}(t)dt$. At $t=0$, we see that this form-invariant Gaussian state co-incides with the ground state of the system. In the adiabatic limit, Eq \eqref{Riccati} can be solved to obtain~\cite{2008Mahajan.PadmanabhanGeneralRelativityandGravitation}:
\begin{equation}
	R\approx \frac{\omega(t)}{2}\quad;\quad \mathscr{N}\approx \mathscr{N}_0\exp{-\frac{i}{2}\int_0^t\omega(t)dt}
\end{equation}
The wave-function therefore takes the following form:
\begin{equation}
	\Psi_{adia}(x,t)=\left(\frac{\omega(t)}{\pi}\right)^{1 / 4}\exp \left[-\frac{\omega(t)}{2} x^2-\frac{i}{2} \int_{0}^t \omega(t') d t'\right].
\end{equation}
We see that in this case the adiabatic Gaussian state $\Psi_{adia}$ coincides with the instantaneous ground state at all times, in agreement with the quantum adiabatic theorem~\cite{2005-Wu.Yang-PRA}. The theorem states that for slow-changing $H(t)$, if the system begins close to an eigenstate, it remains close to that eigenstate throughout. 

Using the adiabatic theorem, we can similarly define the full spectrum of instantaneous eigenstates $\{\phi_n(x,t)\}$ at each time-slice as follows~\cite{2008Mahajan.PadmanabhanGeneralRelativityandGravitation,2018Rajeev.etalGeneralRelativityandGravitation}:
\begin{equation}
	\phi_n=\left(\frac{\omega(t)}{\pi}\right)^{1 / 4} \frac{1}{\sqrt{2^n n !}} H_n(\sqrt{\omega(t)}x) \exp \left[-\frac{\omega(t)}{2} x^2-i\left(n+\frac{1}{2}\right) \int_{0}^t \omega(t') d t'\right]
\end{equation}
With respect to the instantaneous eigenbasis, the general solution $\Psi_{GS}$ for a non-adiabtically evolving Hamiltonian $H(t)$ can be decomposed as follows~\cite{2008Mahajan.PadmanabhanGeneralRelativityandGravitation,2018Rajeev.etalGeneralRelativityandGravitation}:
\begin{equation}
	\Psi_{GS}=\sum_nC_n(t)\phi_n(x,t)
\end{equation}
The state $\Psi_{GS}$ no longer coincides with the instantaneous ground state for $t>0$, i.e., it develops excitations in the instantaneous eigenbasis defined at each time-slice. The above picture is therefore a natural way to describe particle-creation in vacua due to a time-dependent Hamiltonian.

\section{Entanglement entropy of N-HO with time-dependent frequencies}\label{App:A}

Consider the following Hamiltonian\cite{2017Ghosh.etalEPLEurophysicsLetters}:
\begin{equation}
	H=\frac{1}{2}\sum_{i=1}^N p_i^2+\frac{1}{2}\sum_{i,j=1}^N K_{ij}x_{i}x_{j},
\end{equation}
The coupling matrix ($K_{ij}$) here has time-dependent entries and captures all the necessary information about correlations in the system. For the lattice-regularized massive scalar field described in Section \ref{sec:scalarfielddyn}, we see that the time-dependent entries are confined only to the diagonal terms in the matrix for a mass quench. On the other hand, a boundary condition quench can have time-dependent entries that are non-diagonal as well.

The initial values of normal modes, labelled as $\{\omega_i(0)\}$, are the eigenvalues of $K^{1/2}(0)$. The ground state wave-function here is given by:
\begin{equation}
	\Psi_{\rm GS}(\tilde{X},t)=\left(\prod_{n=1}^N\frac{\omega_n(0)}{\pi b_n^2(t)}\right)^{1/4}\exp{-\frac{1}{2}\tilde{X}^T(\Omega_D^{1/2}-iZ_D)\tilde{X}-\frac{i}{2}\sum_{n=1}^N\omega_n(0)\tau_n},
\end{equation}
where $\tilde{X}=M X$ is the normal mode co-ordinate system that diagonalizes the Hamiltonian. We also see that $\Omega_D$ and $Z_D$ are diagonal matrices whose entries are given below:
\begin{equation}
	(\Omega_D)_{nn}=\frac{\omega_n(0)}{b_n^2(t)}\qquad;\qquad(Z_D)_{nn}=\frac{\dot{b}_n(t)}{b_n(t)}
\end{equation}
In the physical co-ordinates, the wave-function is entangled, taking the following form:

\begin{equation}
	\Psi_{\rm GS}(X,t)=\left(\prod_{n=1}^N\frac{\omega_n(0)}{\pi b_n^2(t)}\right)^{1/4}\exp{-\frac{1}{2}\left[X^T(\Omega-iZ)X\right]-\frac{i}{2}\sum_{n=1}^N\omega_n(0)\tau_n},
\end{equation}
where $Z=MZ_DM^T$ and $\Omega=M\Omega_DM^T$. In order to trace out some $m$ degrees of freedom from the system, let us first rewrite the following matrices:
\begin{equation}
	\Omega=\begin{bmatrix}
	(A)_{m\times m}&(B)_{m\times N-m}\\
		(B^T)_{N-m\times m}&(C)_{N-m\times N-m}\end{bmatrix} \quad;\quad Z=\begin{bmatrix}
	(Z_A)_{m\times m}&(Z_B)_{m\times N-m}\\
(Z_B^T)_{N-m\times m}&(Z_C)_{N-m\times N-m}\end{bmatrix}
\end{equation}

 On following the same procedure to calculate entanglement entropy, we first obtain the reduced density matrix as follows:
\begin{equation}\label{nrho}
	\rho_{out}=\sqrt{\frac{\det{\Omega/\pi}}{\det{A/\pi}}}\exp{-\frac{1}{2}X_{out}'^T(\gamma+i\delta)X_{out}'-\frac{1}{2}X_{out}^T(\gamma-i\delta)X_{out}+X_{out}^T\beta X_{out}'}
\end{equation}
where we see that:
\begin{align}\label{nrho2}
	\gamma&=C-\frac{1}{2}B^TA^{-1}B+\frac{1}{2}Z_B^T A^{-1}Z_B\nonumber\\
	\beta&=\frac{1}{2}B^TA^{-1}B+\frac{1}{2}Z_B^TA^{-1}Z_B\nonumber\\
	\delta&=Z_C-Z_B^TA^{-1}B	
\end{align}
Now we perform a series of diagonalizations to simplify the RDM further. Let $V$ be a diagonalizing matrix for $\gamma$ such that $\gamma=V^T \gamma_D V$ and $\tilde{\beta}=\gamma_D^{-1/2}V\beta V^T\gamma_D^{-1/2}$. Let $W$ diagonalize $\tilde{\beta}$ such that
\begin{equation}
\label{nrho3}
\tilde{\beta}=W^T\tilde{\beta}_DW \, \quad \tilde{\delta}=W\gamma_D^{-1/2}V\delta V^T\gamma_D^{-1/2}W^T \, .   
\end{equation}
 In the new co-ordinates $Y=W\gamma_D^{1/2}V X_{out}=\{y_j\}$, the RDM can hence be rewritten as:
\begin{equation}
	\rho_{out}=\frac{1}{\pi^{N-m}}\exp{\frac{i}{2}\left[Y^T\tilde{\delta}Y-Y'^T\tilde{\delta}Y'\right]}\prod_{j=m+1}^N\sqrt{1-\tilde{\beta}_j}\exp{-\frac{1}{2}\left(y_j^2+y_j'^2\right)+\tilde{\beta}_jy_jy_j'},
\end{equation}
where $\tilde{\beta}_j$ are the eignevalues of $\tilde{\beta}$. The integral eigenvalue equation for RDM will therefore have the following solution:
\begin{align}
	f_n(Y,t)&=\left(\prod_{j=m+1}^N H_n(\epsilon^{1/2}y_j)\right)\exp{-Y^T\left(\frac{\epsilon-i\tilde{\delta}}{2}\right)Y}\nonumber\\
	p_n(t)&=\prod_{j=m+1}^N \left(1-\xi_j(t)\right)\xi_j^n(t)\nonumber\\
	\xi_j(t)&=\frac{\tilde{\beta}_j}{1+\sqrt{1-\tilde{\beta}_j^2}}
	\end{align}
The entanglement entropy therefore accumulates contributions from each of the remaining degrees of freedom as $S=\sum_{j=m+1}^N S_j(t)$ where $S_j(t)$ has the familiar form:
\begin{equation}
	S_j(t)=-\log{[1-\xi_j(t)]}-\frac{\xi_j(t)}{1-\xi_j(t)}\log{\xi_j(t)}
\end{equation}

\begin{figure*}[!ht]
			\begin{center}
			\subfloat[]{%
				\includegraphics[width=0.4\textwidth]{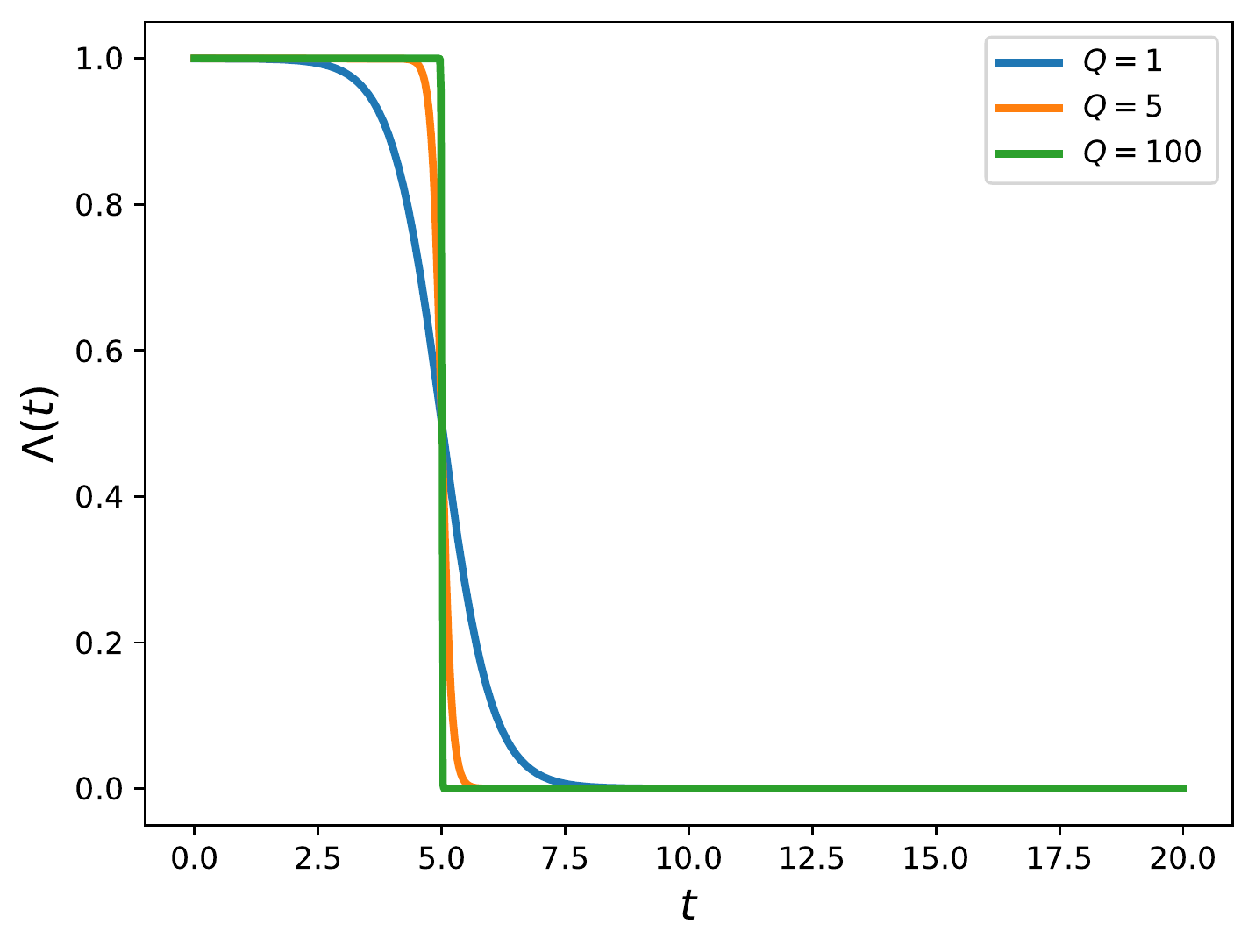}
			}
			\subfloat[]{%
				\includegraphics[width=0.4\textwidth]{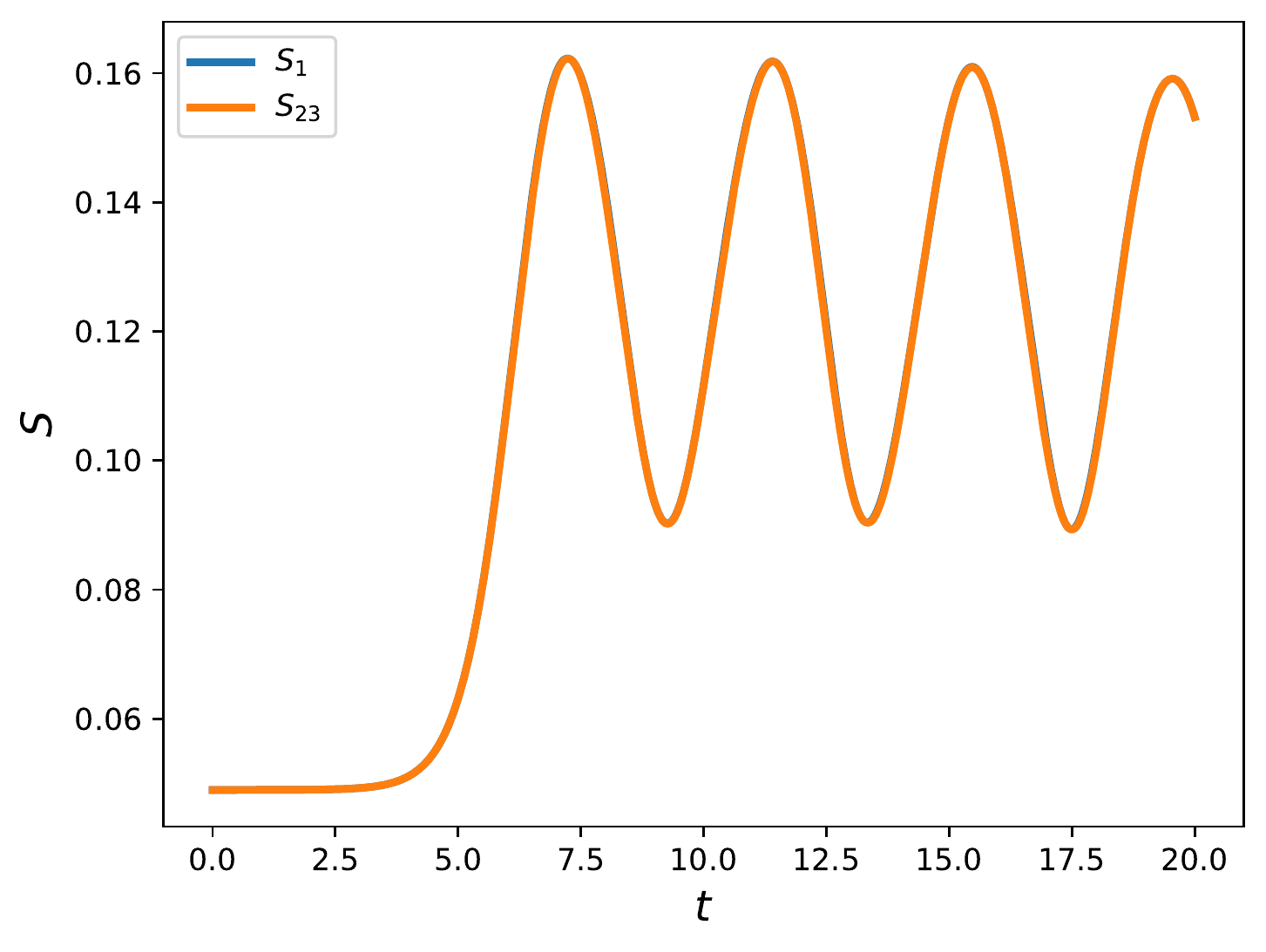}
			}\hfill
			\subfloat[]{%
				\includegraphics[width=0.4\textwidth]{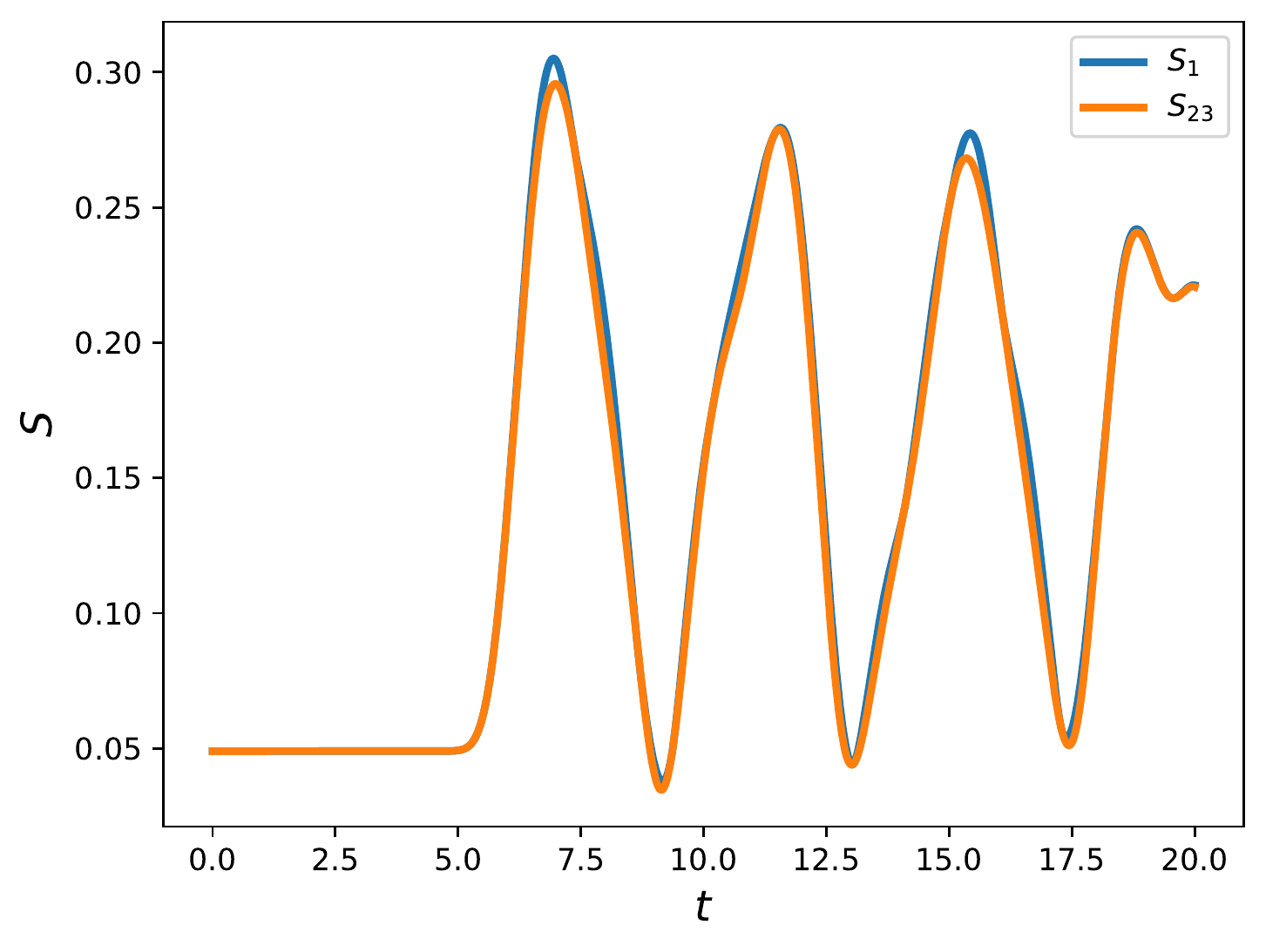}
			}
			\subfloat[]{%
				\includegraphics[width=0.4\textwidth]{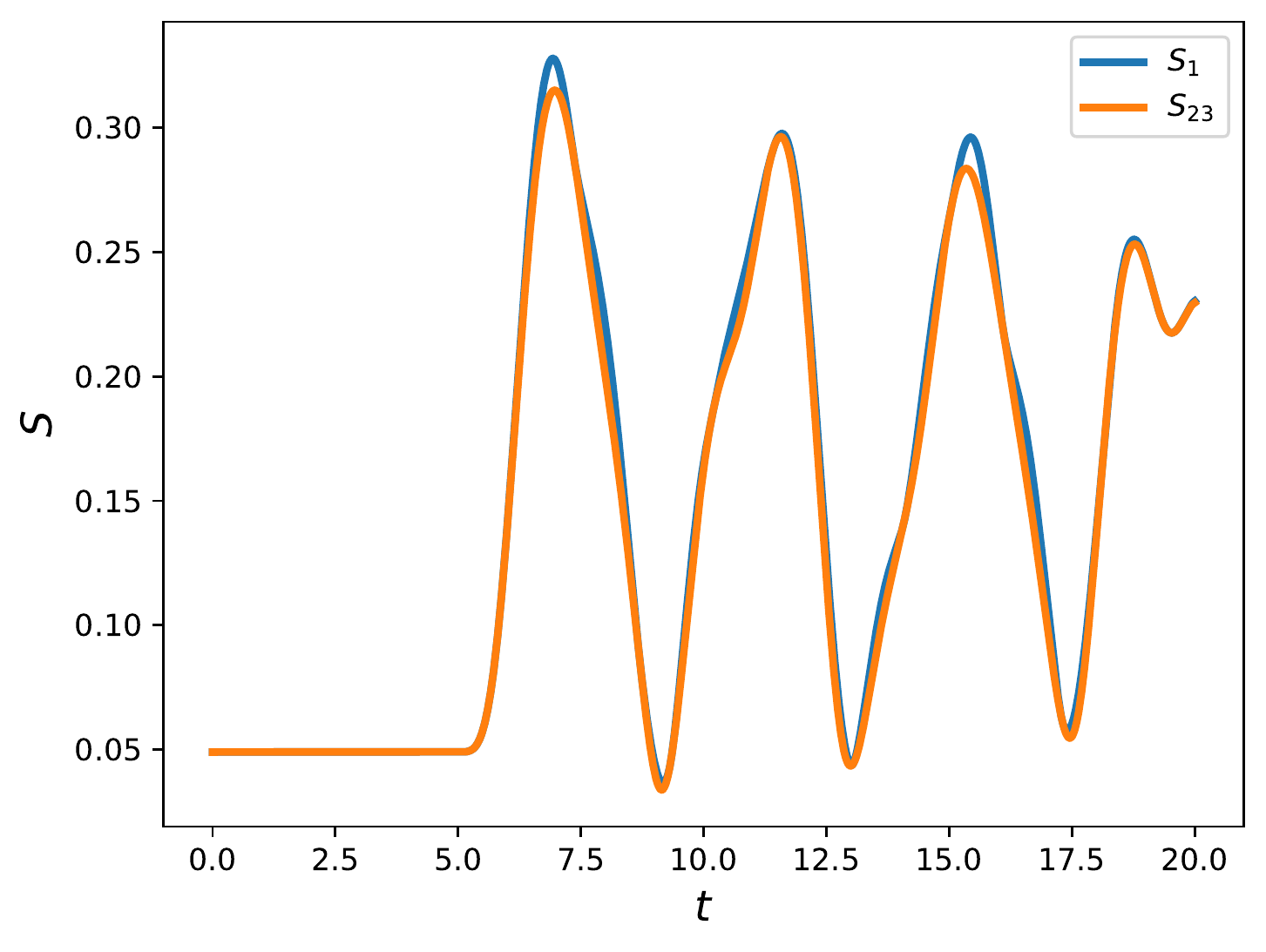}
			}
			\caption{(a) Evolution of rescaled mass $\Lambda(t)$ for quench speeds $Q=1, 5, 100$, and (b-d) Entanglement dynamics ($S_1, S_{23}$) for respective quench speeds. Here we employ Dirichlet BC, with $P=1$ and $N=3$.}
			\label{asym}
			\end{center}
		\end{figure*}

\section{Asymmetry of entanglement across bipartition}\label{App:B}

When we have a pure state $\ket{\Psi}$ that describes $N$-coupled oscillators, a bipartition the system as $\mathscr{H}=\mathscr{H}_A\otimes\mathscr{H}_B$ ensures that $S_A=S_B$. However, in the time-dependent formalism considered here, we see that $S_A\neq S_B$ as we move further away from pure state adiabaticity. We can show this from the Eqs.~\eqref{nrho} and \eqref{nrho2}. From Eq.~\eqref{nrho2}, $\gamma$ and $\beta$ are related by the following relation:
\begin{equation}
    \gamma=C-\beta+Z_B^T A^{-1}Z_B=V^T\gamma_D V
\end{equation}
Substituting this in Eq.~\eqref{nrho3} and after some rearrangement, we obtain the following:
\begin{equation}
    \tilde{\beta}=\gamma_D^{-1/2}V \left[C+Z_B^T A^{-1}Z_B\right] V^T\gamma_D^{-1/2}-\mathbb{1}
\end{equation}
$Z_B\to0$ and the second term in the bracket above vanishes in the time-independent limit. On further diagonalizing $\tilde{\beta}$, we see that the dimensionality as well as the non-zero eigenvalues of $\tilde{\beta}^{(A)}$ and $\tilde{\beta}^{(B)}$ match, and hence, $S_A$ and $S_B$ are identical. However, due to the contribution from $Z$ in the time-dependent case, we see that the spectra of $\tilde{\beta}^{(A)}$ and $\tilde{\beta}{^{(B)}}$ may no longer match. While they seem to match for cases where $A\cup B$ is a symmetric bipartition ($N=2$), we see that $S_A\neq S_B$ for any other kind of bipartition ($N=3$). On considering the evolution in \eqref{quench} for $N=3$, we observe in \ref{asym} that there is an increasing mismatch with an increasing quench speed ($Q$). This is currently under investigation. 
%We discuss the consequence of this feature in Sec. \eqref{sec:scalarfielddyn}.

\section{Hamiltonian of scalar fields in time-dependent space-times}
\label{app:BH-Ham}

In this appendix, we show that the Hamiltonian of the massive scalar field propagating in a time-dependent spherically symmetric space-time, when discretized, reduces to the form \eqref{eq:1DHami-Mink}~\cite{Das2010}. Consider the following time-dependent spherically symmetric metric:
\bea \la{bh-metric}
ds^2 &=& - A(\tau,\xi) \, d\tau^2 +
\fr{d\xi^2}{B(\tau,\xi)} + \r^2(\tau,\xi) d\O^2
\eea
where $A, B, \r$ are continuous, differentiable functions of
$(\tau,\xi)$ and $d\Omega^2 = d\theta^2 + \sin^2 \theta d\phi^2$
is the metric on the unit $2-$sphere. The action for the scalar
field propagating in the above background is given by
\bea
S &=& -\frac{1}{2}
\int d^4 x \, \sqrt{-g}~ \left[g^{\mu\nu}~  \pa_{\mu}\varphi~\pa_{\nu}\varphi + m^2 \varphi^2 \right]\nn \\
%%%%
\label{eq:actgen1}
&=& - \frac{1}{2} \sum_{l m}
\int d\tau d\xi \left[ -\frac{\rho^{2}}{\sqrt{A \, B}}
(\pa_{\tau}\varphi_{_{lm}})^2 \right. \\
%%%%%%%%
& & + \left. \sqrt{A B} \rho^{2} ~(\pa_{\xi}\varphi_{_{lm}})^2
+~ \left[ l(l + 1) + m^2 \right] \sqrt{\frac{A}{B}} \,
\varphi_{_{lm}}^2 \right] \, . \nn
\eea
where we have decomposed $\varphi$ in terms of the real spherical
harmonics ($Z_{lm}(\th, \f)$):
\be \la{tens-sph}
\varphi (x^{\mu}) = \sum_{l m} \varphi_{_{lm}}(\tau,\xi) Z_{l m} (\th, \f) \, .
\ee
Following the standard rules, the canonical momenta and Hamiltonian of
the field are given by
\bea
\la{eq:mom}
{\Pi}_{_{lm}}&=& \frac{\pa \cal{L}}{\pa(\pa_{\tau} \varphi_{lm})} =
\frac{\rho^{2}}{\sqrt{A \, B}} \, \pa_{\tau} \varphi_{_{lm}} \, ,\\
%%%%
\la{eq:gen-Ham}
H_{lm}(\tau) &=& \!\! \frac{1}{2} \int_{\tau}^{\infty} \!\!\!\!\!\!
d\xi \! \le[\! \frac{\sqrt{A B}}{\rho^{2}} \Pi_{_{lm}}^2
+ \sqrt{A B} \, \rho^{2} (\pa_{\xi} \varphi_{_{lm}})^2 \ri. \\
%%%%
& +& \le.(l(l + 1)  + m^2)\sqrt{\frac{A}{B}} \, \varphi_{_{lm}}^2 \ri] \, ,
\qquad H = \sum_{lm} H_{lm} \, . \nn
\eea
The canonical variables $(\varphi_{_{lm}}, \Pi_{_{lm}})$ satisfy the
Poisson brackets
\bea
\label{eq:gen-PB}
& & \{\varphi_{_{lm}}(\tau,\xi), \Pi_{_{lm}}(\tau,\xi')\} = \delta(\xi - \xi') \\
%%%%
& & \{\varphi_{_{lm}}(\tau,\xi), \varphi_{_{lm}}(\tau,\xi')\} = 0 =
\{\Pi_{_{lm}}(\tau,\xi), \Pi_{_{lm}}(\tau,\xi')\} \, .\nn
\eea
When discretized in the radial direction, the above Hamiltonian \eqref{eq:gen-Ham} reduces to \eqref{eq:1DHami-Mink}.

%\bibliography{ref,refshanki}

\end{document}